\documentclass{aa}  

\usepackage{graphicx}
\usepackage{txfonts}
\usepackage{color}

\begin{document}

   \title{Detailed abundances from integrated-light spectroscopy: Milky Way globular clusters\thanks{
Based on observations collected at the European Organisation for Astronomical Research in the Southern Hemisphere 
under ESO programme(s) 095.B-0677(A). } 
   }


   \author{S. S. Larsen
          \inst{1}
          \and
          J. P. Brodie\inst{2}
          \and
          J. Strader\inst{3}
   }

   \institute{Department of Astrophysics/IMAPP,
              Radboud University, PO Box 9010, 6500 GL Nijmegen, The Netherlands\\
              \email{s.larsen@astro.ru.nl}
         \and
             UCO/Lick Observatory, University of California, Santa Cruz, CA 95064, USA
         \and
            Department of Physics and Astronomy, Michigan State University, East Lansing, Michigan 48824, USA
             }

   \date{Received November 25, 2016; accepted February 21, 2017}

 
  \abstract
   {Integrated-light spectroscopy at high spectral resolution is rapidly maturing as a powerful way to measure detailed chemical abundances for extragalactic globular clusters (GCs).}
   {We test the performance of our analysis technique for integrated-light spectra by applying it to seven well-studied Galactic GCs that span a wide range of metallicities.}
   {Integrated-light spectra were obtained by scanning the slit of the UVES spectrograph on the ESO \emph{Very Large Telescope} across the half-light diameters of the clusters. We modelled the spectra using resolved \emph{Hubble Space Telescope} colour-magnitude diagrams (CMDs), as well as theoretical isochrones, in combination with standard stellar atmosphere and spectral synthesis codes. The abundances of Fe, Na, Mg, Ca, Ti, Cr, and Ba were compared with literature data for individual stars in the clusters.}
   {The typical differences between iron abundances derived from our integrated-light spectra and those compiled from the literature are less than $\sim0.1$ dex.  A larger difference is found for one cluster (NGC~6752), and is most likely caused primarily by stochastic fluctuations in the numbers of bright red giants within the scanned area. As expected, the $\alpha$-elements (Ca, Ti) are enhanced by about 0.3 dex compared to the Solar-scaled composition, while the $\mathrm{[Cr/Fe]}$ ratios are close to Solar. When using up-to-date line lists, our $\mathrm{[Mg/Fe]}$ ratios also agree well with literature data. Our $\mathrm{[Na/Fe]}$ ratios are, on average, 0.08--0.14 dex lower than average values quoted in the literature, and our $\mathrm{[Ba/Fe]}$ ratios may be overestimated by 0.20--0.35 dex at the lowest metallicities. 
We find that analyses based on theoretical isochrones give very similar results to those based on resolved CMDs.
}
{Overall, the agreement between our integrated-light abundance measurements and the literature data is satisfactory.  Refinements of the modelling procedure, such as corrections for stellar evolutionary and non-LTE effects, might further reduce some of the remaining offsets. 
}


\maketitle
%

\section{Introduction}

A typical globular cluster (GC) has an integrated absolute magnitude of $M_V\approx-7.5$ \citep[e.g.][]{Harris1991}, about 5 magnitudes brighter than an individual star at the tip of the red giant branch (RGB). For a given apparent magnitude limit, such a GC is thus observable at a 10 times greater distance compared to the brightest individual RGB stars, which corresponds to a boost of a factor of 1000 in the accessible volume of space. It is therefore not surprising that there has long been a strong interest in utilising GCs as tracers of stellar populations in external galaxies, where the main alternative (at least for early-type galaxies) is observing the integrated diffuse light. The latter approach, however, faces the significant challenge of deconstructing a potentially complex mix of stellar populations with different ages, metallicities, and kinematics, a challenge which is much more easily overcome with GCs. A particularly useful application of GCs is to use them as tracers of the (metal-poor) halos, which account for only a small fraction of the total stellar mass (and luminosity) in most galaxies, but often have large numbers of GCs associated with them. 
GCs have, indeed, been shown to trace coherent structures in phase-space that are likely related to hierarchical galaxy build-up \citep[e.g. in the halos of M31 and M87;][]{Mackey2010,Romanowsky2012}.  
While photometry of the brightest individual halo giants can currently be obtained out to distances of $\sim10$ Mpc with the \emph{Hubble Space Telescope} \citep{Harris2007,Peacock2015}, and the \emph{James Webb Space Telescope} will soon push the boundary even further, detailed spectroscopy of individual RGB stars at such distances will remain beyond the capabilities even of future 30--40 m telescopes. Measurements of ages and chemical composition for globular clusters thus hold great potential for studying the assembly histories of galaxies, especially when combined with other information such as kinematics and spatial distributions of the GCs and/or resolved imaging of individual stars.

The advent of efficient multi-object spectrographs on 8--10 m telescopes made it possible to carry out systematic spectroscopic studies of extragalactic GCs as far away as the Virgo and Fornax galaxy clusters \citep[see review in][]{Brodie2006}, although investigations of smaller samples started much earlier \citep[e.g.,][]{VandenBergh1969,Racine1978,Brodie1991}. The analysis of such observations has traditionally relied on methods that are similar to those employed in the study of integrated galaxy light, typically involving measurements of line strength indices on relatively low resolution spectra. Physical properties, such as ages, metallicities, and even some information on individual abundances (e.g., the level of alpha-element enrichment), can then be inferred either via empirical calibrations \citep{Brodie1990,Strader2004,Caldwell2011} or by comparison with simple stellar population (SSP) models \citep{Proctor2004,Puzia2005,Graves2008}. This type of work has shown that GC populations around most external galaxies resemble the Milky Way GC system by being predominantly old ($\ga$ 10 Gyr) and enhanced in the $\alpha$-elements, with metallicities spanning a wide range from $\mathrm{[Fe/H]}\approx-2$ to about solar  \citep{Beasley2008,Puzia2005,Strader2005,Cenarro2007}.

While studies such as those outlined above have provided important insights into the nature of extragalactic GC populations, a large gap remains between the relatively crude constraints on chemical composition that are possible from line index measurements and the wealth of detail that can be extracted from high-dispersion spectroscopy of individual stars \citep{McWilliam1997,Venn2004,Ishigaki2013,Mikolaitis2014,Roederer2014}. Because of the modest internal velocity  dispersions of GCs, it is potentially possible to harvest far more information from their integrated-light spectra  
than can be provided by the classical techniques designed for spectroscopy of galaxy light. 
However, the analysis must still account for the fact that the spectra are of a composite nature with significant contributions from stars across the Hertzsprung-Russell diagram. 

Several recent studies have used high-dispersion, integrated-light spectra to perform detailed chemical abundance analysis of GCs. \citet{McWilliam2008} found that abundances derived from an integrated-light spectrum of the Galactic GC NGC~104 (47 Tuc) agreed with data for individual stars in the cluster within typically $\sim0.1$ dex, although larger differences were found for some elements (e.g., Mg, see below). Subsequent comparisons involving larger samples (5--11) of Galactic GCs have found similar results \citep{Sakari2013,Sakari2014,Colucci2017}.
 Integrated-light spectra have now also been used to measure abundances  for GCs in several external galaxies, including the Large Magellanic Cloud, M31, and NGC~5128 \citep{Colucci2009,Colucci2012,Colucci2013,Colucci2014,Sakari2015} and the Fornax and Wolf-Lundmark-Melotte dwarf galaxies  \citep[][hereafter L12 and L14]{Larsen2012a,Larsen2014}. These studies have found abundance patterns that mostly match those seen in Milky Way GCs and halo stars, with similar Fe-peak element abundances ($\mathrm{[Cr/Fe]}\approx0$, $\mathrm{[Sc/Fe]}$ slightly super-solar) and generally super-solar $\mathrm{[\alpha/Fe]}$ ratios. However, potentially interesting differences have also emerged. Several of the above studies have found Mg to be depleted with respect to other $\alpha$-elements (such as Ca and Ti), in some cases even reaching sub-solar values of $\mathrm{[Mg/Fe]}$ \citep{Colucci2014}. At the same time,  $\mathrm{[Na/Fe]}$ is commonly found to be elevated compared to the abundances observed in Galactic halo stars of similar metallicity \citep[L14,][]{Colucci2014,Sakari2015}.  These patterns are reminiscent of the Na/O and Mg/Al anti-correlations observed in Galactic GCs \citep{Cohen1978,Shetrone1996,Sneden1997,Gratton2004,Carretta2009,Gratton2012}, and may simply reflect the prevalence of the ``multiple populations'' phenomenon also in these extragalactic clusters. 

While $[\mathrm{Na/Fe}]$ spreads up to $\sim0.5$ dex or more are common in Galactic GCs, the range in $\mathrm{[Mg/Fe]}$ is usually quite small and very few stars reach sub-solar $\mathrm{[Mg/Fe]}$ values in most GCs \citep{Carretta2009}. It is therefore not clear that integrated-light Mg abundances are expected to be strongly affected by the presence of multiple populations (unless much larger Mg variations are common in the extragalactic clusters studied so far). This raises the question whether the integrated-light Mg abundances may be subject to unknown systematic differences compared to observations of individual stars. Indeed, \citet{Colucci2017} found their integrated-light $\mathrm{[Mg/Fe]}$ ratios for eleven Galactic GCs to be about 0.3 dex lower on average compared to measurements of individual stars in the same clusters. On the other hand, \citet{Sakari2013} found integrated-light $\mathrm{[Mg/Fe]}$ ratios in good agreement with the average of literature data for individual stars in five GCs. Clearly, obtaining a better understanding of these differences would be desirable.
 
Our technique for abundance analysis from integrated light was introduced in our study of the GCs in the Fornax dwarf galaxy (L12). 
Compared to the techniques for integrated-light abundance measurements that have been developed and tested by other groups \citep{McWilliam2008,Colucci2012,Sakari2013}, our approach is more heavily based on spectral synthesis and full spectral fitting. For elements with many lines (such as Fe, Ti, Ca) we generally obtain our abundance measurements by fitting relatively broad spectral ranges that contain multiple features, rather than by measuring single lines individually. In principle, we can fit for abundances of multiple elements simultaneously, and the full spectral fitting approach automatically accounts for line blending and hyperfine structure (provided, of course, that adequate input line lists are used). Conceptually, as well as in its practical implementation, our approach thus differs sufficiently from those of other groups that separate testing is warranted.

In L12 we carried out a number of basic comparisons of our integrated-light abundance measurements for the GC Fornax 3 with data for metal-poor stars in dwarf galaxies and in the cluster itself. However, a more stringent test is to compare abundances derived from integrated-light observations with measurements of individual stars in well-studied Galactic globular clusters. This is particularly important at higher metallicities, where some aspects of the analysis (e.g., line blending and continuum placement) become more challenging. In this paper we present new integrated-light spectroscopy of seven Galactic GCs, selected to span nearly two decades in metallicity from $\mathrm{[Fe/H]}\approx-2.3$ (NGC~7078, NGC~7099) to $\mathrm{[Fe/H]}\approx-0.4$ (NGC~6388). By comparing our integrated-light abundance determinations with measurements of individual stellar abundances compiled from the literature, we test whether our procedure correctly recovers the overall metallicities, as well as individual abundance ratios. 


\section{Sample selection and observations}

\begin{table*}
\caption{Basic data for the observed clusters}             
\label{tab:gcs}      
\centering          
\begin{tabular}{l c c c c c c c c}     
\hline\hline       
  & NGC~104 & NGC~362 & NGC~6254 & NGC~6388 & NGC~6752 & NGC~7078 & NGC~7099 & Ref \\ 
\hline                    
$V$ (mag) & 3.95 & 6.40 & 6.60 & 6.72 & 5.40 & 6.20 & 7.19 & 1 \\
$r_h$ & 168\arcsec & 49\arcsec & 108\arcsec & 39\arcsec & 138\arcsec & 63\arcsec & 69\arcsec & 1 \\
$\mu_h$ (mag arcsec$^{-2}$) & 17.1 & 16.8 & 18.8 & 16.7 & 18.1 & 17.2 & 18.4 & \\
$D$ (kpc) & 4.7 & 8.6 & 5.0 & 9.9 & 4.0 & 10.3 & 8.1 & 1,4,5 \\
$E(B-V)$ (mag) & 0.04 & 0.05 & 0.28 & 0.37 & 0.04 & 0.10 & 0.03 & 1 \\
$M_V$ (mag) & $-9.5$ & $-8.4$ & $-7.8$ & $-9.4$ & $-7.7$ & $-9.2$ & $-7.4$ \\
$\mathrm{[Fe/H]}$ & $-0.77$ & $-1.17$ & $-1.58$ & $-0.44$ & $-1.56$ & $-2.32$ & $-2.34$ & 2,3 \\
Scan length & 336\arcsec & 98\arcsec & 216\arcsec & 78\arcsec & 138\arcsec & 126\arcsec & 100\arcsec & \\
$T_\mathrm{exp}$ (s) & $2\times1500$ & $1\times1500$ & $8\times1800$ & $2\times1200$ & $4\times1800$ & $2\times1800$ & $4\times1800$ & \\
$M_V$(scan) & $-6.8$ & $-6.2$ & $-6.3$ & $-7.7$ & $-5.4$ & $-7.2$ & $-5.8$ \\
$V_\mathrm{helio}$ (km s$^{-1}$) & $-16.7\pm0.1$ & $224.6\pm0.1$ & $75.4\pm0.1$ & $83.2\pm0.1$ & $-28.2\pm0.2$ & $-105.0\pm0.2$ & $-184.3\pm0.1$ & 6 \\
$\sigma_\mathrm{1D}$ (km s$^{-1}$) &  $11.9\pm0.1$ & $7.9\pm0.3$ & $6.7\pm0.2$ & $17.8\pm0.3$ & $8.2\pm0.1$ & $12.9\pm0.3$ & $5.4\pm0.1$ & 6 \\
\hline                  
\end{tabular}
\tablefoot{$V$ = apparent visual magnitude, $r_h$ = half-light radius, $\mu_h =$ mean visual surface brightness within $r_h$, $D=$ distance. $M_V$(scan) is the estimated luminosity of the area covered by the scans, $V_\mathrm{helio}$ is the heliocentric radial velocity derived from our observations, and $\sigma_\mathrm{1D}$ is the line-of-sight velocity dispersion.
}
\tablebib{(1)~\citet{Harris1996}, 2010 edition;
(2) \citet{Carretta2009c}; 
(3) \citet{Carretta2013}; 
(4) \citet{Woodley2012};
(5) \citet{VandenBosch2006}; 
(6) This work. See text for details.
}
\end{table*}

   \begin{figure}
   
   \noindent \begin{minipage}{44mm}
   \noindent NGC 104 (47 Tuc) \\
   \includegraphics[width=44mm]{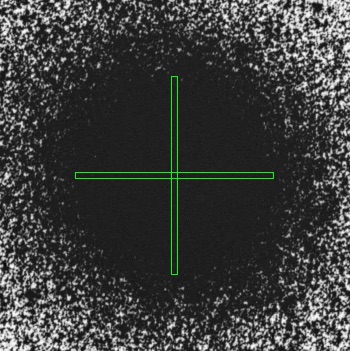}
  \end{minipage}
  \begin{minipage}{44mm}
  \noindent NGC 362 \\
   \includegraphics[width=44mm]{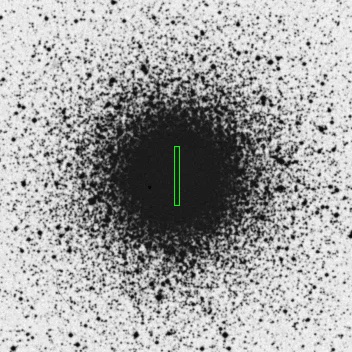}
  \end{minipage}
   \noindent \begin{minipage}{44mm}
   \noindent NGC 6254 (M10) \\
   \includegraphics[width=44mm]{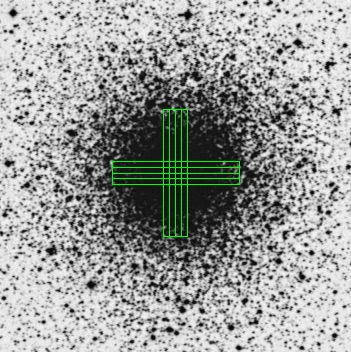}
  \end{minipage}
  \begin{minipage}{44mm}
  \noindent NGC 6388 \\
   \includegraphics[width=44mm]{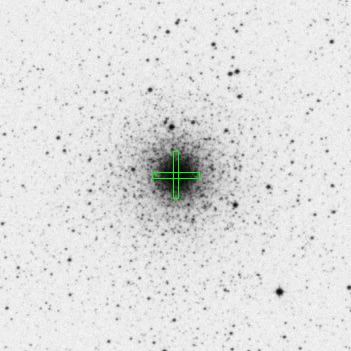}
  \end{minipage}
   \noindent \begin{minipage}{44mm}
   \noindent NGC 6752 \\
   \includegraphics[width=44mm]{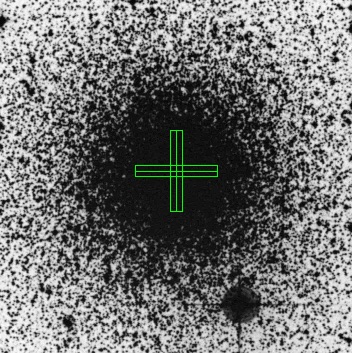}
  \end{minipage}
  \begin{minipage}{44mm}
  \noindent NGC 7078 (M15) \\
   \includegraphics[width=44mm]{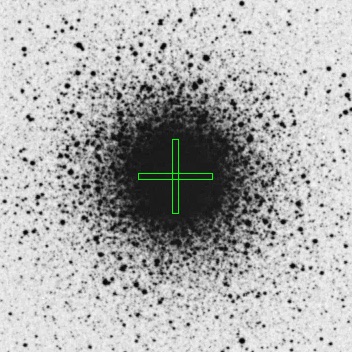}
  \end{minipage}
   \noindent \begin{minipage}{44mm}
   \noindent NGC 7099 (M30) \\
   \includegraphics[width=44mm]{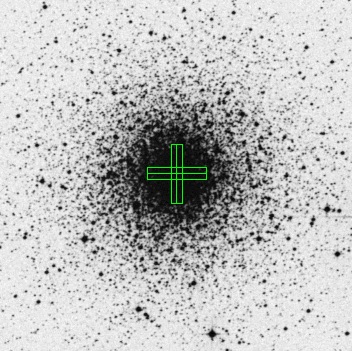}
  \end{minipage}
   \caption{Digitized Sky Survey images of the globular clusters observed in this work. The regions covered by our drift-scan observations are indicated. The size of each panel is $10\arcmin\times10\arcmin$.
              }
         \label{fig:gcscans}
   \end{figure}

For the selection of our sample we used the list of clusters observed by \citet{Carretta2009} as a starting point. It includes 19 Galactic GCs, for which the abundances of several light elements (including Na and Mg) were measured for individual stars in the clusters. As in our previous work, integrated-light spectra were obtained with the UVES spectrograph \citep{Dekker2000} on the ESO Very Large Telescope by scanning the slit across the target clusters.
According to the UVES Exposure Time Calculator, the exposure times required to obtain good integrated-light spectra (with signal-to-noise ratios of S/N $>$ 150--200 per \AA) become excessively long (many hours) for surface brightnesses fainter than $\mu \approx 19$ mag arcsec$^{-2}$. We therefore calculated the mean surface brightness within the half-light radius ($\mu_h$) for each cluster in the \citet{Carretta2009} sample, using the apparent visual magnitudes and half-light radii from \citet{Harris1996}, and eliminated clusters with $\mu_h > 19$ mag arcsec$^{-2}$ from the list. Among the remaining clusters, we selected two at the metal-poor end of the range (NGC~7078, NGC~7099), two at the metal-rich end (NGC~104, NGC~6388), and two at intermediate metallicities (NGC~6254, NGC~6752). Among several back-up targets, we also included NGC~362 \citep{Carretta2013}.

The clusters were observed on July 22/23 and July 23/24, 2015. We used the standard Cross Disperser \#3 setting centred at 520 nm with a slit width of 1\farcs0, which yields a resolving power of $R\sim40\,000$ over the wavelength range 4200 \AA\ -- 6200 \AA.  To sample the integrated light of the clusters we employed the same basic drift-scan technique as in our previous observations of the Fornax GCs, whereby the UVES slit was scanned across the half-light diameter of the clusters during the exposures.  Each science exposure was bracketed by two sky exposures with half the exposure time, which were co-added and used for sky subtraction.
Thin cirrus clouds were present throughout the run but did not significantly affect the observations, except towards the end of the second night. The seeing varied between 1\arcsec\ and 2\arcsec,  of little consequence for our observations. 

The large apparent sizes of Galactic GCs on the sky, combined with the relatively short UVES slit (10\arcsec) pose special challenges for this type of observations. For a given total exposure time, the highest S/N ratio would be achieved by scanning only the cluster cores, where the surface brightness is highest. However, this has to be balanced against other considerations. The stellar mass function might be affected by mass segregation in the cores, leading to possible systematic effects in the integrated light, and to minimise stochastic fluctuations in the number of bright RGB stars it is desirable to include the largest possible fraction of the total cluster light. A further practical limitation is that the total area covered by the scans is given by the product of the exposure time, the slit length, and the differential tracking rate.  For large, high surface brightness clusters (such as NGC 104), where only relatively short exposures are necessary, a very large differential tracking rate would be required in order to cover a significant fraction of the cluster. Each observation would then need to be split up into a large number of short exposures, which would result in a substantially reduced observing efficiency and a more dominant contribution of detector read noise. For these reasons, we did not use integrations shorter than 20--30 min.

Table~\ref{tab:gcs} lists basic data for the seven clusters in our sample, along with the exposure times and number of scans obtained for each cluster. In general, we scanned the clusters in both the north-south and east-west directions and for the lower surface brightness targets we obtained multiple parallel scans in each direction. In some cases, slight departures from our standard strategy were adopted: for NGC~6752, the large differential tracking rate required for a scan across the full half-light diameter made guide star acquisition problematic, so the scan length was reduced by a factor of two. For NGC~7099, the exposure time had to be reduced from the planned $4\times2500$ s to $4\times1800$ s because of time constraints, and the scan length was therefore decreased slightly in order to preserve S/N. The north-south scans of NGC~7099 were affected by the thickening cloud cover towards the end of the second night, but still proved useful. For NGC~362, the available time only allowed for a single scan. 

The regions covered by the scans are indicated on Figure~\ref{fig:gcscans}, which shows images of the clusters from the \emph{Digitized Sky Survey}. In Table~\ref{tab:gcs} we also list the approximate integrated absolute magnitudes of the areas covered by the scans, $M_V$(scan). These were estimated by selecting all stars from the ACS Survey of Galactic Globular Clusters \citep[ACSGCS;][]{Sarajedini2007} within the approximate areas covered by the scans and adding up the luminosities of these stars. The $M_V$(scan) magnitudes should only be treated as rough estimates as we have not attempted to correct the ACSGCS photometry for incompleteness, and there is some uncertainty in determining which stars exactly fall within the scanned areas. Nevertheless, comparison of the $M_V$(scan) values with the total integrated magnitudes shows that our observations include between 8\% (for NGC~104) and 25\% (NGC~6254) of the integrated light, a significantly smaller fraction than would typically be included for extragalactic GCs (as is also evident from Fig.~\ref{fig:gcscans}). 
   
 \section{Data reduction}  
  
   \begin{figure}
   \centering
   \includegraphics[width=\columnwidth]{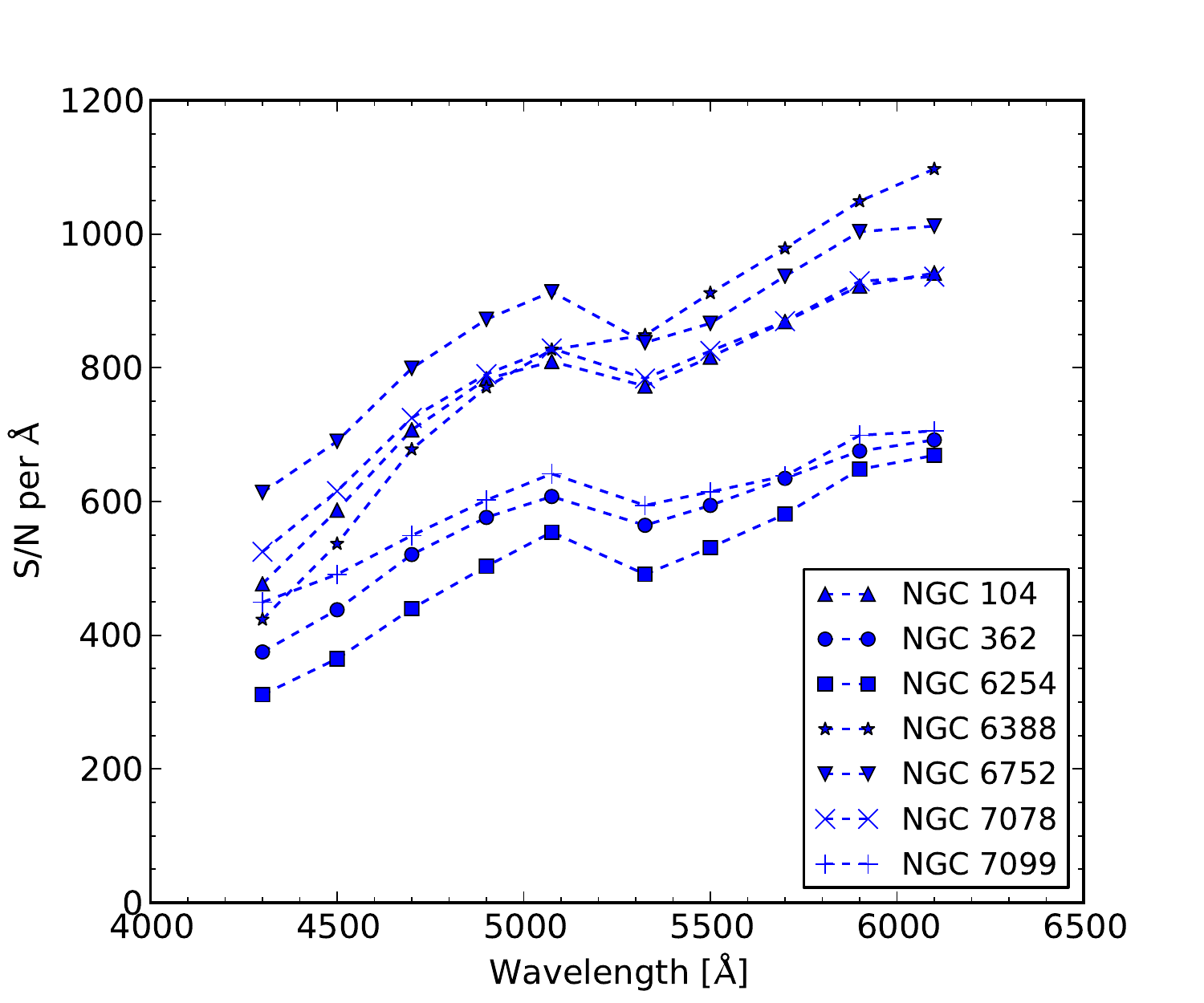}
      \caption{Signal-to-noise (per \AA ) for the final reduced one-dimensional integrated-light spectra.}
         \label{fig:s2n}
   \end{figure}

 The initial processing of the data (following interpolation over a few bad columns with the \texttt{fixpix} task in \texttt{IRAF}) was done with version 5.5.7 of the UVES pipeline, executed within the \texttt{Esorex} environment.
 The pipeline processing involved bias subtraction, flat-fielding, and wavelength calibration. It also included merging of the echelle orders to a single two-dimensional spectrum for each of the two CCD detectors, with the UVES slit mapped onto 22 pixels in the spatial direction.
 
The two sky exposures bracketing each science exposure were median filtered in the spatial direction to eliminate cosmic ray hits and were then co-added and subtracted from the science exposure. The 22 lines along the spatial direction were then extracted from each of the sky-subtracted 2-D spectra of a given cluster and co-added to a single one-dimensional spectrum, using a custom-made programme that rejected bad pixels by means of a sigma-clipping algorithm. The S/N ratios of the final spectra were estimated based on the variance of the individual co-added pixels at each wavelength sampling point.

Fig.~\ref{fig:s2n} shows the S/N ratio for each cluster as a function of wavelength. The S/N ratios (per \AA) range from 300--600 at the blue end to 650--1100 at the red end.
The jump at 5200 \AA\ occurs because the blue and red parts of the spectra are recorded on two separate CCD detectors. 

\section{Analysis}

\subsection{General overview} 

Here we provide only a brief summary of our technique for abundance analysis from integrated-light spectra; the reader is referred to L12 and L14 for more details. In essence, chemical abundances are determined by fitting synthetic model spectra to the observed spectra, iteratively adjusting the abundances of individual elements until the best match to the data is obtained. The model spectra are computed by dividing the Hertzsprung-Russell diagrams (HRD) of the underlying stellar populations into a large number of bins (typically about 100), computing a model atmosphere and a synthetic spectrum for each bin, and then co-adding the spectra for each bin with appropriate weights (given by the number of stars in each bin). The model spectra are then smoothed to match the resolution of the observed spectra, a scaling is applied, and the $\chi^2$ of the fit is evaluated. 

In our previous work we have used the \texttt{ATLAS9} and \texttt{SYNTHE} codes written by R.\ Kurucz to compute the model atmospheres and synthetic spectra \citep{Kurucz1970,Kurucz1979,Kurucz1981,Kurucz2005,Sbordone2004}. However, these models are less suitable for the coolest giants, which are present in the more metal-rich clusters, and for stars with $T_\mathrm{eff}< 4000\,\mathrm{K}$ we instead use \texttt{MARCS} atmospheres \citep{Gustafsson2008} and the \texttt{TurboSpectrum} code \citep{Alvarez1998,Plez2012} to compute the synthetic spectra. This allows for spherical (rather than plane-parallel) symmetry of the atmospheres. We downloaded the full grid of spherical model atmospheres from the \texttt{MARCS} website\footnote{\texttt{http://marcs.astro.uu.se}} and selected the closest model from the grid for each bin of the HRD.
Our reference abundance scale remains that of \citet{Grevesse1998}.

The input HRDs can be based on empirical data (e.g., observed colour-magnitude diagrams, CMDs), on theoretical isochrones, or on a combination of both. When using theoretical isochrones, the stellar parameters ($\log g$, $T_\mathrm{eff}$, luminosity) can usually be taken directly from the isochrone tables, although assumptions need to be made about age and the appropriate weights (e.g., one needs to assume a distribution of stellar masses). When using empirical CMDs, stellar parameters can be derived from photometry, once a set of transformations from observables to physical properties have been adopted. For extragalactic GCs it is usually not feasible to obtain CMDs of sufficient depth to rely on a purely empirical modelling of the HRDs, although observations can still provide useful constraints on, e.g., horizontal branch morphology. In our study of the WLM GC, we found that purely isochrone-based modelling gave very similar results to models involving empirical CMDs (L14). We will revisit this point in more detail below.

\subsection{Modelling of the Hertzsprung-Russell diagrams}
\label{sec:hrmod}

\begin{figure}
   \includegraphics[width=\columnwidth]{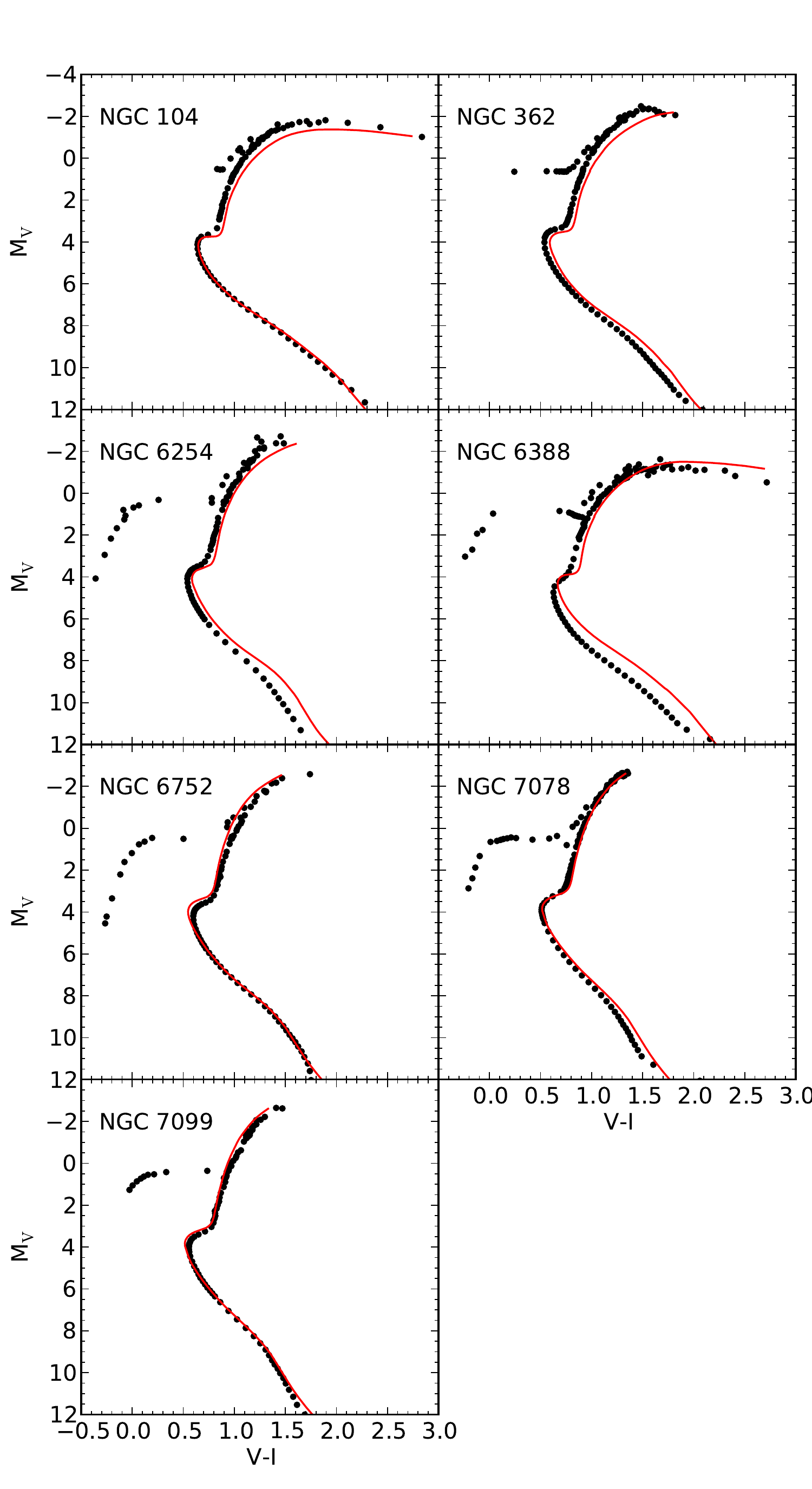}
      \caption{Binned colour-magnitude diagrams (black dots) and theoretical isochrones \citep[red lines]{Dotter2007}. The isochrones are those corresponding to the best-fitting metallicities in Table~\ref{tab:abun3} (see also Table~\ref{tab:abdiff}).
              }
         \label{fig:cmds}
\end{figure}

In this work we based the HRD modelling on photometry from the Advanced Camera for Surveys (ACS) on board the \emph{Hubble Space Telescope}, obtained as part of the ACSGCS \citep{Sarajedini2007,Anderson2008}. 
Following the same procedure as in our previous work, the colour-magnitude diagrams were arranged into about 100 bins.  For each bin, the median $V-I$ colour and $V$ magnitude were recorded, along with the number of stars in the bin. Most of the bins were distributed along the red giant branch and main sequence (MS), but we also included bins along the horizontal branch (HB) and the asymptotic giant branch (AGB). For most clusters, the very tip of the RGB is relatively sparsely sampled and the brightest 10--20 RGB stars were then included individually without any binning. 
In some clusters, sequences of blue stragglers (BS) were evident, but these were not included in our general modelling procedure. However, the effect of including BS stars was investigated and found to be very minor (Sec.~\ref{sec:imfeh}). 

The photometry was corrected for foreground dust extinction using $E(B-V)$ values from the 2010 edition of the \citet{Harris1996} catalogue (Table~\ref{tab:gcs}).  Most of the distances were also taken from that source. For NGC~104 we assumed $D=4.7$ kpc \citep{Woodley2012}, which is quite similar to the 4.5 kpc in the Harris catalogue. For NGC~7078 we used $D=10.3$ kpc \citep{VandenBosch2006}, again very close to the distance in the Harris catalogue (10.4 kpc). 
For NGC~6254, the distance of 4.4 kpc listed in the Harris catalogue led to a poor match to isochrones and the CMD of NGC~6752 (which has a similar age and metallicity), whereas an increase of about 0.3 mag in the distance modulus resulted in a much better agreement.  We therefore assumed a distance of 5.0 kpc for NGC~6254. 
The adopted distances are listed in Table~\ref{tab:gcs} and the binned CMDs are shown in Fig.~\ref{fig:cmds}.

The case of NGC~6388 warrants some remarks. The cluster is subject to substantial differential reddening across the ACS field of view \citep{Busso2007}, which is clearly noticeable as a ``blurring'' of the ACSGCS CMD. We did not attempt to correct the photometry for this effect but expect that its consequences are relatively limited for our purpose, since we only use the median colours and magnitudes in each CMD-bin.  In addition, NGC~6388 has a rather peculiar HB morphology for its metallicity, with an extended blue tail and a strongly sloped red stub. These features have been extensively discussed in the literature and are not fully understood, but may be at least partly due to variations in He abundance \citep{Rich1997,Busso2007,Bellini2013}.

For the brightest bins, including those along the HB, weights were assigned simply by using the number of stars present in these bins in the ACS CMDs. At fainter magnitudes ($M_V > +1$), we relied on theoretical predictions for the luminosity functions, based on alpha-enhanced isochrones from the Dartmouth group \citep{Dotter2007}. For these we assumed the metallicities in Table~\ref{tab:gcs} and ages of 13 Gyr, except for NGC~104 and NGC~362 where slightly younger ages of 11 Gyr were assumed \citep{Bellazzini2001,Grundahl2002a,Gratton2003}.
The theoretical LFs were normalised to give the same number of stars as the empirical CMDs at the bright end. The distribution of stellar masses was assumed to follow the \citet{Salpeter1955} law ($\mathrm{d}N/\mathrm{d}M \propto M^{-2.35}$). The choice of mass function hardly affects the LFs above the MS turn-off, since the post-MS phases contain stars spanning only a small range in (initial) mass and the LFs are therefore determined mainly by stellar evolution (i.e., by how much time a star spends at a given location in the HRD).
For stars with masses less than $0.4 M_\odot$--$0.5 M_\odot$ (corresponding to $M_V\ga+8$), the stellar mass function becomes more uncertain and may have been significantly modified from its initial value by dynamical evolution, which typically leads to depletion of the number of low-mass stars \citep{Spitzer1987}.
For a Salpeter mass function and an age of 13 Gyr, stars fainter than $M_V=+9$ contribute about 4\%--5\% of the total $V$-band light, but this is almost certainly an overestimate of the actual contribution since even the initial slope of the MF was likely significantly shallower at these low masses \citep{Bastian2010}. Following L12, we simply leave out stars fainter than $M_V=+9$, noting that their effect on the overall metallicities, as well as on individual abundance ratios, was found to be small ($<0.1$ dex) by L12 even if the MF was assumed to follow the Salpeter law down to 0.1 $M_\odot$.
  
The last step was to convert the photometry to physical parameters. Again, we mostly followed the approach described in L12, where temperatures and bolometric corrections were computed from the photometry, using transformations based on Kurucz models \citep{Castelli1999}. To calculate the surface gravities, stellar masses were obtained by interpolation in the same isochrones used for the theoretical LFs. For the most metal-rich clusters in our sample (NGC~104 and NGC~6388), the coolest giants fall outside the Castelli model grid and thus had to be dealt with separately. For these stars, we defined a  colour-$T_\mathrm{eff}$ relation based on the colours and temperatures tabulated in the Dartmouth isochrones, which use synthetic photometry based on PHOENIX model atmospheres \citep{Hauschildt1999,Hauschildt1999a}.

\subsection{Spectral fitting - standard analysis}
\label{sec:stdanal}

   \begin{figure*}
   \centering
   \includegraphics[width=19cm]{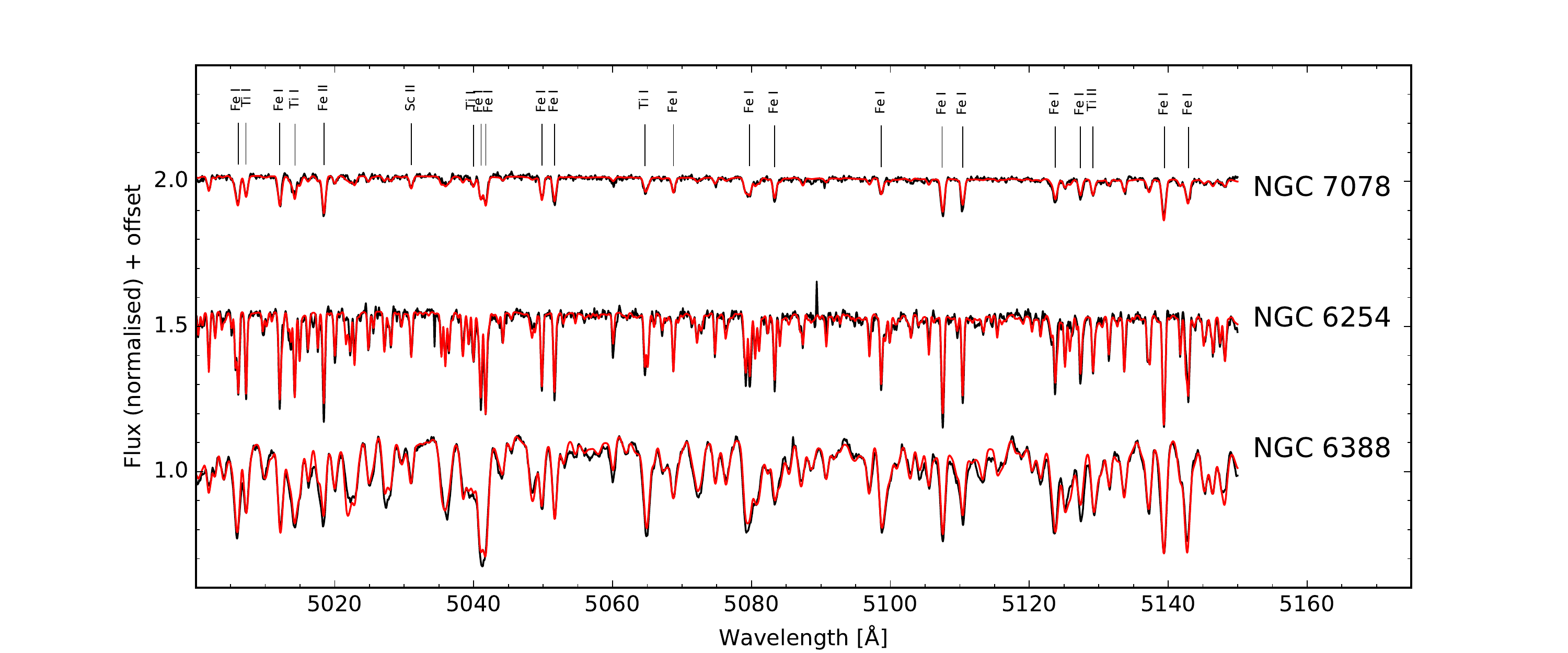}
      \caption{Observed (black lines) and best-fitting model spectra (red lines) for three GCs spanning the full range of metallicities.}
         \label{fig:speccmp}
   \end{figure*}
   
   \begin{figure}
   \centering
   \includegraphics[width=\columnwidth]{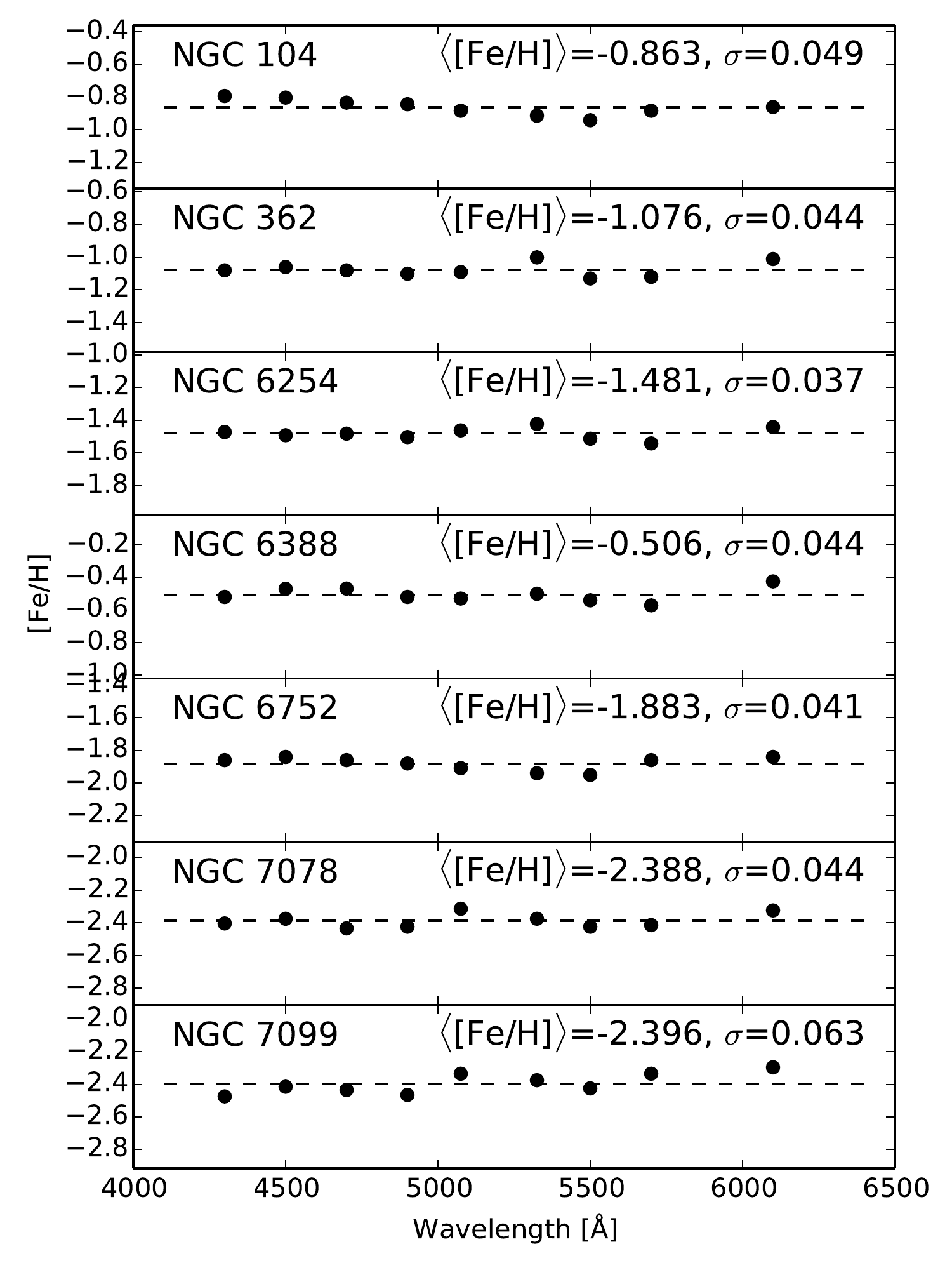}
      \caption{Iron abundance vs.\ wavelength. The horizontal dashed line in each panel indicates the average [Fe/H] value (also given in the legend).}
         \label{fig:lamfe}
   \end{figure}

The spectral fits were carried out mostly as described in L12 and L14.  Radial velocity shifts and the appropriate broadening of the model spectra required to match the observations were determined in an initial set of fits (using 200~\AA\ spectral windows between 4200 \AA\ and 6200 \AA) during which the overall scaling of all abundances was also allowed to vary. The iron abundances were then determined using the same 200 \AA\ windows and for other elements we used smaller spectral windows around specific features. In general, the spectral windows used were the same as in our previous work. 

The radial velocities (corrected to the heliocentric reference frame) and velocity broadenings are included in Table~\ref{tab:gcs}. 
The formal errors on the radial velocities (based on the bin-to-bin dispersions) are  0.1--0.2 km~s$^{-1}$, but this only reflects the uncertainties on the raw wavelength shifts required to match the observations to the synthetic spectra.
The mean difference between our radial velocities and those in the Harris catalogue is 0.9 km~s$^{-1}$ with a standard deviation of 1.5 km~s$^{-1}$. These differences exceed the formal errors on our measurements, as well as the internal errors of 0.1--0.8 km~s$^{-1}$ quoted in the Harris catalogue, but according to the notes accompanying the catalogue the true errors may be at least a factor of 2 higher. For our velocities, an additional source of uncertainty comes from variations in the internal UVES temperature of about $\sim1^\circ$ C during the run (according to the image headers), which can introduce shifts of $\sim0.5$ km~s$^{-1}$ in the wavelength scale \citep{DOdorico2000}. We do not explore these issues in further detail here, but note that the radial velocities in Table~\ref{tab:gcs} are probably accurate to within $\sim1$ km~s$^{-1}$.

For the velocity dispersions, the instrumental broadening, which corresponds to 3 km s$^{-1}$ (L12) has been subtracted in quadrature. The errors, once again, reflect only the bin-to-bin dispersions. We have not attempted to convert the velocity dispersions to central or global values, nor has any correction been made for other effects that may contribute to line broadening such as stellar rotation and/or macroturbulence \citep{Gray1982,Carney2008}. Nevertheless, our velocity dispersions generally agree with those in the Harris catalogue within about 1 km~s$^{-1}$, although NGC~6752 deviates more significantly from the central velocity dispersion of $\sigma_0=4.9\pm0.4$ km~s$^{-1}$ in the Harris catalogue.
However, recent studies have found larger velocity dispersions for NGC~6752. \citet{Kimmig2015} find $\sigma_0=6.1\pm0.2$ km~s$^{-1}$ while \citet{Lardo2015} find $\sigma_0=8.2$ km~s$^{-1}$, the latter value being identical to our integrated-light estimate. From HST proper motions, \citet{Drukier2003} find an even higher $\sigma_0 = 12$ km~s$^{-1}$ (which would, however, imply a very high mass-to-light ratio for a GC, $M/L_V \sim 9 \, M_\odot/L_{V,\odot}$).

While the UVES pipeline reduction in principle accounts for the blaze function, this does not work perfectly and a ``wavy'' structure on scales of 30--40 \AA\ (corresponding to the wavelength range of the echelle orders) was still evident in the reduced spectra. We found that these variations in the continuum level (at the level of 1\%--2\%) could be removed by using a spline function when scaling the observed spectra to match the models; for a 200 \AA\ range we found that typically 25 knots (i.e., corresponding to 26 spline segments) were required to adequately sample the variations.

Compared to our previous work, the modelling procedure was modified in a few relatively minor ways. For the modelling of CN and CH, we replaced the older Kurucz line lists for these molecules with the more recent line lists by \citet{Masseron2014} and \citet{Brooke2014}, which are also used by \texttt{TurboSpectrum}. With the new molecular line lists, we found that C abundances of $\mathrm{[C/Fe]}\sim-0.3$  yielded better fits in the blue, instead of $\mathrm{[C/Fe]}\sim-0.6$ for the old line lists. Once this was taken into account, the effect of using the new CH/CN line lists on other abundance ratios was negligible. We remind the reader that the interpretation of the carbon abundances in the integrated-light spectra is not straight forward, due to significant effects of extra mixing along the RGB \citep{Gratton2000,Martell2008}.
For the atomic lines we continue using the line list of \citet[][hereafter CH04]{Castelli2004} as our main source. We have included hyperfine splitting for some Mn and Sc lines from the line list at the Kurucz website. Hyperfine splitting is already included in the CH04 list for Ba, but we have adopted an oscillator strength of $\log gf = -0.15$ instead of $\log gf=-0.458$ for the 4934 \AA\ \ion{Ba}{II} line. The new value is taken from the VALD database \citep{Piskunov1995,Kupka1999} and results in much better agreement with the Ba abundances inferred from other lines. In addition, we have modified the Ba isotopic ratios according to the $r$-process mixture in \citet{McWilliam1998}, which may be more appropriate for GCs \citep{Straniero2014} than the default $s$-process dominated mixture in the CH04 list. This implies that a larger fraction of the Ba is in the form of $^{135}$Ba and $^{137}$Ba (in contrast to the $^{138}$Ba-dominated $s$-process mixture) and hyperfine structure thus becomes more important, leading to lower Ba abundances.

For stars fainter than the Sun ($M_V>4.8$) we assume a microturbulent velocity of $\xi=0.5$ km s$^{-1}$ \citep{Takeda2002}. For brighter stars we adopt the same relation as in L12, i.e., $\xi=1$ km s$^{-1}$ for $3.64 < M_V<4.8$, $\xi=2$ km s$^{-1}$ for $M_V<-1.81$, and linear interpolation in $\xi(M_V)$ between $M_V=3.64$ and $M_V=-1.81$. For horizontal branch stars we assume $\xi=1.8$ km s$^{-1}$.

Figure~\ref{fig:speccmp} shows example fits to the spectra of NGC~7078, NGC~6254, and NGC~6388, representing GCs of low, intermediate, and high metallicity, respectively. The increase in the strength of the spectral features with increasing metallicity is evident, and the fits are quite satisfactory over the full metallicity range. In Fig.~\ref{fig:lamfe} we show the individual [Fe/H] values derived for each 200 \AA\ bin as a function of wavelength. The dispersion around the mean values is 0.04--0.06 dex and there are no obvious trends with wavelength. 

Throughout this paper, we will refer to the analysis described above as the \emph{standard analysis}. To assess the sensitivity of our results to the details of the analysis, we also carried out a number of analyses where the standard analysis was modified in various ways.

\subsection{Spectral fitting - modified analyses}

\subsubsection{The effect of stochasticity}

\begin{figure}
\centering
\includegraphics[width=\columnwidth]{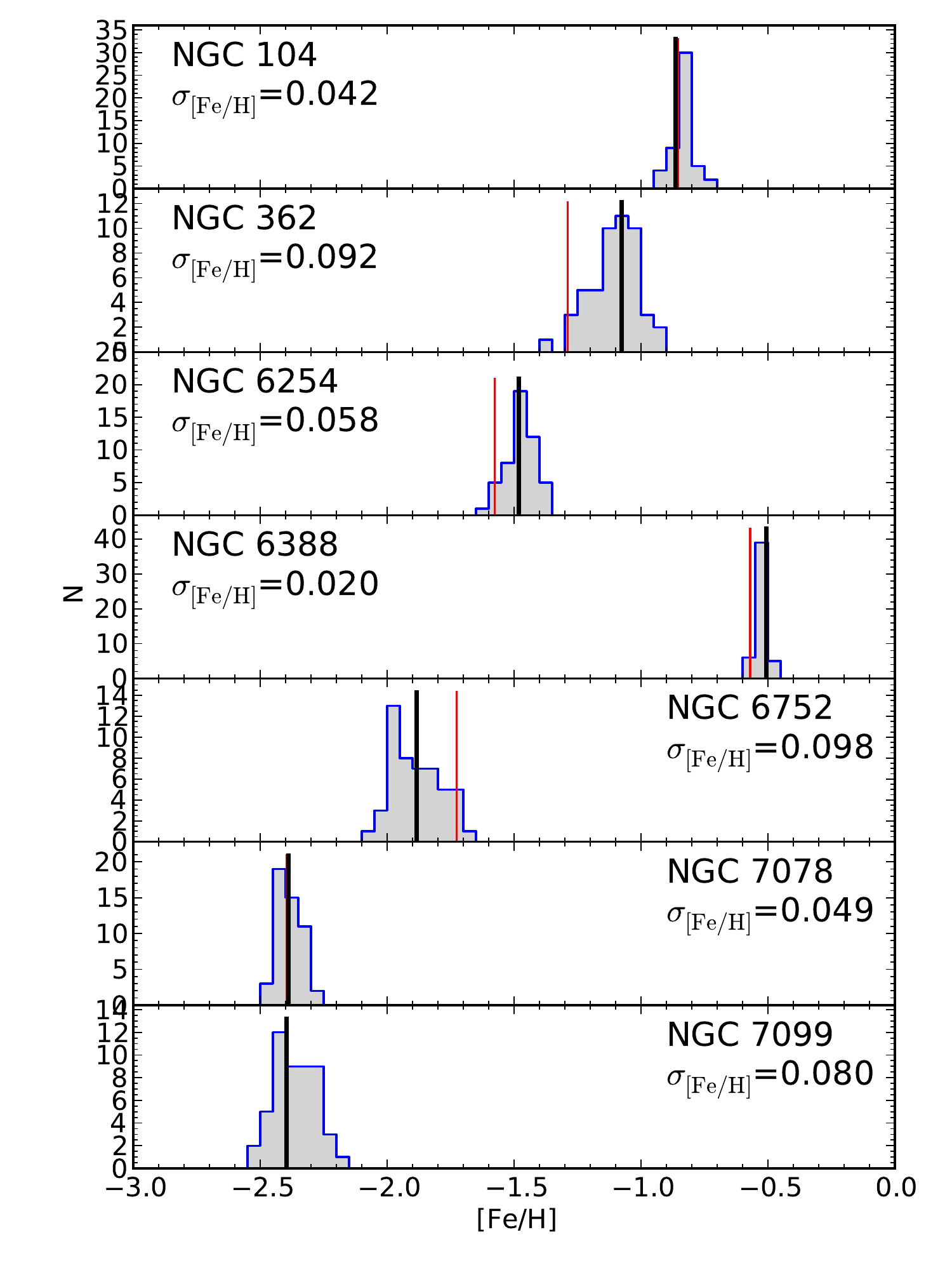}
\caption{Distributions of Fe abundances for 50 random sub-samples of the cluster CMDs. Each subsample contains a similar number of stars to the areas covered by the slit scans. Vertical thin (red) lines: Fe abundances based on stars located \emph{approximately} within the slit scan areas. Vertical thick (black) lines: Fe abundances based on all stars.}
\label{fig:stocfe}
\end{figure}

\begin{figure*}
\centering
\includegraphics[width=\textwidth]{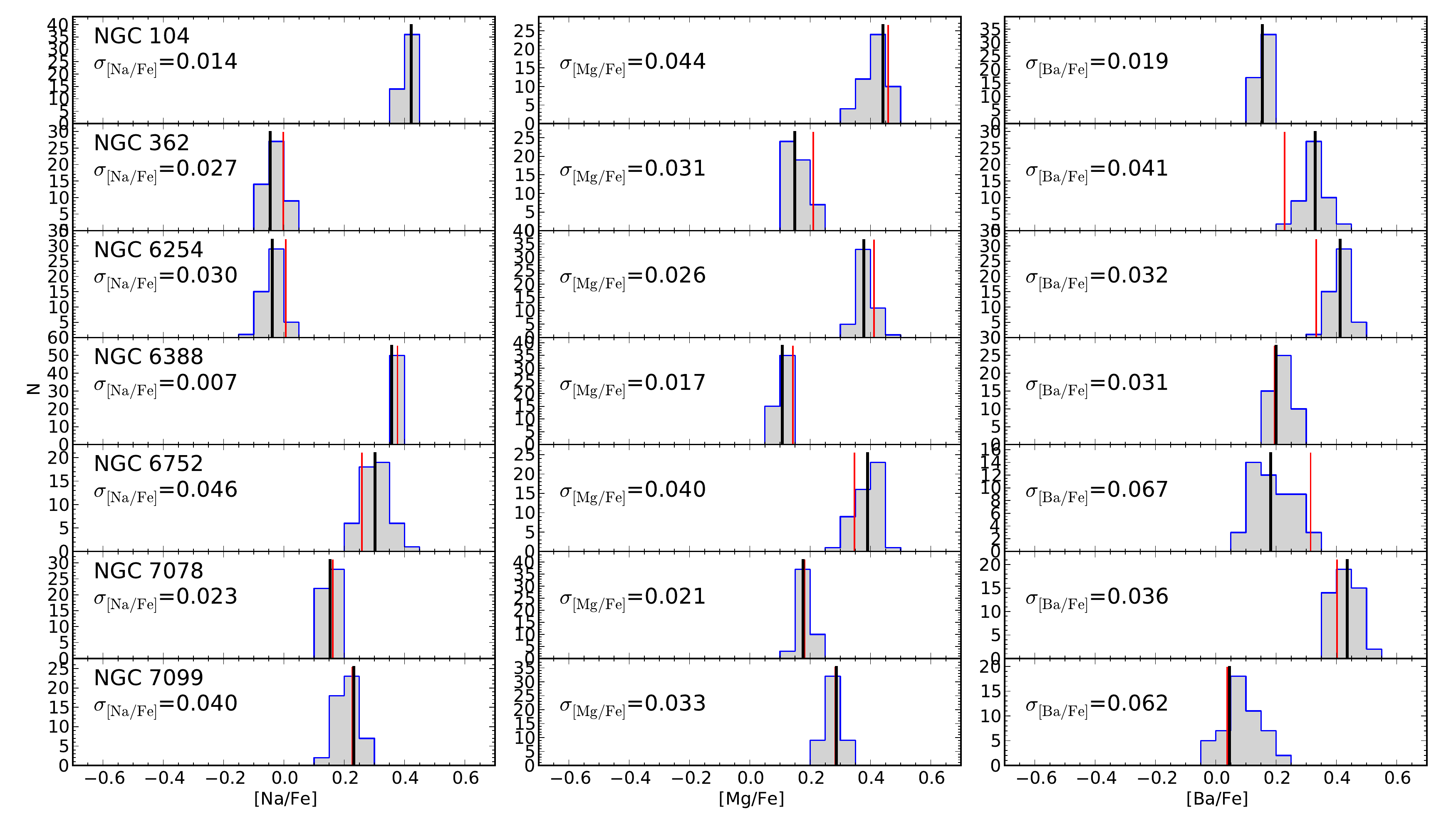}
\caption{As Fig.~\ref{fig:stocfe}, but for the $\mathrm{[Na/Fe]}$, $\mathrm{[Mg/Fe]}$, and $\mathrm{[Ba/Fe]}$ abundance ratios.}
\label{fig:stocxfe}
\end{figure*}

\begin{figure*}
\centering
\includegraphics[width=\textwidth]{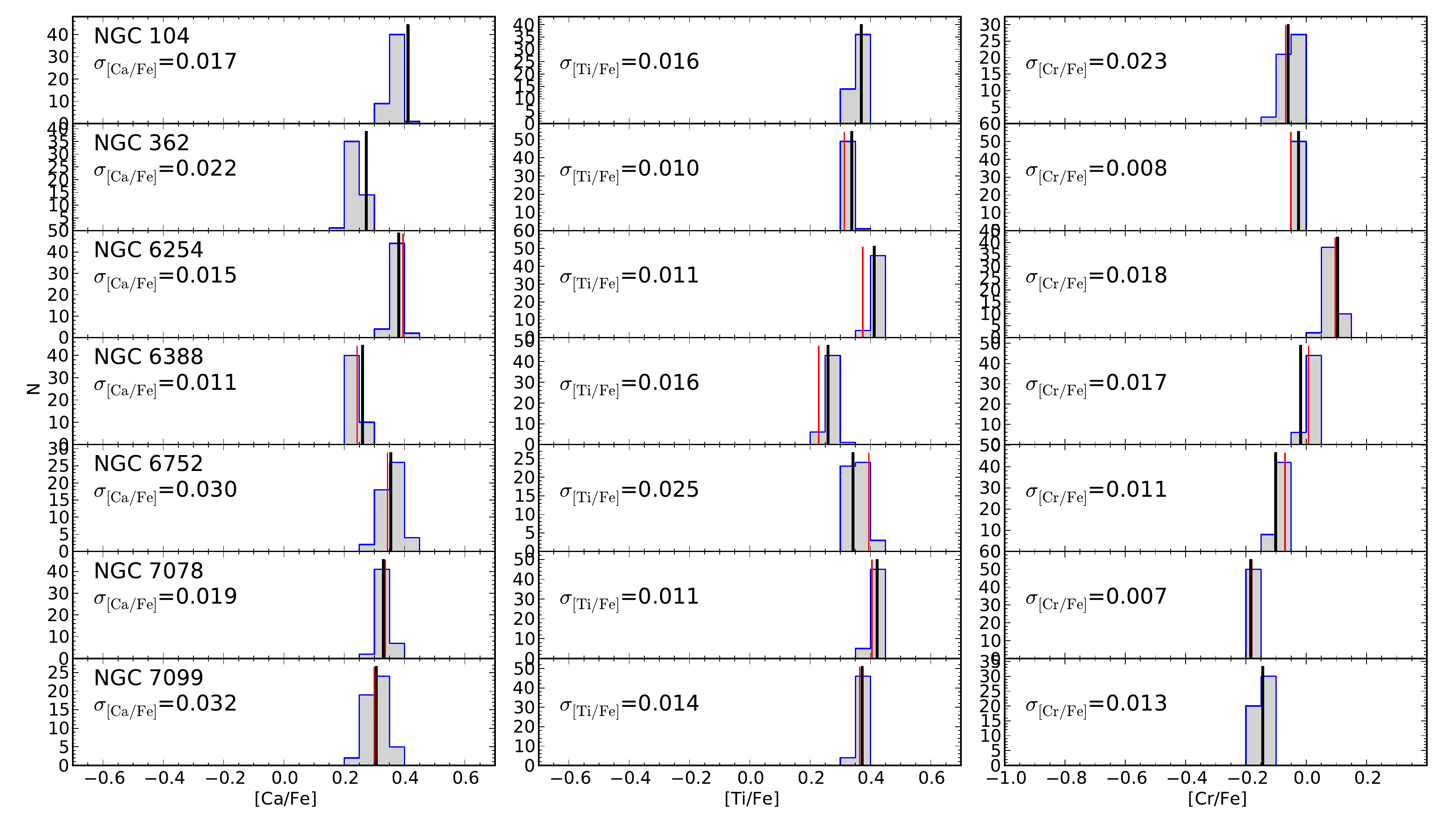}
\caption{As Fig.~\ref{fig:stocfe}, but for the $\mathrm{[Ca/Fe]}$, $\mathrm{[Ti/Fe]}$, and $\mathrm{[Cr/Fe]}$ abundance ratios.}
\label{fig:stocx2fe}
\end{figure*}

As pointed out above, our integrated-light spectra only sample a fraction of $\sim$8\%--25\% of the total cluster light. While the spectra are of high S/N, they are still subject to stochastic fluctuations in the number of stars in a given evolutionary stage that happen to fall within the slit scan areas. It is well-known that such stochastic fluctuations can lead to large random differences between the colours of star clusters with otherwise similar properties (age, metallicity, mass), although for clusters with GC-like masses the effect becomes less dramatic for ages older than $10^8$--$10^9$ years   \citep[e.g.][]{Girardi1995,Bruzual2002,Piskunov2009,Fouesneau2010,Popescu2010b,Silva-Villa2011}.

To test how much uncertainty stochastic sampling of the HRD introduces into our analysis, we carried out a set of Monte-Carlo simulations in which the same number of stars as those present within the slit scan areas were sampled at random from the full ACSGCS CMDs. The corresponding random realisations of the CMDs were then used to redetermine the abundances from our spectra. For each cluster, we generated 50 random realisations of the CMD, and thus obtained 50 sets of abundance measurements.

In Fig.~\ref{fig:stocfe} we show the distributions of $\mathrm{[Fe/H]}$ values derived from the different stochastic realisations of the CMDs for each cluster. 
The corresponding distributions for the main element abundance ratios discussed in this paper are shown in Figs.~\ref{fig:stocxfe} and \ref{fig:stocx2fe}. In each of these figures
we also indicate the values derived for the full CMDs and for stars located within the estimated areas actually covered by the slit scans (see below). We see that the dispersion in the metallicity measurements varies considerably from one cluster to another, and there is a strong inverse correlation with the luminosity of the sampled area. However, even for NGC~6752, which has the least luminous sampled area, the dispersion on the metallicity measurements does not exceed 0.1 dex. 
For abundance ratios, the dispersions are generally smaller than for the $\mathrm{[Fe/H]}$ determinations and exceed 0.05 dex only for the $\mathrm{[Ba/Fe]}$ ratios of NGC~6752 and NGC~7099. This is because variations in the abundances  derived for different elements in the stochastic trials tend to correlate with each other, although there are exceptions. For example, the abundances of Mg correlate only weakly with those of Fe for the metal-rich clusters (NGC~104, NGC~6388), so for NGC~104 the dispersion on $\mathrm{[Mg/Fe]}$ actually exceeds that on $\mathrm{[Fe/H]}$.

\subsubsection{Using only stars in the slit scan areas}

In principle, the issue of stochasticity would be moot for our observations if we knew \emph{exactly} which regions of the clusters were covered by the slit scans. We could then simply select the stars from the ACSGCS data that fall within those regions and use them in the CMD modelling. In practice, however, it is still uncertain which stars contribute to the light. Stars near the edge of the slit are not exactly ``in'' or ``out'', but contribute by some fractional amount that depends on the seeing, guiding errors, etc. In addition, since each exposure started with the slit located at about one half-light radius from the centre, and the telescope was already drifting by the differential tracking rate, this introduced a substantial uncertainty in the exact centring of the slit at the beginning of the exposure. When multiple parallel scans were made, there may also be slight gaps or overlaps between the  scans.

With these caveats in mind, we carried out an additional set of abundance determinations where only stars falling within the estimated slit scan areas were used for the CMD modelling. In combination with the fully stochastic tests, the difference between this analysis and that based on the full CMDs should help us quantify the uncertainties associated with sampling only part of the cluster light. The $\mathrm{[Fe/H]}$ values derived from these fits are indicated with the vertical thin red lines in Figs.~\ref{fig:stocfe}-\ref{fig:stocx2fe}.

\subsubsection{Using the most recent Kurucz line list}

\begin{table}
\caption{Oscillator strengths ($\log gf$) for \ion{Mg}{I} lines from various sources.}
\label{tab:loggmg}
\centering
\begin{tabular}{c r r r r r}
\hline\hline
$\lambda$  & Kurucz & CH04  & G2003  & NIST  & VALD \\
(\AA) & & (1) & (2) & (3) & (4,5) \\
\hline
4352\phantom{\tablefootmark{a}} & $-0.583$ & $-0.833$ & \ldots      & $-0.583$ & $-0.583$ \\
4571\phantom{\tablefootmark{a}} & $-5.623$ & $-5.691$ & \ldots      & $-5.623$ & $-5.623$ \\
4703\phantom{\tablefootmark{a}} & $-0.440$ & $-0.374$ & $-0.471$ & $-0.440$ & $-0.440$ \\
4730\phantom{\tablefootmark{a}} & $-2.347$ & $-2.340$ & $-2.389$ & $-2.347$ & $-2.347$ \\
5167\tablefootmark{a} & $-0.870$ & $-0.856$ & $-0.952$ & $-0.870$ & $-0.931$ \\
5173\tablefootmark{a} & $-0.393$ & $-0.380$ & $-0.324$ & $-0.393$ & $-0.450$ \\
5184\tablefootmark{a} & $-0.167$ & $-0.158$ & $-0.102$ & $-0.167$ & $-0.239$ \\
5528\phantom{\tablefootmark{a}} & $-0.498$ & $-0.341$ & $-0.522$ & $-0.498$ & $-0.498$ \\
5711\phantom{\tablefootmark{a}} & $-1.724$ & $-1.833$ & $-1.729$ & $-1.724$ & $-1.724$ \\
\hline
\end{tabular}
\tablefoot{\tablefoottext{a}{These lines (Mg $b$ triplet) are not used in our analysis.}
}

\tablebib{
(1)~\citet{Castelli2004};
(2)~\citet{Gratton2003a};
(3)~\citet{NIST}; 
(4)~\citet{Piskunov1995};
(5)~\citet{Kupka1999} 
}
\end{table}

In addition to our analysis based on the modified CH04 line list, we carried out a set of fits using the more recent atomic line list that is available from the Kurucz web site\footnote{\texttt{http://kurucz.harvard.edu}} (the version used here was downloaded on 18 April 2016 and contains data updated on 18 February 2016). The new Kurucz list includes 244719 atomic lines in the wavelength range from 4200~\AA\ -- 6200~\AA, as compared to 69516 in the CH04 list. Part of this difference is due to the inclusion of hyperfine components for many of the odd-$Z$ elements, such as Na, Mn, and Sc (as well as a large number of V transitions, although we do not measure these).  The Kurucz list also includes more lines for other elements, e.g. 30541 Fe lines in this wavelength range as compared to  12888 Fe lines in the CH04 list. Of course, the majority of these lines are too weak in most stars to make a substantial difference for the synthetic spectra. 

For many lines in common between the two lists, oscillator strengths have been updated. In particular, we call attention to  Mg, which has been found in our previous studies to be under-abundant compared to other $\alpha$-elements (Ca, Ti). 
The $\log gf$ values for the \ion{Mg}{I} lines according to various sources are summarised in Table~\ref{tab:loggmg}. The CH04 and ``Kurucz'' columns give the $\log gf$ values in the CH04 and Kurucz lists. The ``G2003'' column gives the values according to \citet{Gratton2003a}, which were used by \citet{Carretta2009}. The remaining two columns give the values listed in the NIST and VALD databases. We see that the $\log gf$ values in the CH04 list differ by up to 0.25 dex from those in the other lists. We will return to the implications of these differences for our abundance determinations below.

We note that the Kurucz list does not include hyperfine structure for Ba. We have added this information from the CH04 list, but kept the $s$-process dominated isotopic mixture. By comparison with the $r$-process dominated mixture in our modified CH04 list, this allows us to assess the sensitivity of the Ba abundance measurements to the assumed isotopic ratios.

\subsubsection{Isochrone-based analysis}
\label{sec:iso}

In more distant GCs, colour-magnitude diagrams may not always be available, and one may need to rely on theoretical isochrones for the analysis.
To test how much this would change our results, we carried out a set of fits based purely on theoretical isochrones, again using the Dartmouth set. The metallicities of the isochrones were chosen to self-consistently match those derived from the spectroscopy; this sometimes required a couple of iterations to reach convergence. 
The isochrones were combined with the empirical CMDs for the horizontal branch (HB) and asymptotic giant branch (AGB), taken from ACSGCS. 

The appropriate scalings of the weights of the empirical data were determined by requiring the isochrone-based CMDs and the empirical ones to have the same number of RGB stars in the range $1 < M_V < 2$. We assumed the same ages as for the luminosity functions (Sec.~\ref{sec:hrmod}). 

The isochrones used for these fits are included in Fig.~\ref{fig:cmds}. We stress that these isochrones are not necessarily the same as those that would provide the best fits to the empirical CMDs. In some cases, it might be possible to improve the fits by adjusting the reddening and distance, but with the exception of NGC~6388 (see below) we have not attempted to do so.

\section{Results}

\begin{table*}
\caption{Abundance measurements for the standard analysis.}
\label{tab:abun}
\centering
\begin{tabular}{l r r r r r r r}
\hline\hline
 & NGC 104  & NGC 362  & NGC 6254  & NGC 6388  & NGC 6752  & NGC 7078  & NGC 7099 \\
\hline
\mbox{[Fe/H]}  & $-0.863$ \phantom{(9)}  & $-1.076$ \phantom{(9)}  & $-1.481$ \phantom{(9)}  & $-0.506$ \phantom{(9)}  & $-1.883$ \phantom{(9)}  & $-2.388$ \phantom{(9)}  & $-2.396$ \phantom{(9)} \\
 \; rms$_w$ (N)   & $0.046$ (9)  & $0.042$ (9)  & $0.034$ (9)  & $0.042$ (9)  & $0.039$ (9)  & $0.042$ (9)  & $0.059$ (9) \\
\mbox{[Na/Fe]}  & $0.422$ \phantom{(2)}  & $-0.045$ \phantom{(2)}  & $-0.038$ \phantom{(2)}  & $0.357$ \phantom{(2)}  & $0.302$ \phantom{(2)}  & $0.154$ \phantom{(2)}  & $0.232$ \phantom{(2)} \\
 \; rms$_w$ (N)   & $0.008$ (2)  & $0.000$ (2)  & $0.003$ (2)  & $0.013$ (2)  & $0.023$ (2)  & $\ldots$ (1)  & $0.006$ (2) \\
\mbox{[Mg/Fe]}  & $0.442$ \phantom{(5)}  & $0.149$ \phantom{(5)}  & $0.378$ \phantom{(5)}  & $0.108$ \phantom{(5)}  & $0.391$ \phantom{(5)}  & $0.177$ \phantom{(5)}  & $0.286$ \phantom{(5)} \\
 \; rms$_w$ (N)   & $0.095$ (5)  & $0.088$ (5)  & $0.067$ (5)  & $0.092$ (5)  & $0.130$ (5)  & $0.126$ (5)  & $0.144$ (5) \\
\mbox{[Ca/Fe]}  & $0.412$ \phantom{(7)}  & $0.273$ \phantom{(8)}  & $0.380$ \phantom{(8)}  & $0.260$ \phantom{(7)}  & $0.355$ \phantom{(8)}  & $0.330$ \phantom{(8)}  & $0.306$ \phantom{(8)} \\
 \; rms$_w$ (N)   & $0.116$ (7)  & $0.128$ (8)  & $0.137$ (8)  & $0.182$ (7)  & $0.114$ (8)  & $0.069$ (8)  & $0.096$ (8) \\
\mbox{[Sc/Fe]}  & $0.219$ \phantom{(6)}  & $0.117$ \phantom{(6)}  & $0.130$ \phantom{(6)}  & $0.118$ \phantom{(6)}  & $0.120$ \phantom{(6)}  & $0.221$ \phantom{(6)}  & $0.192$ \phantom{(6)} \\
 \; rms$_w$ (N)   & $0.178$ (6)  & $0.098$ (6)  & $0.068$ (6)  & $0.190$ (6)  & $0.067$ (6)  & $0.107$ (6)  & $0.099$ (6) \\
\mbox{[Ti/Fe]}  & $0.370$ \phantom{(8)}  & $0.338$ \phantom{(9)}  & $0.412$ \phantom{(9)}  & $0.259$ \phantom{(8)}  & $0.342$ \phantom{(8)}  & $0.422$ \phantom{(9)}  & $0.372$ \phantom{(9)} \\
 \; rms$_w$ (N)   & $0.115$ (8)  & $0.065$ (9)  & $0.097$ (9)  & $0.114$ (8)  & $0.128$ (8)  & $0.100$ (9)  & $0.077$ (9) \\
\mbox{[Cr/Fe]}  & $-0.060$ \phantom{(6)}  & $-0.026$ \phantom{(6)}  & $0.104$ \phantom{(6)}  & $-0.019$ \phantom{(6)}  & $-0.101$ \phantom{(6)}  & $-0.185$ \phantom{(6)}  & $-0.144$ \phantom{(6)} \\
 \; rms$_w$ (N)   & $0.135$ (6)  & $0.083$ (6)  & $0.411$ (6)  & $0.220$ (6)  & $0.070$ (6)  & $0.067$ (6)  & $0.125$ (6) \\
\mbox{[Mn/Fe]}  & $-0.229$ \phantom{(2)}  & $-0.371$ \phantom{(2)}  & $-0.430$ \phantom{(2)}  & $-0.154$ \phantom{(2)}  & $-0.366$ \phantom{(2)}  & $-0.347$ \phantom{(2)}  & $-0.344$ \phantom{(2)} \\
 \; rms$_w$ (N)   & $0.087$ (2)  & $0.005$ (2)  & $0.127$ (2)  & $0.149$ (2)  & $0.055$ (2)  & $0.094$ (2)  & $0.026$ (2) \\
\mbox{[Ba/Fe]}  & $0.155$ \phantom{(4)}  & $0.330$ \phantom{(4)}  & $0.413$ \phantom{(4)}  & $0.200$ \phantom{(4)}  & $0.182$ \phantom{(4)}  & $0.436$ \phantom{(4)}  & $0.045$ \phantom{(4)} \\
 \; rms$_w$ (N)   & $0.078$ (4)  & $0.051$ (4)  & $0.087$ (4)  & $0.252$ (4)  & $0.109$ (4)  & $0.053$ (4)  & $0.171$ (4) \\
\hline
\end{tabular}
\tablefoot{For each abundance ratio, the second line lists the weighted r.m.s. and the number of individual measurements in parentheses.
}
\end{table*}

\begin{table}
\caption{Literature metallicities.}
\label{tab:litabun}
\centering
\begin{tabular}{lrl}
\hline\hline
Cluster & [Fe/H] & Ref. \\
\hline
NGC~104 & $-0.743\pm0.003$ & 1 (G) \\
                 & $-0.768\pm0.016$ & 1 (U) \\
                & $-0.76\pm0.01$ & 2 \\
                & $-0.62$ & 3 \\
                & $-0.72\pm0.08$ & 4 \\
                & $-0.83\pm0.01$ & 5 \\
NGC~362 & $-1.174\pm0.004$ & 6 (G) \\
               & $-1.168\pm0.014$ & 6 (U) \\
               & $-1.33$ & 7 \\
               & $-1.21\pm0.09$ & 8 \\
NGC~6254 & $-1.556\pm0.004$ & 1 (G) \\
                 & $-1.575\pm0.016$ & 1 (U) \\
               & $-1.52$ & 3 \\
               & $-1.53\pm0.06$ & 4 \\
NGC~6388 & $-0.406\pm0.013$ & 1 (G) \\
                & $-0.441\pm0.014$ & 1 (U) \\
               & $-0.55\pm0.15$ & 4 \\
               & $-0.79$/$-0.58$ & 9 \\
               & $-0.60\pm0.02$ & 8 \\
NGC~6752 & $-1.561\pm0.004$ & 1 (G) \\
             & $-1.555\pm0.014$ & 1 (U) \\
             & $-1.63\pm0.01$ & 10 \\
             & $-1.54$ & 3 \\
             & $-1.53\pm0.16$ & 4 \\
NGC~7078 & $-2.341\pm0.007$ & 1 (G) \\
            & $-2.320\pm0.016$ & 1 (U) \\
            & $-2.38$ & 3 \\
            & $-2.39\pm0.14$ & 4 \\
NGC~7099 & $-2.359\pm0.006$ & 1 (G) \\
         & $-2.344\pm0.015$ & 1 (U) \\
        & $-2.31$ & 3 \\
\hline
\end{tabular}
\tablefoot{For \citet{Carretta2009c} we list the $\mathrm{[Fe/H]}$ values derived from their GIRAFFE (G) and UVES (U) observations.}
\tablebib{
(1)~\citet{Carretta2009c};
(2)~\citet{Koch2008};
(3)~\citet{Pritzl2005};
(4)~\citet{Roediger2014}; 
(5)~\citet{Lapenna2014}; 
(6)~\citet{Carretta2013}; 
(7)~\citet{Pritzl2005}; 
(8)~\citet{Worley2010}; 
(9)~\citet{Wallerstein2007}; 
(10)~\citet{Yong2008}; 
}  

\end{table}

The individual abundance measurements from the standard analysis are listed in Tables~\ref{tab:abN0104}--\ref{tab:abN7099}. For each spectral bin, we give the best-fitting abundance of the corresponding element, along with the formal error from the $\chi^2$ minimisation. By comparison with Fig.~\ref{fig:lamfe}, it is clear that the true uncertainties are larger than the random errors on the fits; the bin-to-bin dispersions on the Fe abundances are 0.04--0.06 dex, whereas the formal errors are generally $< 0.01$ dex. Comparing, for example, NGC~7078 and NGC~7099 (both of which have very similar overall metallicities), the variations in [Fe/H] with wavelength do indeed appear non-random. Possible systematic effects that could cause such variations include uncertainties in the atomic parameters of the lines, as well as in the continuum scaling procedure. 

Table~\ref{tab:abun} lists the weighted average of the individual abundance measurements for each cluster for the standard analysis. The  weights $w_i$ are based on the random uncertainties in Tables~\ref{tab:abN0104}--\ref{tab:abN7099}, but we have added a ``floor'' of 0.01 dex in quadrature to avoid having bins with very small formal uncertainties become too dominant. Hence, the weights were computed as
\begin{equation}
  w_i = \left(\sigma_i^2 + (0.01\,\mathrm{dex})^2\right)^{-1}
  \label{eq:wi}
\end{equation}
As in L14, we also list the bin-to-bin r.m.s., again weighted using Eq.~(\ref{eq:wi}):
\begin{equation}
 \mathrm{rms}_w = \left(\frac{\sum w_i \left(x_i - \langle x \rangle\right)^2}{\sum w_i}\right)^{1/2}
\end{equation}
The numbers in parentheses after rms$_w$ give the number of individual fits for each entry, $N$. 

Simple propagation of the formal errors on the individual measurements will likely underestimate the uncertainties on the mean values in Table~\ref{tab:abun}.  Taking the rms$_w$ as indicative of the true uncertainties on the individual measurements,  we instead estimate the uncertainties on the mean values as
\begin{equation}
\sigma = \mathrm{rms_w}/\sqrt{N-1}
\label{eq:sigma}
\end{equation} 
Whenever we refer to uncertainties on the mean abundance ratios, these will have been calculated using Eq.~(\ref{eq:sigma}).
The equivalents of Table~\ref{tab:abun}, but for the modified analysis methods, are included as Tables~\ref{tab:abun1}--\ref{tab:abun3} in the appendix.

\subsection{Iron abundances}

\begin{figure}
\centering
\includegraphics[width=\columnwidth]{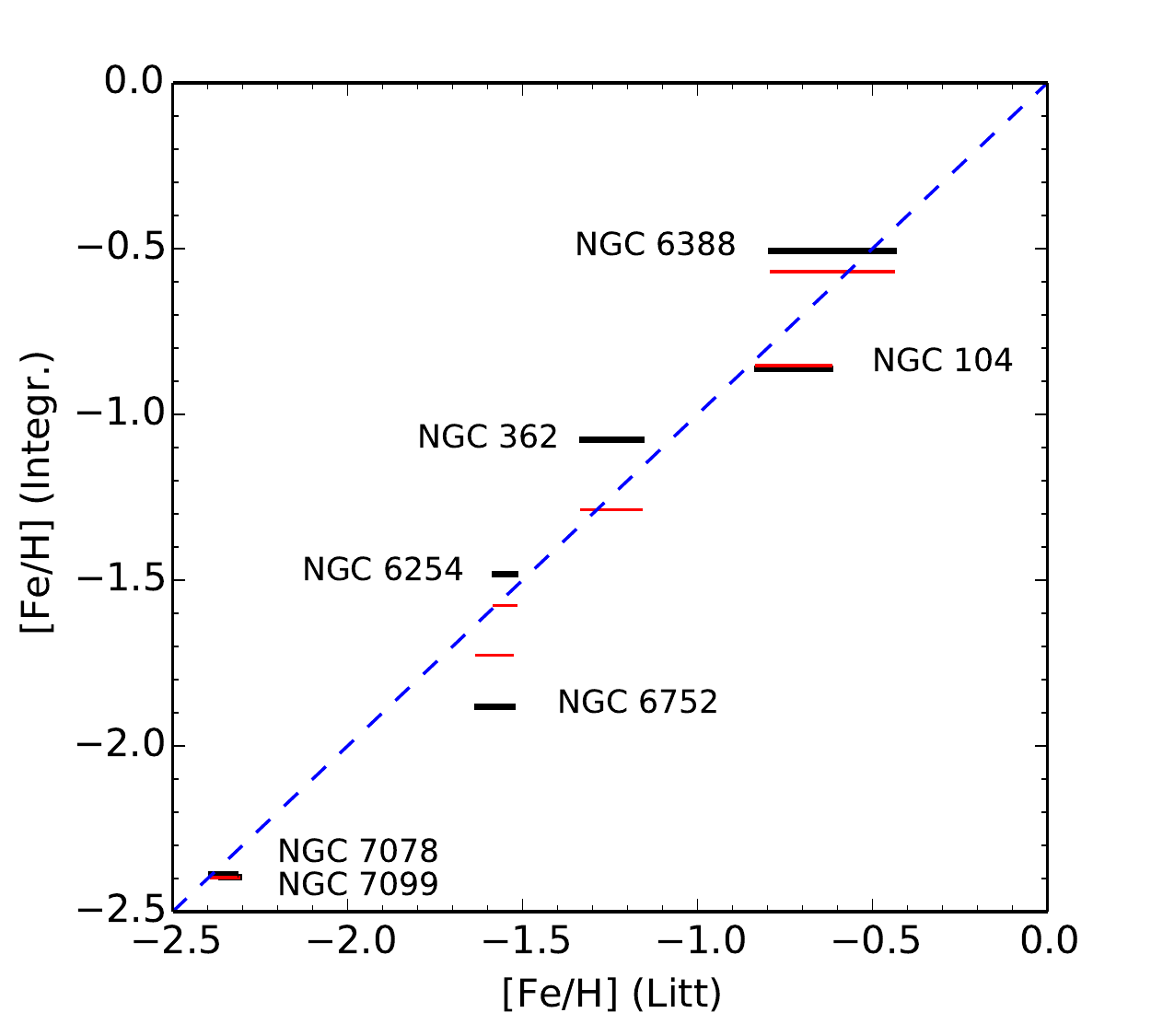}
\caption{Integrated-light iron abundances versus literature values. The horizontal bars represent the range of $\mathrm{[Fe/H]}$ values quoted in the literature (Table~\ref{tab:litabun}). Black bars indicate the values derived from our standard analysis and red bars are for the stars within the slit scan areas. The blue (dashed) line is the 1:1 relation, not a fit.
}
\label{fig:fecmp}
\end{figure}

We first compare our measurements of the iron abundances with data from the literature. We have not attempted to carry out a complete literature search for abundance measurements of all the clusters in our sample, but rely on three main sources: the VLT/FLAMES survey by \citet{Carretta2009c}, for which we include both the UVES and GIRAFFE iron abundances, and the compilations by \citet{Pritzl2005} and \citet{Roediger2014}. The iron abundances from these studies, as well as a few others, are summarised in Table~\ref{tab:litabun}.  The differences between different studies often exceed the formal random uncertainties on the measurements, which are usually small ($\sim0.01$ dex). For the \citet{Roediger2014} compilation, the quoted uncertainties represent the variation among the different studies in the compilation. Note that the overlap with \citet{Pritzl2005} and \citet{Roediger2014} is only partial, with NGC~6388 missing from the former and NGC~362 and NGC~7099 from the latter.

In Figure~\ref{fig:fecmp} we plot our integrated-light iron abundance measurements against the literature values. The horizontal bars (black for the standard analysis, red for stars in the slit scan areas) represent the range of values quoted by the literature sources and the dashed line shows the 1:1 relation. From this figure, and by comparing Tables~\ref{tab:abun} and \ref{tab:litabun}, we see that our integrated-light iron abundances generally agree well with the literature values. NGC~6752 is the most conspicuous outlier in the standard analysis, but moves closer to the 1:1 line when the stars in the slit scan areas are used. Comparing with Figure~\ref{fig:stocfe}, we see that this cluster also has the largest spread in the iron abundances derived from the random CMD realisations. 
This suggests that stochastic fluctuations, caused by the limited coverage of the Galactic GCs by our scans, may indeed be an important contributor to the scatter around the 1:1 line Fig.~\ref{fig:fecmp}. 

Next, we discuss the comparison with individual clusters in more detail.

\subsubsection{Metal-rich clusters: NGC~104, NGC~362, NGC~6388}
\label{sec:mrgcs}

\begin{figure}
\centering
\includegraphics[width=\columnwidth]{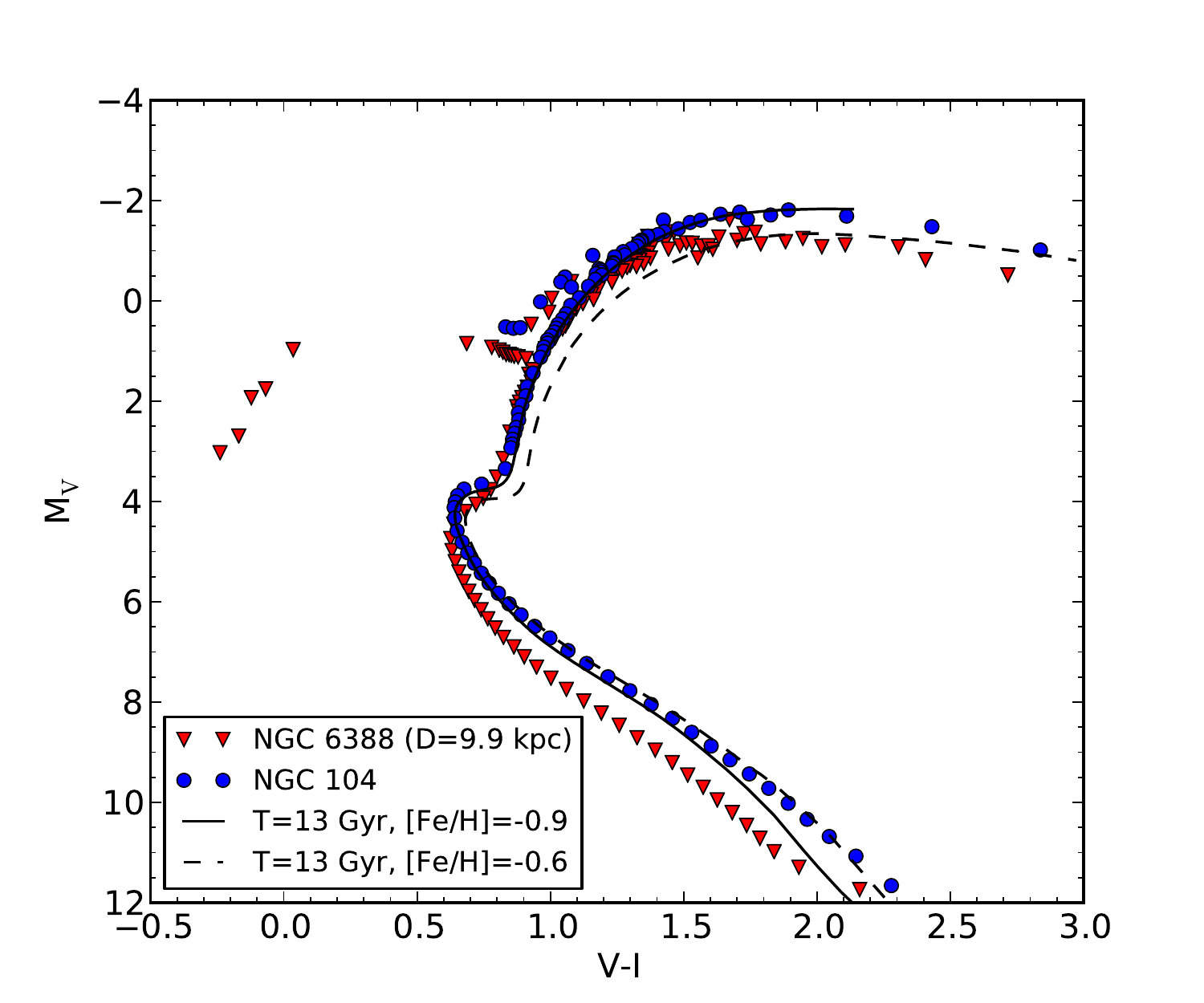}
\caption{Binned colour-magnitude diagrams of NGC~6388 and NGC~104.}
\label{fig:cmd6388}
\end{figure}

According to the literature, as well as our measurements, the most metal-rich cluster in our sample is NGC~6388. 
\citet{Carretta2009c} find $\mathrm{[Fe/H]}=-0.41$ (GIRAFFE) or $ \mathrm{[Fe/H]}=-0.44$ (UVES).
\citet{Wallerstein2007} found a metallicity as low as $\mathrm{[Fe/H]}=-0.79$ when using spectroscopically determined surface gravities, but noted that these surface gravities seemed too low for the locations of the stars in the CMDs. When using photometrically inferred surface gravities, \citet{Wallerstein2007} instead found $\mathrm{[Fe/H]}=-0.58\pm0.03$, which is closer to other recent determinations. Our integrated-light measurement of $\mathrm{[Fe/H]}=-0.506\pm0.015$ from the standard analysis thus falls well within the range found by studies of individual stars. 
We note that while NGC~6388 has a somewhat higher iron abundance than NGC~104, it is probably less $\alpha$-enhanced \citep[e.g.][]{Wallerstein2007} so the total metallicities of the two clusters may not be very different.

NGC~6388 is quite massive and compact, and the uncertainty on our integrated-light analysis that arises from stochastic CMD sampling is small, with $\sigma_{\mathrm{[Fe/H]}} = 0.020$ dex (Fig.~\ref{fig:stocfe}). A more significant source of uncertainty for this cluster may lie in the modelling of the colour-magnitude diagram. In Fig.~\ref{fig:cmd6388} we show the binned CMD of NGC~6388 together with that of NGC~104. We also include theoretical isochrones for metallicities of $\mathrm{[Fe/H]}=-0.6$ and $\mathrm{[Fe/H]}=-0.9$ with $[\alpha/\mathrm{Fe}]=+0.2$. Isochrones with $\mathrm{[Fe/H]}=-0.8$ and $\mathrm{[Fe/H]}=-1.1$ for $[\alpha/\mathrm{Fe}]=+0.4$ look almost identical. Two problems are apparent from this comparison: First, while the lower parts of the RGBs in NGC~6388 and NGC~104 overlap almost perfectly (as would be expected for similar overall metallicities) there is a substantial mismatch elsewhere in the CMD. 
Second, the isochrones that best fit the lower RGB have a lower metallicity 
($\mathrm{[Fe/H]}=-0.9$ for $[\alpha/\mathrm{Fe}]=+0.2$, or $\mathrm{[Fe/H]}=-1.1$ for $[\alpha/\mathrm{Fe}]=+0.4$)
than the values quoted in the literature for these clusters, and do not reach the coolest parts of the RGB (although some of the coolest stars might be asymptotic giant branch stars). For NGC~6388, neither isochrone fits the MS.

Considering only the RGB, we can get a better match between the two clusters if the assumed distance of NGC~6388 is increased. For a distance of $D=11.5$ kpc (as listed in the 1996 version of the Harris catalogue) we find that the RGBs of the two clusters (and the HBs) match well, but an offset remains for the MS. This larger distance is also in better agreement with the dynamical distance estimate of $10.9^{+0.40}_{-0.45}$ kpc by \citet{Watkins2015}.
The effect of a larger distance on the integrated-light abundances is, in any case, quite modest: it leads to a decrease of 0.13 dex in the surface gravities, and the iron abundance then decreases by 0.03 dex. Thus, irrespective of the assumed distance, the integrated-light analysis based on the observed CMD of NGC~6388 gives an iron abundance in good agreement with that derived from measurements of individual stars.
If we also allow the reddening to vary, it may be possible to find combinations of metallicity, age, alpha-enhancement, distance, and reddening that allow acceptable fits to both CMDs. Such an exercise is beyond the scope of this work, but in Sec.~\ref{sec:isocmd} we will briefly consider the effect of changing $E(B-V)$ for NGC~6388.

For NGC~104 our standard analysis yields $\mathrm{[Fe/H]}=-0.863\pm0.016$, and for stars within the slit scan areas we find an only marginally higher value of $\mathrm{[Fe/H]}=-0.853$. 
Our integrated-light Fe abundance is thus about 0.1 dex lower than those measured by \citet{Koch2008} and \citet{Carretta2009c} from individual stars.  In their integrated-light analysis, \citet{McWilliam2008} found $\mathrm{[Fe/H]}=-0.75$ (based on 102 \ion{Fe}{I} lines) with an estimated systematic error of 0.045 dex.  A lower iron abundance of $\mathrm{[Fe/H]}=-0.83\pm0.01$ has recently been measured by \citet{Lapenna2014} for 11 individual RGB stars, differing only slightly from that derived here. It is interesting to note that \citet{Lapenna2014} found a significantly lower metallicity of $\mathrm{[Fe/H]}=-0.94$ from AGB stars when using only \ion{Fe}{I} lines, whereas measurements of \ion{Fe}{II} lines in the AGB stars gave an iron abundance similar to that measured for the RGB stars, $\mathrm{[\ion{Fe}{II}/H]} = -0.83\pm0.01$.
While the reason for this difference remains unclear, discrepancies between Fe abundances derived from \ion{Fe}{I} and \ion{Fe}{II} lines have been observed in other clusters, e.g. M22 \citep{Mucciarelli2015a}. In our integrated-light analysis we do not treat \ion{Fe}{I} and \ion{Fe}{II} lines separately, and it is  possible that our abundances for NGC 104 are affected at some level by the anomalous behaviour of the AGB stars. However, in their analysis of NGC~104, \citet{McWilliam2008} found no significant difference between the iron abundances based on \ion{Fe}{I} lines (quoted above) and \ion{Fe}{II} lines ([\ion{Fe}{II}/H]=$-0.72$, based on 7 lines with a dispersion of 0.15 dex).

Our iron abundance for NGC~362, $\mathrm{[Fe/H]}=-1.076\pm0.015$ for the standard analysis, is about 0.1 dex higher than the \citet{Carretta2013} estimate of $\mathrm{[Fe/H]}=-1.17$. However, the dispersion in the stochastic trials is relatively large for this cluster ($\sigma_\mathrm{[Fe/H]}=0.092$ dex) and the full range of Fe abundances encountered in the 50 stochastic trials goes from $\mathrm{[Fe/H]}=-1.35$ to $\mathrm{[Fe/H]}=-0.91$. When using stars within the slit scan area to model the CMD we  get $\mathrm{[Fe/H]} = -1.29$.
Thus, within the uncertainties introduced by stochastic sampling, our metallicity determination for NGC~362 is in agreement with most of the values quoted in the literature.

\subsubsection{Intermediate metallicities: NGC~6254, NGC~6752}
\label{sec:imfeh}

\begin{figure}
\centering
\includegraphics[width=\columnwidth]{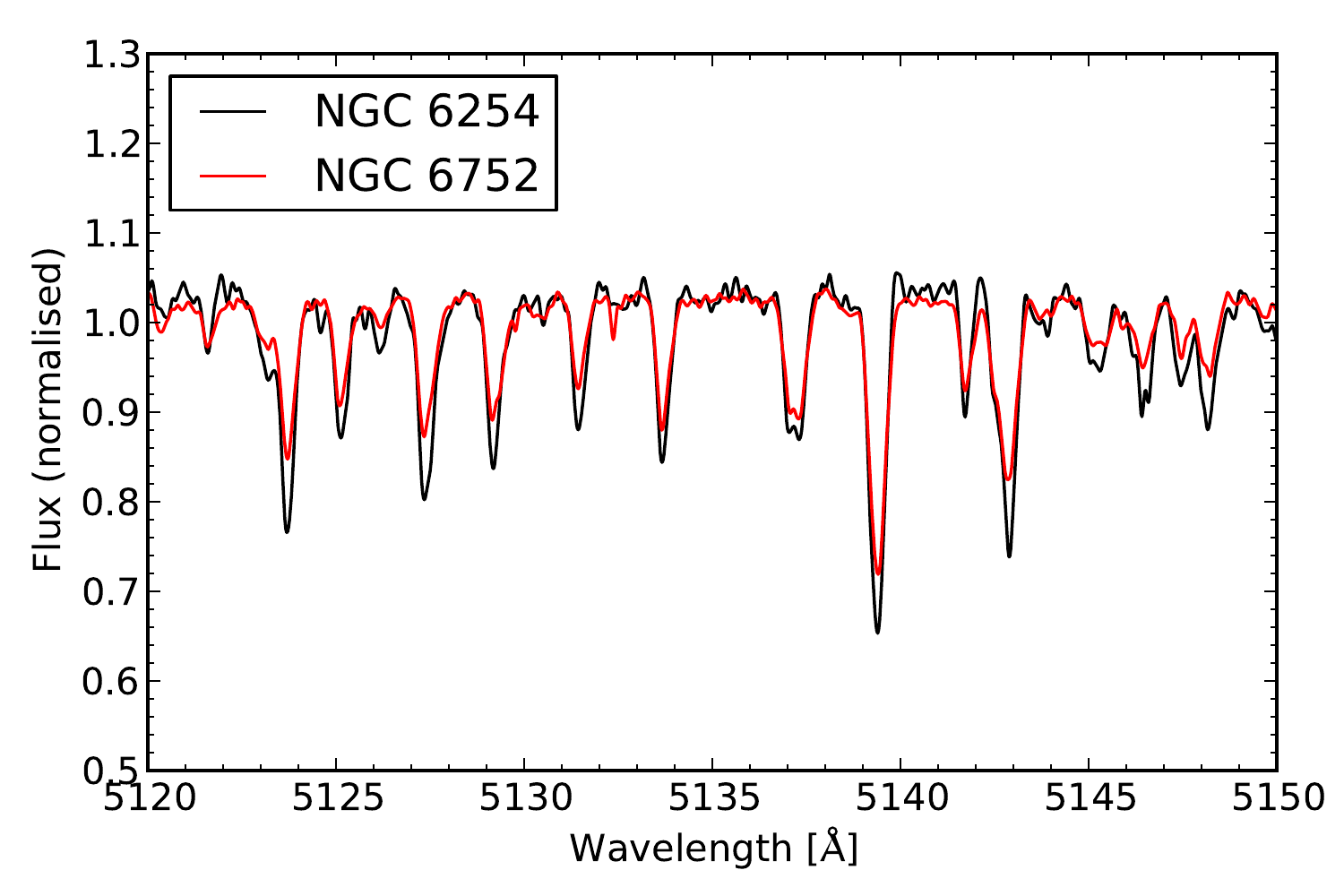}
\caption{Spectra of NGC~6254 and NGC~6752. The difference in the strengths of the spectral features in the two integrated-light spectra is evident, despite their similar metallicities according to measurements of individual stars.}
\label{fig:n6752_n6254}
\end{figure}

Most literature sources agree that these two clusters have very similar metallicities, with [Fe/H] values in the range $-1.58$ to $-1.52$ for NGC~6254 and between $-1.63$ and $-1.53$ for NGC~6752 (Table~\ref{tab:litabun}). In spite of this, our integrated-light iron abundances for the two clusters differ by more than 0.4 dex, a difference that appears far too large to be explained simply by measurement uncertainties.

For NGC~6254, our standard analysis gives $\mathrm{[Fe/H]} = -1.481\pm0.012$, which is only slightly higher than the typical literature values. From the stochastic trials we find a dispersion of 0.058 dex and a full range from $-1.62$ to $-1.35$ in [Fe/H], suggesting that the small difference with respect to the literature may well be caused by stochastic fluctuations. Indeed, the analysis based on stars in the estimated slit scan area gives $\mathrm{[Fe/H]} = -1.58$, in excellent agreement with the literature values. 

NGC~6752, on the other hand, is the most outlying of the clusters in our sample with an integrated-light iron abundance of $\mathrm{[Fe/H]}=-1.883\pm0.014$ from the standard analysis, which is about 0.3 dex lower than that found by other studies. A direct comparison of our spectra of NGC~6254 and NGC~6752  (Fig.~\ref{fig:n6752_n6254}) shows that the spectral features are indeed substantially weaker in the spectrum of NGC~6752. From this we conclude that the difference in the derived metallicities stems from a real difference in our integrated-light spectra of the two clusters, and not from some failure of the fitting/modelling procedure. 
NGC~6752 is the cluster in our sample for which the scanned area has the lowest luminosity, and it is therefore expected that the effects of stochastic CMD sampling manifest themselves more strongly in this cluster. Fig.~\ref{fig:stocfe} shows that the dispersion in [Fe/H] for the stochastic trials is indeed the largest among the clusters in our sample ($\sigma_{\mathrm{[Fe/H]}} = 0.098$), but the difference between our integrated-light iron abundance and the literature is still about 3 $\sigma_{\mathrm{[Fe/H]}}$. By using only stars in the estimated slit scan area we get $\mathrm{[Fe/H]}=-1.73$, which reduces (but does not entirely eliminate) the discrepancy with respect to the literature measurements. The highest iron abundance reached in the 50 stochastic realisations of the CMD is $\mathrm{[Fe/H]}=-1.69$, still not quite reaching the literature values, although even higher values might conceivably be reached if  more trials were carried out.

A number of other uncertainties in our modelling procedure have been considered in our previous papers (L12, L14). These include the adopted microturbulent velocities, luminosity functions, colour-$T_\mathrm{eff}$ transformations and extinction corrections. Uncertainties in the treatment of each of these quantities may add up to an uncertainty of $\sim0.1$ dex on the iron abundance derived from integrated-light spectra (see also \citet{Sakari2014} and \citet{Colucci2017}). One source of uncertainty that we have not previously discussed is the role of blue stragglers (BSs). Although these are visible in some of the CMDs (including that of NGC~6752), they have been omitted from our standard analysis. To test how much the inclusion of BSs would affect our results, we carried out an extra set of fits in which three extra bins were introduced along the BS sequence in the slit scan CMD of NGC~6752. The BSs were treated in the same way as the other stars, i.e., their gravities and temperatures were estimated from the location in the CMD, and they were assumed to have the same composition as other stars in the cluster.  From these fits we found the effect of including the BSs to be relatively minor, with the iron abundance increasing by $\sim0.03$ dex to $\mathrm{[Fe/H]}=-1.70$. 
A similar conclusion was reached by \citet{Sakari2014}, who estimated that BSs contributed by at most 0.07 dex to the uncertainty on $\mathrm{[Fe/H]}$ in their analysis.
This still leaves a difference of $\sim0.1$ dex compared to the iron abundances found by most other authors for NGC~6752.

\subsubsection{Metal-poor clusters: NGC~7078, NGC~7099}

NGC~7078 (M15) and NGC~7099 (M30) are among the most metal-poor GCs in our Galaxy. Most recent studies find iron abundances in the range $\mathrm{[Fe/H]}=-2.4$ to $-2.3$ for both clusters, although values as low as $\mathrm{[Fe/H]}=-2.6$ to $-2.5$ have been reported for M15 \citep[][and references therein]{Sobeck2011}.
In many ways, metal-poor clusters such as these represent the most straight-forward cases for integrated-light abundance analysis. In more metal-rich clusters, complications arise from the presence of cooler giants where molecular features become more important, and line blending in general becomes more severe (e.g. Fig.~\ref{fig:speccmp}). 
From our standard analysis we find $\mathrm{[Fe/H]}=-2.39$ for NGC~7078 and $\mathrm{[Fe/H]}=-2.40$ for NGC~7099, in very good agreement with the literature values. The scatter in the stochastic CMD realisations is 
$\sigma_{\mathrm{[Fe/H]}} = 0.049$ and $\sigma_{\mathrm{[Fe/H]}} = 0.080$ for NGC~7078 and NGC~7099, respectively, but the iron abundances derived using stars in the slit scan areas are nearly identical to those based on the full CMDs ($\mathrm{[Fe/H]}=-2.40$ for both clusters). 
NGC~7078 has also been studied in integrated light by \citet{Sakari2013}, who found $\mathrm{[Fe/H]}=-2.30\pm0.03$ (from \ion{Fe}{I} lines) and $\mathrm{[Fe/H]}=-2.38\pm0.10$ (from one \ion{Fe}{II} line), which is again in good agreement with our measurements.

\subsection{Abundance ratios} 

\subsubsection{Light elements: Na, Mg}
\label{sec:light}

   \begin{figure}
   \centering
   \includegraphics[width=\columnwidth]{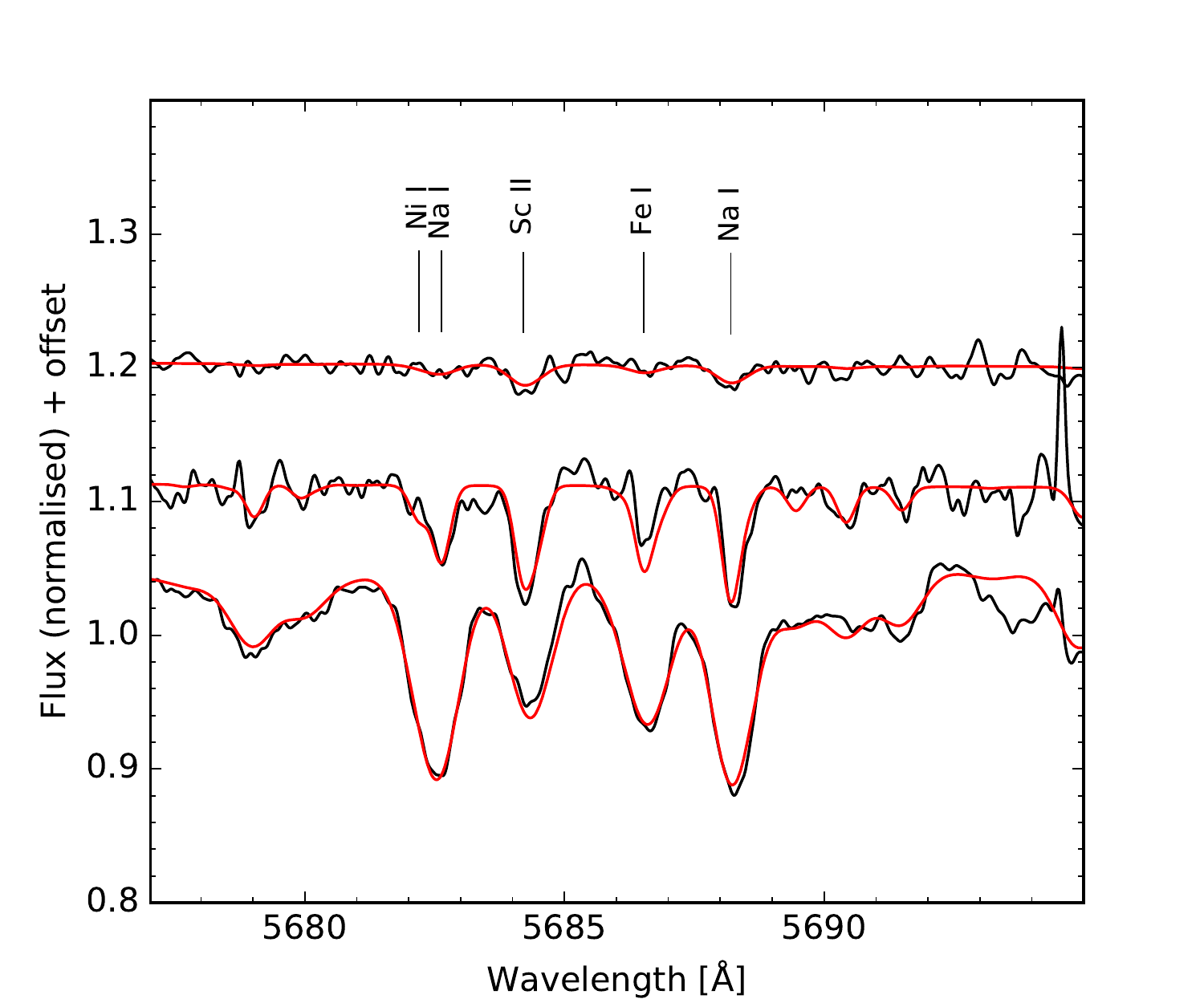}
   \includegraphics[width=\columnwidth]{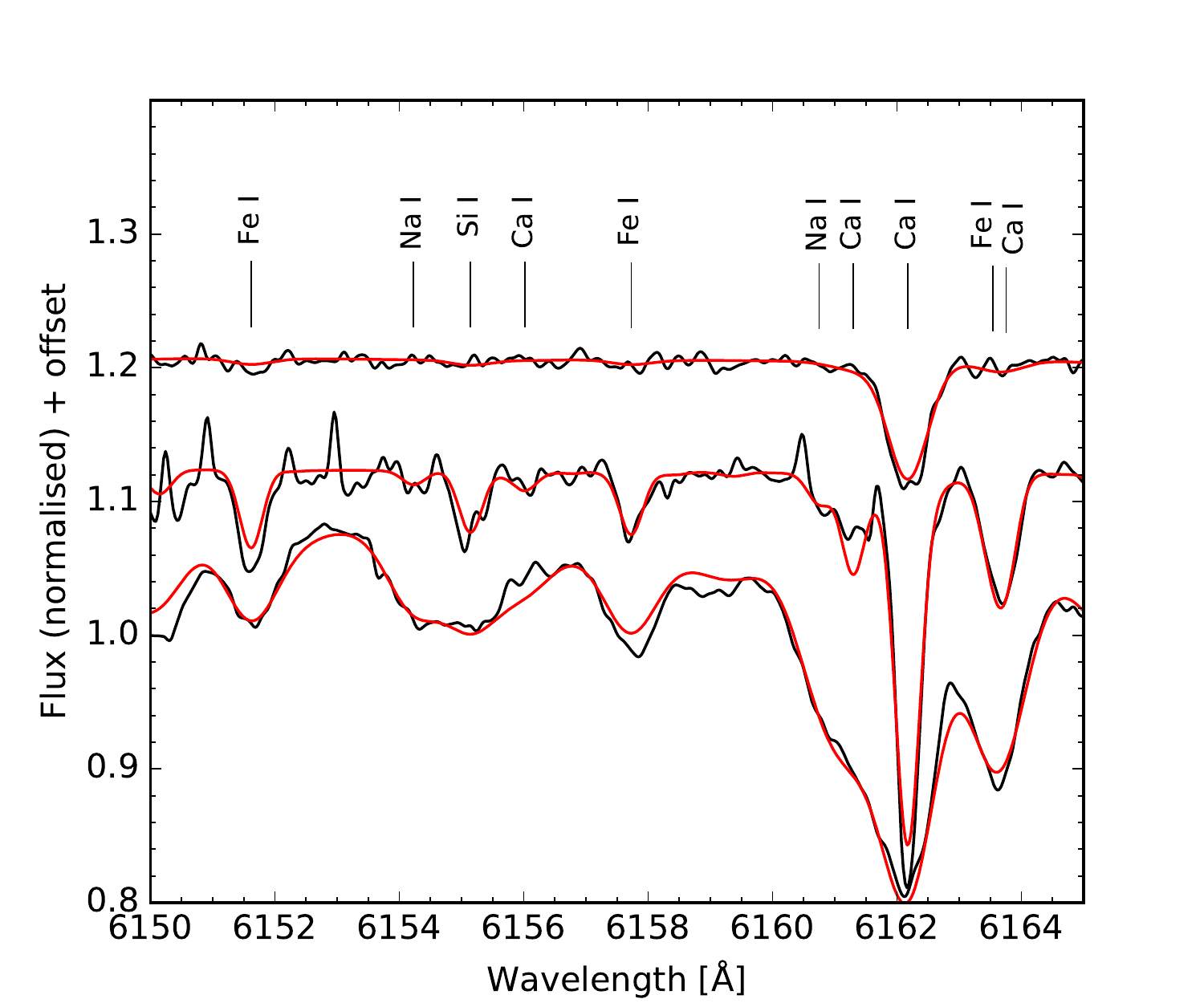}
      \caption{Spectral fits to the sodium doublets at 5683/5688 \AA\ (top panel) and 6154/6161 \AA\ (bottom panel). Top to bottom, in each panel: NGC~7078, NGC~6254, and NGC~6388.}
         \label{fig:speccmp_na}
   \end{figure}

   \begin{figure}
   \centering
   \includegraphics[width=\columnwidth]{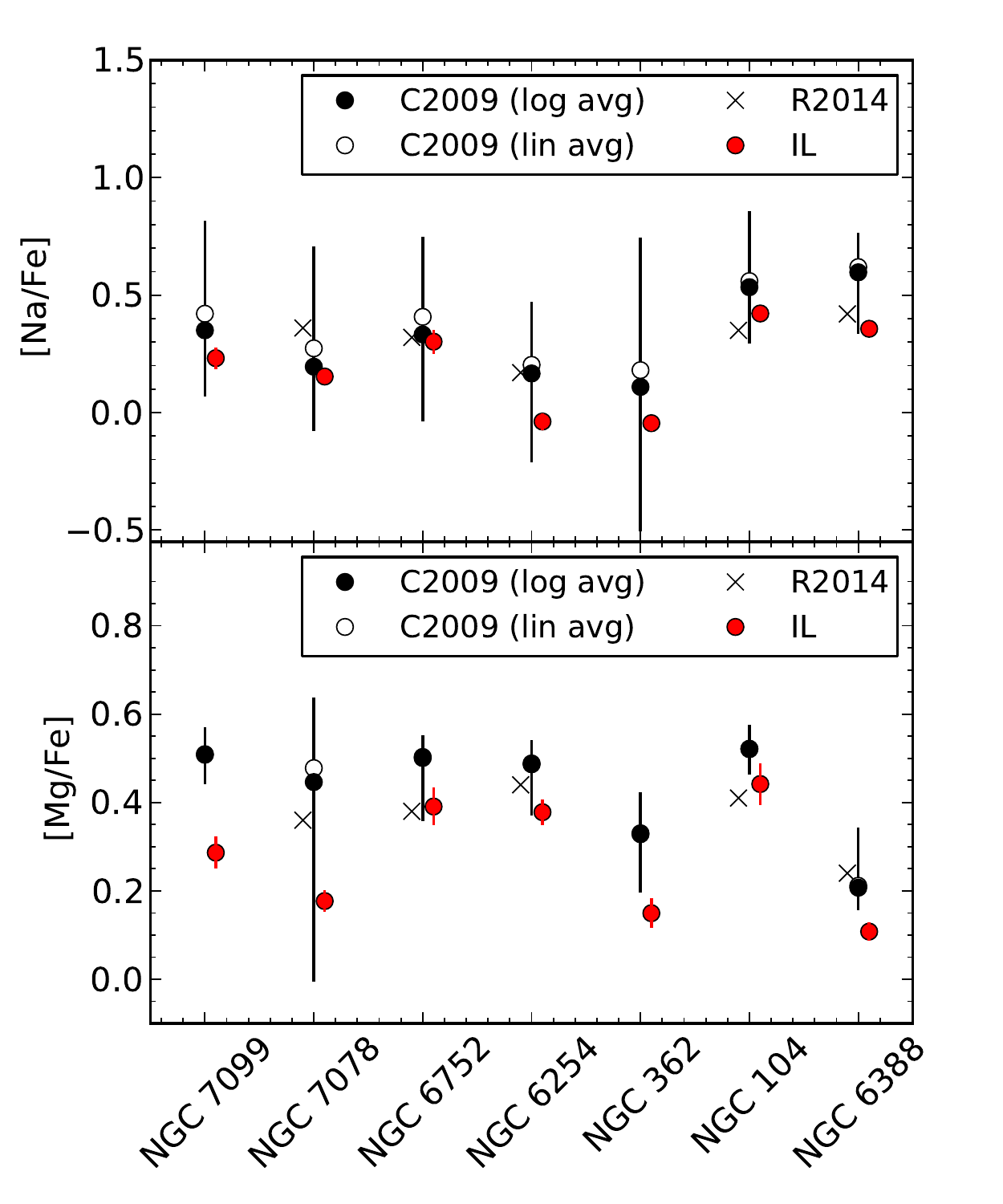}
      \caption{Comparison of our integrated-light [Na/Fe] and [Mg/Fe] abundance ratios with measurements of individual stars from \citet[C2009]{Carretta2009} and \citet[R2014]{Roediger2014}. Slight offsets to the right and left have been applied to our integrated-light data points and those of \citet{Roediger2014} for clarity.}
         \label{fig:lightcmp}
   \end{figure}

   \begin{figure}
   \centering
   \includegraphics[width=\columnwidth]{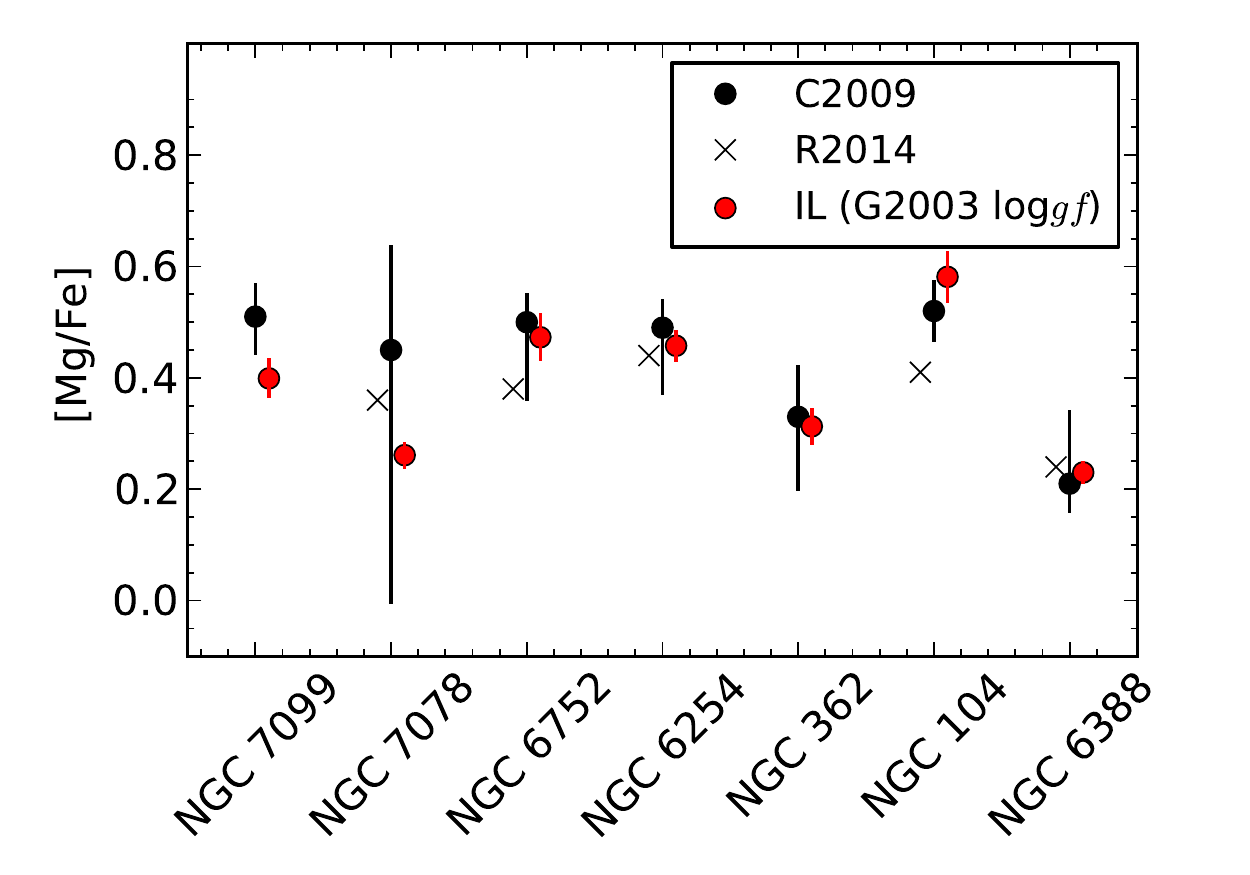}
      \caption{As the bottom panel of Fig.~\ref{fig:lightcmp}, but using $\log gf$ values for \ion{Mg}{I} lines from \citet{Gratton2003a} and basing the modelling of other lines on the Kurucz line list.}
         \label{fig:mgcmp}
   \end{figure}

We next turn to the element abundance ratios. In the context of globular clusters, the light elements (C, N, Na, O, Mg, Al) are of special interest because of the star-to-star variations in the abundances of these elements and their relation to the phenomenon of multiple stellar populations in GCs \citep[e.g.][]{Gratton2012}. In our spectra, we can measure sodium (via the \ion{Na}{I} lines at 5683/5688 \AA\ and 6154/6161 \AA) and magnesium (via several lines, see Table~\ref{tab:loggmg}). Example fits to the sodium  lines are shown in Fig.~\ref{fig:speccmp_na} for the same three GCs as in Fig.~\ref{fig:speccmp}. For all three clusters, the regions around these lines are well fitted by our model spectra. The 5683/5688~\AA\ lines are relatively clean (albeit weak at low metallicities), while the 6154/6161~\AA\ lines are more challenging to measure in our integrated-light spectra. At low metallicities these lines are very weak indeed, and at high metallicities they are blended with \ion{Si}{I} and \ion{Ca}{I} lines, the effect of which is exacerbated by the high velocity dispersion of NGC~6388. As such, these lines represent an interesting test of our full spectral fitting approach.  
As can be appreciated from Tables~\ref{tab:abN0104}-\ref{tab:abN7099}, the agreement between the sodium abundances inferred from the two sets of lines is, in fact, remarkably good, with a mean difference of only 0.02 dex and an r.m.s. dispersion of 0.02 dex. For NGC~7078, the 6154/6161~\AA\ lines only yielded an upper limit of $\mathrm{[Na/Fe]} < +0.22$ (one sigma; not included in Table~\ref{tab:abN7078}), which is consistent with the measurement based on the 5683/5688~\AA\ lines. A key feature of our approach is that the abundances of other elements that have lines within the fitted spectral windows are constrained based on other regions of the spectrum, too, and can be kept fixed with only the Na abundance (in this case) varying as a free parameter. Clearly, a requirement for this to work successfully is that the input line list contains accurate atomic parameters of the other lines.

In Fig.~\ref{fig:lightcmp} we compare our integrated-light measurements of sodium and magnesium with the results for individual stars from \citet{Carretta2009}. The clusters are ordered  according to their metallicities,  increasing from left to right.
For each cluster, we show the average of the logarithmic abundances for the individual stars (filled black markers) as well as the abundances averaged on a linear scale (open markers). The full range of $\mathrm{[Na/Fe]}$ and $\mathrm{[Mg/Fe]}$ ratios measured for individual stars are indicated by the vertical black lines. We also indicate the average literature values from \citet{Roediger2014} where available (`x'-markers).

We see that, in general, the integrated-light $\mathrm{[Na/Fe]}$ values fall within the range determined from individual stars by \citet{Carretta2009}. At the lowest metallicities the sodium lines still lie on the linear part of the curve-of-growth for a typical GC giant, so in this regime we may expect the integrated-light abundances to most closely reflect the linear average of individual stellar abundances (i.e., the open markers). At higher metallicities the sodium lines begin to saturate, and the integrated-light measurements may then yield abundances  that are closer to the logarithmic average abundances of individual stars. The actual relations will be more complicated as the lines will be on different parts of the curve-of-growth for different stars in the cluster. Regardless of which average is considered, there is a tendency for our integrated-light measurements to yield slightly lower $\mathrm{[Na/Fe]}$ ratios than the average of the individual stars. From Table~\ref{tab:abun1}-\ref{tab:abdiff2}, we see that the sodium abundances are not very sensitive to the details of the analysis; if we model the CMDs using stars within the slit scan areas the $\mathrm{[Na/Fe]}$ ratios generally change by less than 0.05 dex, and the same is true if we use theoretical isochrones instead of the empirical CMDs.

We note that the Na abundance measurements from \citet{Carretta2009} have been corrected for non-LTE effects, whereas no such correction has been attempted for our integrated-light measurements. These corrections depend on the surface temperature and gravity of the stars, but for cool giants the corrections tend to be positive \citep{Gratton1999} and might thus account for part of the offset. A full investigation of this issue is beyond the scope of this work and will be left for a future paper.

The lower panel in Fig.~\ref{fig:lightcmp} shows that the $\mathrm{[Mg/Fe]}$ values derived from our integrated-light spectra are all super-solar. This is also the conclusion reached from the observations of individual stars in the clusters, and is not unexpected for old stellar populations such as GCs. 
There is, nonetheless, again a noticeable offset between the $\mathrm{[Mg/Fe]}$ ratios derived from our standard analysis and those of \citet{Carretta2009}. The latter find [Mg/Fe] ratios as high as $\sim+0.5$ for five out of the seven clusters (with somewhat lower values for NGC~362 and NGC~6388), while our Mg abundances for the same five clusters fall in the range $\mathrm{[Mg/Fe]}=0.18$ to 0.44. 
The mean difference between our $\mathrm{[Mg/Fe]}$ ratios and those of \citet{Carretta2009} is about 0.15 dex.
The picture remains the same if we base our modelling on stars within the slit scan areas or on theoretical isochrones. However,  we recall that the $\log gf$ values for several of the Mg lines in the \citet{Castelli2004} line list differ significantly from those listed by other sources (Table~\ref{tab:loggmg}). For our observations, the lines at 4703~\AA\ and 5528~\AA\ tend to have the smallest errors and thus carry the most weight in the average values. For these two lines, the $\log gf$ values in the CH04 list are about 0.07 dex and 0.15 dex higher than those listed by most of the other sources, and our average Mg abundances increase by about 0.07 dex when the Kurucz line list is used instead. 
The $\log gf$ values in the list of \citet{Gratton2003a}, on which Carretta et al.\ based their analysis, are lower still. If we derive Mg abundances from the three lines that we have in common with G2003 (4703~\AA, 5528~\AA, 5711~\AA) and use their $\log gf$ values, while otherwise following the standard analysis, then the mean difference between our [Mg/Fe] ratios and those of \citet{Carretta2009} is reduced to $-0.07$ dex (with an r.m.s. of 0.06 dex). If we use the G2003 $\log gf$ values together with the Kurucz line list, the difference is reduced further to $-0.04$ dex (Fig.~\ref{fig:mgcmp}). (But we note that only the 5711~\AA\ line is in common with \citet{Carretta2009}, who also used redder lines (near 6319~\AA) that are not included in our analysis.).

In any case, differences at the level of 0.1 dex are not uncommon when comparing measurements of individual stars from different studies, and other studies have found lower $\mathrm{[Mg/Fe]}$ ratios for some clusters than those quoted by \citet{Carretta2009}. For example, \citet{Carretta2009} find a mean $\mathrm{[Mg/Fe]} = +0.52$ for NGC~104 (with a star-to-star r.m.s. of 0.03 dex), whereas \citet{Koch2008} find  $\mathrm{[Mg/Fe]} = +0.46$ (with an r.m.s. of 0.05 dex) and 
\citet{Thygesen2014} find a median $\mathrm{[Mg/Fe]} = +0.44$ for the same cluster. For NGC~7078, \citet{Sneden1997} found an average $\langle\mathrm{[Mg/Fe]}\rangle = +0.35\pm0.06$, compared with $\mathrm{[Mg/Fe]}=+0.45$ (with an r.m.s. of 0.19 dex for 13 stars) from \citet{Carretta2009}. 
In a recent study, \citet{Dias2016} find average [Mg/Fe] values close to $\sim +0.4$ for metal-poor Galactic GCs, which is close to the abundance ratios derived from our integrated-light analysis.

We note that a similar issue regarding the $\log gf$ values does not exist for the Na lines, where all the line lists contain essentially identical values. However, Fig.~\ref{fig:lightcmp} shows that small differences between literature averages and \citet{Carretta2009} are present also for $\mathrm{[Na/Fe]}$.

\subsubsection{Other elements}
\label{sec:otherz}

   \begin{figure}
   \centering
   \includegraphics[width=\columnwidth]{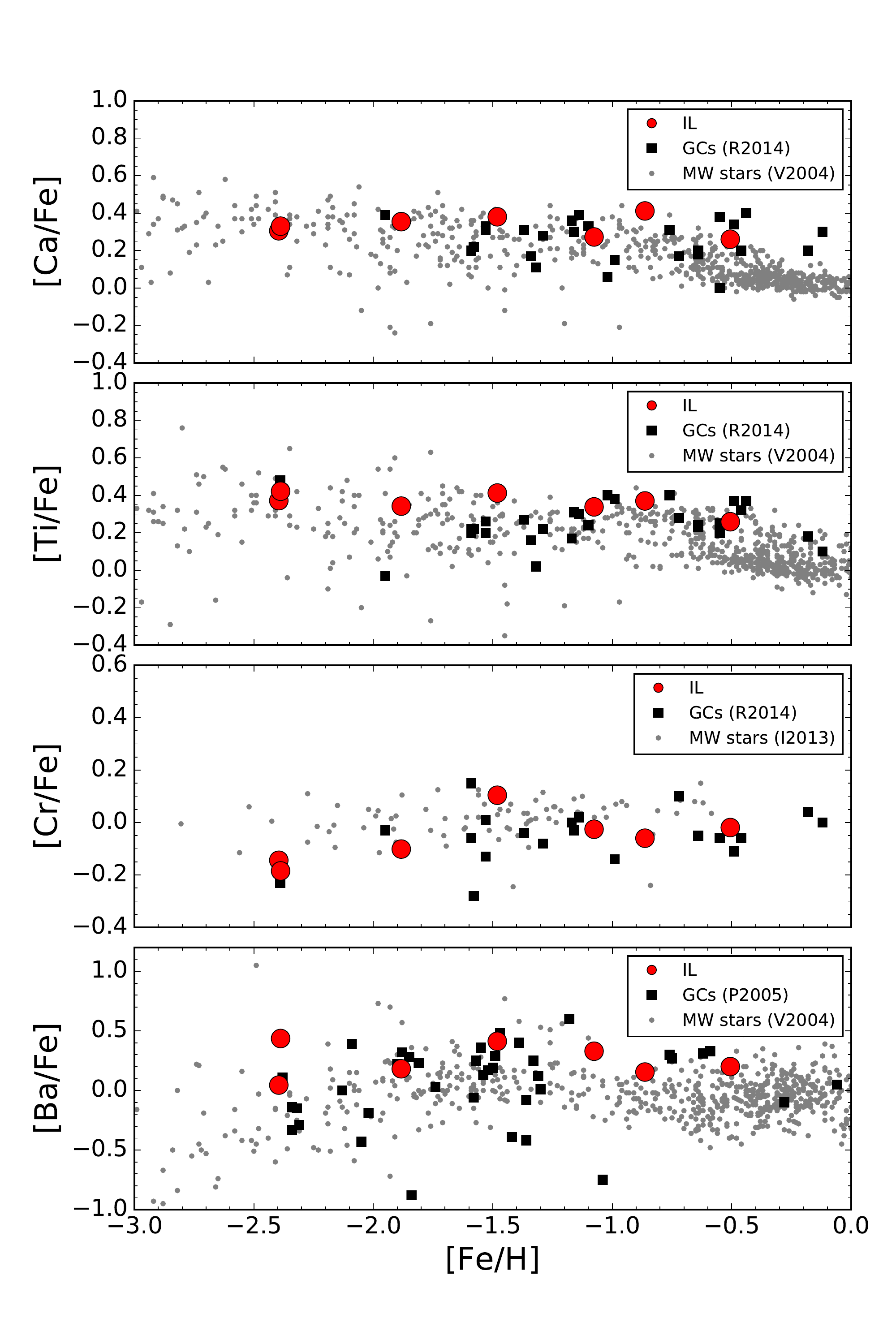}
      \caption{Comparison of our integrated-light [Ca/Fe], [Ti/Fe], [Cr/Fe], and [Ba/Fe] abundance ratios with data for individual Milky Way stars \citep[V2014, I2013:][]{Venn2004,Ishigaki2013} and literature data for globular clusters \citep[P2005, R2014: ][]{Pritzl2005,Roediger2014}.}
         \label{fig:other}
   \end{figure}

Apart from the abundances of sodium and magnesium discussed above, there is no homogeneous source of abundance measurements that we can use as a reference for a detailed one-to-one comparison with our integrated-light abundances. Instead, we compare the general trends of elemental abundance ratios as a function of metallicity with data for individual Milky Way stars \citep{Venn2004,Ishigaki2013} and the compilations of GC data \citep{Pritzl2005,Roediger2014}. The result of this comparison is shown in Fig.~\ref{fig:other} for elements that we have in common with the GC compilations: Ca, Ti, Cr, and Ba.
It should be noted that the stars in the sample of \citet{Ishigaki2013} belong primarily to the thick disc and halo components of the Milky Way, whereas the \citet{Venn2004} sample also includes thin disc stars. Hence, differences in the abundance patterns of GCs and field stars may be expected, especially at higher metallicities. 

Overall, our integrated-light abundances agree well with the literature compilations for GCs, as well as with the abundance measurements for individual stars.  At metallicities of $\mathrm{[Fe/H]}\approx-0.5$ and above, the different sequences followed by the GCs and Galactic (disc) stars become apparent.
We find the $\alpha$-elements (Ca, Ti) to be enhanced at about 0.3 dex relative to Solar-scaled abundances, with formal (unweighted) mean values of $\langle\mathrm{[Ca/Fe]}\rangle=+0.33$ and $\langle\mathrm{[Ti/Fe]}\rangle=+0.36$. The r.m.s. dispersions are 0.05 dex for both elements. 
The mean values change only slightly if the Kurucz line list is used, with the mean Ca abundance increasing to $\langle\mathrm{[Ca/Fe]}\rangle=+0.34$ and with Ti remaining unchanged at $\langle\mathrm{[Ti/Fe]}\rangle=+0.36$. 
Some of the scatter is probably real; in particular, other studies have found the metal-rich cluster NGC~6388 to be less $\alpha$-enhanced than a typical GC, with \citet{Roediger2014} quoting mean values of $\mathrm{[Ca/Fe]}=0.0\pm0.13$ and $\mathrm{[Ti/Fe]}=+0.20\pm0.19$ (the original sources being \citet{Carretta2007} and \citet{Wallerstein2007}). Our Ti abundance for this cluster ($\mathrm{[Ti/Fe]}=+0.26\pm0.04$) agrees well with the literature average, although our Ca  abundance measurement ($\mathrm{[Ca/Fe]}=+0.26\pm0.08$) is not as depleted as suggested by the literature compilation. In fact, it matches the Ti abundance, as observed in other GCs.

Cr belongs to the group of Fe-peak elements, and both Milky Way field stars \citep{Ishigaki2013} and GC literature data indicate a mean [Cr/Fe] ratio close to zero over a wide metallicity range. This is confirmed by our integrated-light abundances, for which the standard analysis yields $\langle\mathrm{[Cr/Fe]}\rangle=-0.06$ with a dispersion of 0.09 dex. The analysis based on the Kurucz line list gives $\langle\mathrm{[Cr/Fe]}\rangle=-0.04$, with a slightly larger dispersion (0.10 dex). The other two iron-peak elements that we measure (Sc, Mn) are not included in the literature compilations for GCs, but our slightly super-solar Sc abundance ratios ($\mathrm{[Sc/Fe]}=0.1-0.2$) agree well with data for individual stars with similar metallicities in the Milky Way \citep{Nissen2000,Ishigaki2013,Battistini2015}. Our $\mathrm{[Mn/Fe]}$ ratios are generally negative, decreasing from $\mathrm{[Mn/Fe]}\approx-0.2$ in NGC~104 and NGC~6388 to $\mathrm{[Mn/Fe]}\approx-0.3$ to $-0.4$ in the more metal-poor clusters. Again, this is consistent with the trend observed in Milky Way stars \citep{Nissen2000,Ishigaki2013}, although recent work suggests that this trend may be due to non-LTE effects; \citet{Battistini2015} find an essentially flat behaviour with $\mathrm{[Mn/Fe]}\approx0$ for $\mathrm{[Fe/H]}\ga-1.5$ when non-LTE corrections are applied.

\begin{table}
\caption{Barium abundances. }
\label{tab:bafe}
\centering
\begin{tabular}{lll}
\hline\hline
  & [Ba/Fe] & Ref. \\ \hline
NGC~104 & $+0.25 \pm 0.24$ & 1 \\
            & $+0.64\pm0.09$ & 2 \\
            & $+0.35\pm0.12$ (SG) & 3 \\
            & $+0.20\pm0.12$ (TO) & 3 \\
            & $+0.32\pm0.11$ (IL) & 4 \\
            & $-0.01\pm0.08$ (IL) & 5 \\
            & $+0.16\pm0.05$ & Our analysis \\
NGC~362 & $+0.42\pm0.02$ & 6 \\
      & $+0.56\pm0.30$ & 7 \\
      & $+0.28\pm0.22$ & 8 \\
      & $+0.25\pm0.07$ (IL) & 4 \\
      & $+0.33\pm0.03$ & Our analysis \\
NGC~6254 & $+0.17\pm0.09$ & 9 \\
      & $+0.41\pm0.05$ & Our analysis \\
NGC~6388 & $+0.96\pm0.21$ & 7 \\
   & $+0.21\pm0.10$ & 10 \\
   & $+0.21\pm0.07$ & 11 \\
   & $+0.00\pm0.10$ (IL) & 4 \\
   & $+0.20\pm0.15$ & Our analysis \\
NGC~6752 & $+0.18\pm0.11$ & 12 \\
  & $+0.08\pm0.06$ (IL) & 4 \\
  & $+0.18\pm0.06$ & Our analysis \\
NGC~7078 & $+0.08\pm0.02$ & 13 \\   
 & $-0.21\pm0.06$ (IL) & 5 \\
 & $+0.44\pm0.03$ & Our analysis \\
NGC~7099 & $-0.29\pm0.11$ & 14 \\ 
 & $+0.05\pm0.10$ & Our analysis \\
\hline
\end{tabular}        
\tablefoot{(IL) refers to integrated-light measurements, whereas (SG) and (TO) indicate measurements for sub-giants and turn-off stars.
}     
\tablebib{
(1)~\citet{Thygesen2014};
(2)~\citet{Worley2010a};
(3)~\citet{James2004a};
(4)~\citet{Colucci2017}; 
(5)~\citet{Sakari2014}; 
(6)~\citet{Carretta2013}; 
(7)~\citet{Worley2010}; 
(8)~\citet{Shetrone2000a}; 
(9)~\citet{Mishenina2003}; 
(10)~\citet{Carretta2007}; 
(11)~\citet{Wallerstein2007}; 
(12)~\citet{James2004}; 
(13)~\citet{Sobeck2011}; 
(14)~\citet{Shetrone2003}
}  
\end{table}

The bottom panel of Fig.~\ref{fig:other} shows that our integrated-light [Ba/Fe] ratios also appear consistent with the general trends seen for Milky Way GCs. 
Since the compilation by \citet{Pritzl2005} was published, a number of more recent determinations of Ba abundances have appeared for several of the clusters in our sample. These are summarised in Table~\ref{tab:bafe}, which also includes the integrated-light measurements from \citet{Colucci2017} and \citet{Sakari2014} for the clusters that we have in common.
For most clusters, our [Ba/Fe] ratios agree fairly well with the literature data for individual stars, but for the most metal-poor clusters (NGC~7078, NGC~7099) we find higher [Ba/Fe] ratios than those quoted in the literature (although the data for NGC~7099 are based on just a single star). 
The integrated-light Ba abundances measured by \citet{Sakari2014}, on the other hand, are lower than our values as well as those by the other literature sources for both NGC~104 and NGC~7078.

NGC~7078 is one of a few GCs that are known to exhibit a substantial star-to-star spread in Ba abundance, and the value 
of $\mathrm{[Ba/Fe]} = +0.08\pm0.02$ for \citet{Sobeck2011} in Table~\ref{tab:bafe} is the weighted mean of their measurements for nine individual stars. \citet{Sneden2000} found a mean Ba/Ca ratio of $\mathrm{[Ba/Ca]} = -0.1$ for 31 stars in this cluster or $\mathrm{[Ba/Fe]} \sim 0.25$ (if we assume $\mathrm{[Ca/Fe]} = 0.35$), somewhat higher than the value quoted by \citet{Sobeck2011}. Both values, however, are lower than our integrated-light value of $\mathrm{[Ba/Fe]} =+0.44\pm0.03$. 

As noted above, our Ba abundances are sensitive to the isotopic ratios assumed. If we use the Kurucz line list with its $s$-process dominated mixture, the [Ba/Fe] ratios increase by about 0.06 dex for NGC~104, by 0.09 dex for NGC~362, and by 0.03 dex for NGC~6388.
The [Ba/Eu] ratios in these three clusters tend to be close to the Solar value \citep{Carretta2007,Wallerstein2007,Worley2010,Thygesen2014}, which suggests an $s$-process dominated origin of the heavy elements in these clusters, although other authors have found evidence of an $r$-process dominated mixture in NGC~104 \citep{Cordero2014}.  Even if we adopt the $s$-process mixture for these clusters, our [Ba/Fe] ratios still agree well with the literature \citep[apart from the very high $\mathrm{[Ba/Fe]}$ ratio found for NGC~6388 by][]{Worley2010}.
In the low-metallicity cluster NGC~7078, the [Ba/Eu] ratio is about $-0.4$ \citep{Sneden1997}, suggesting a more $r$-process dominated origin.

\subsection{Isochrones vs.\ CMDs}
\label{sec:isocmd}

\begin{figure}
\centering
\includegraphics[width=\columnwidth]{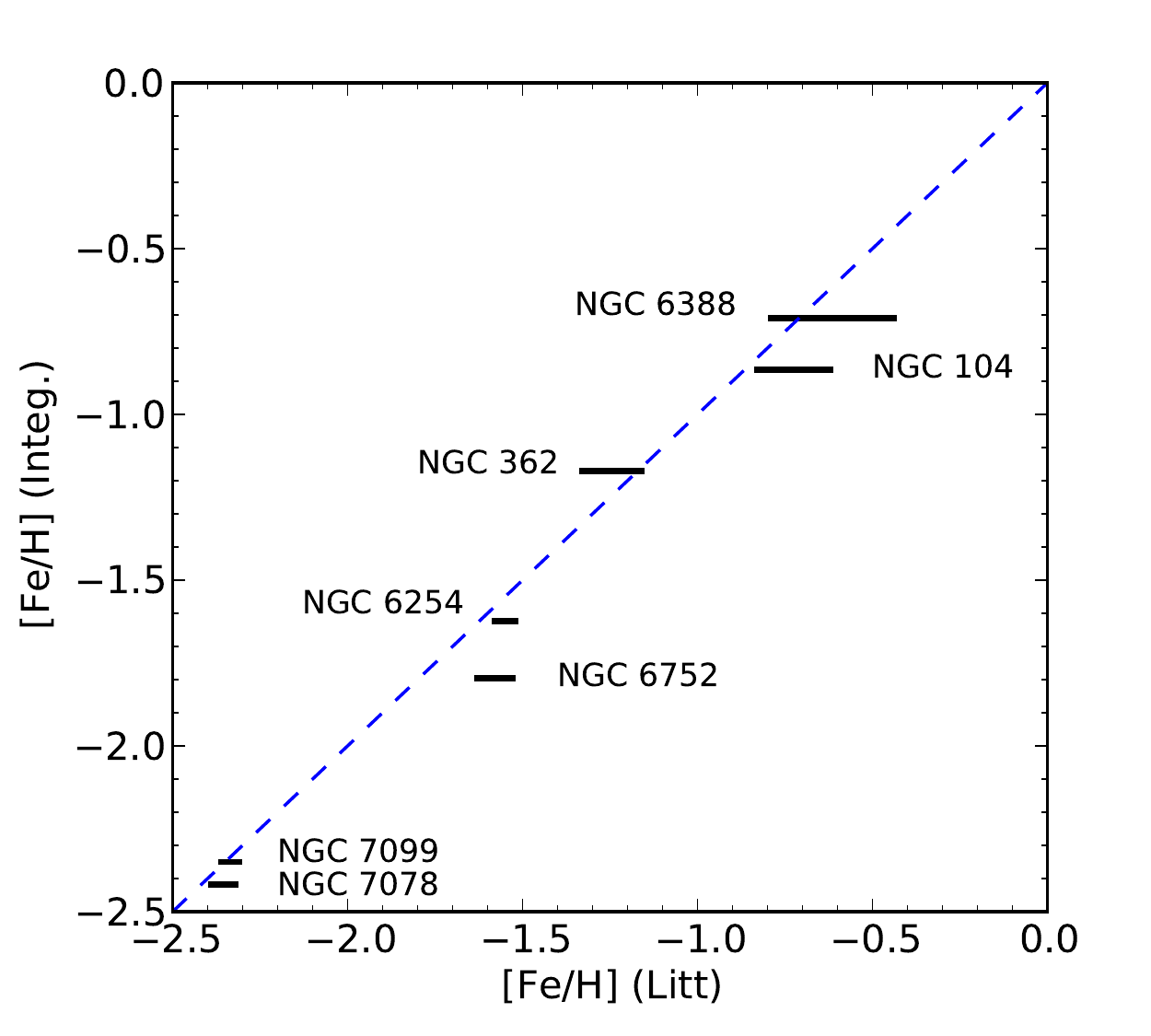}
\caption{Integrated-light iron abundances for isochrone-based analysis versus versus literature values. Symbols are the same as in Fig.~\ref{fig:fecmp}.
}
\label{fig:fecmp_iso}
\end{figure}

\begin{table*}
\caption{Differences between CMD-based and isochrone-based abundances.}
\label{tab:abdiff}
\centering
\begin{tabular}{l r r r r r r r}
\hline\hline
 & NGC 104  & NGC 362  & NGC 6254  & NGC 6388  & NGC 6752  & NGC 7078  & NGC 7099 \\
\hline
$t_\mathrm{iso}$ (Gyr) & 11 & 11 & 13 & 13 & 13 & 13 & 13 \\
\mbox{[Fe/H]}$_\mathrm{iso}$ & $-0.80$ & $-1.20$ & $-1.50$ & $-0.70$ & $-1.90$ & $-2.40$ & $-2.40$ \\
\mbox{[$\alpha$/Fe]}$_\mathrm{iso}$ & $+0.4$ & $+0.4$ & $+0.40$ & $+0.20$ & $+0.40$ & $+0.40$ & $+0.40$ \\
$\Delta_{i-c}$\mbox{[Fe/H]}  & $-0.001$  & $-0.094$  & $-0.142$  & $-0.204$  & $0.086$  & $-0.029$  & $0.045$ \\
$\Delta_{i-c}$\mbox{[Na/Fe]}  & $0.004$  & $0.028$  & $0.050$  & $0.046$  & $-0.015$  & $0.021$  & $-0.012$ \\
$\Delta_{i-c}$\mbox{[Mg/Fe]}  & $-0.005$  & $0.040$  & $-0.008$  & $0.056$  & $-0.022$  & $0.000$  & $-0.023$ \\
$\Delta_{i-c}$\mbox{[Ca/Fe]}  & $-0.036$  & $-0.066$  & $-0.036$  & $-0.054$  & $0.007$  & $-0.008$  & $-0.010$ \\
$\Delta_{i-c}$\mbox{[Sc/Fe]}  & $0.011$  & $0.019$  & $0.046$  & $-0.013$  & $-0.080$  & $0.022$  & $-0.054$ \\
$\Delta_{i-c}$\mbox{[Ti/Fe]}  & $-0.001$  & $-0.011$  & $-0.020$  & $-0.004$  & $-0.026$  & $-0.013$  & $-0.030$ \\
$\Delta_{i-c}$\mbox{[Cr/Fe]}  & $-0.007$  & $-0.024$  & $-0.027$  & $-0.022$  & $0.013$  & $0.001$  & $0.014$ \\
$\Delta_{i-c}$\mbox{[Mn/Fe]}  & $0.011$  & $-0.002$  & $0.011$  & $0.023$  & $0.013$  & $0.011$  & $-0.002$ \\
$\Delta_{i-c}$\mbox{[Ba/Fe]}  & $0.023$  & $0.046$  & $-0.045$  & $-0.008$  & $-0.032$  & $0.002$  & $-0.059$ \\
\hline
\end{tabular}
\tablefoot{The first three lines give the parameters of the isochrones \citep[from][]{Dotter2007} used in the modelling of the integrated-light spectra.
}
\end{table*}

Fig.~\ref{fig:fecmp_iso} shows the isochrone-based metallicities as a function of the literature values and in Table~\ref{tab:abdiff} we list the differences between the isochrone-based and CMD-based abundances for each cluster.
(Equivalent difference tables for the other modified analysis techniques are in Tables~\ref{tab:abdiff1}-\ref{tab:abdiff2}).
In terms of overall metallicities, the largest difference between CMD- and isochrone based results occurs for NGC~6388, for which the isochrone-based metallicity is about 0.2 dex lower than that based on the CMD. The isochrone-based metallicity remains within the range quoted in the literature, albeit towards the lower-metallicity end.
Relatively large shifts are also seen for NGC~362 ($-0.09$ dex) and NGC~6254 ($-0.14$ dex), but for both clusters the isochrone-based metallicities remain close to those found by literature studies for individual stars. For the most part, the changes to individual abundance ratios are small, typically less than about 0.05 dex, even for NGC~6388.

For NGC~362, the RGB of the $\mathrm{[Fe/H]}=-1.2$ isochrone is offset noticeably towards the red compared to the empirical CMD (Fig.~\ref{fig:cmds}). There is a hint that this cluster may be slightly less $\alpha$-enhanced than most of the others (with the exception of NGC~6388) and an isochrone with $[\alpha/\mathrm{Fe}]=+0.2$ (instead of $+0.4$) would indeed provide a better match to the observed RGB. The effect of using the less $\alpha$-enhanced isochrone is very small, however, with the iron abundance changing by just $-0.01$ dex to $\mathrm{[Fe/H]}=-1.18$.

The peculiar CMD of NGC~6388 has already been discussed above (Sec.~\ref{sec:mrgcs}), and the isochrone is indeed a relatively poor fit to the observed CMD. A better match can be obtained by decreasing the reddening correction by about 0.06 mag to $E(B-V) \approx 0.31$ (and simultaneously increasing the distance modulus by $\sim0.20$ mag). This would also lead to a decrease in the effective temperatures of the cluster stars, in turn causing a decrease in the metallicity obtained from the CMD-based analysis, and better agreement with the isochrone-based result (but then giving a somewhat lower metallicity than most recent literature estimates). However, such a low reddening would be well outside the range of $E(B-V)$ values found by other studies. In addition to the $E(B-V)=0.37$ given in the Harris catalogue, \citet{Pritzl2002} find $E(B-V)=0.40$ (from RR Lyrae stars) and quote $E(B-V)$ between 0.35 and 0.41 from other sources. The NED gives $E(B-V)=0.355$ \citep{Schlafly2011} or $E(B-V)=0.386$ \citep{Schlegel1998}. A more detailed analysis of the NGC~6388 CMD in order to constrain the reddening, distance, and other parameters appears worthwhile, but for the present work we note that reconciling the ACSGCS CMD with the Dartmouth isochrones remains problematic. 

We have chosen to base our analysis on the Dartmouth isochrones, since these are available for the full range of compositions needed here. Other commonly used grids include PARSEC \citep{Bressan2012} and BaSTI \citep{Pietrinferni2009}, as well as the recent MIST models \cite[MESA Isochrones and Stellar Tracks;][]{Choi2016,Dotter2016}. MIST has a more complete coverage of stellar evolutionary phases than the Dartmouth models (it also includes the HB and AGB), and would therefore in principle not require us to supplement the model isochrones with empirical data for the post-RGB phases. However, these models are currently only available for solar-scaled chemical composition. We nevertheless carried out a few tests using the MIST isochrones. At low metallicities we found very similar results to those obtained from our standard analysis, with $\mathrm{[Fe/H]} = -2.39$ for NGC~7078 and $\mathrm{[Fe/H]} = -2.37$ for NGC~7099 and very similar abundance ratios to those presented in this paper. At the metal-rich end, the situation is more complicated. Using the MIST isochrones by themselves, we tended to get lower metallicities than from the other methods adopted in this paper ($\mathrm{[Fe/H]}=-0.94$ for NGC~104 and $\mathrm{[Fe/H]}=-0.81$ for NGC~6388). The modelling of the post-RGB phases is still quite uncertain, and \citet{Choi2016} note that the MIST models have difficulties reproducing the observed luminosity functions of AGB stars in the Magellanic Clouds. Additionally, these late phases may be affected by the anomalous properties of stellar populations in GCs. If, instead, we only used the MIST isochrones up to the tip of the RGB and combined them with empirical HB and AGB data, then we found $\mathrm{[Fe/H]} = -0.87$ for NGC~104 (i.e., very similar to our standard analysis) and $\mathrm{[Fe/H]}=-0.65$ for NGC~6388, slightly higher than the value based on the Dartmouth isochrone. Given the mismatch between the composition of the MIST isochrones and that of the GCs, it would be premature to draw strong conclusions from this comparison, but we note that the results appear to be very robust at the metal-poor end, whereas somewhat larger model dependencies may exist in the more metal-rich regime.

\section{Discussion}

Our main aim in this paper has been to assess how reliably the chemical composition of globular clusters can be determined from spectroscopy of their integrated light. The fundamental assumption underlying the work presented here is that the ``best'' way of doing so is to compare with measurements of individual stars in the clusters. While this is probably a reasonable assumption, it should be kept in mind that even high-dispersion spectroscopy of individual stars has  associated systematic uncertainties at the level of $\sim$0.1 dex depending on details of the analysis. Hence, at this stage agreement within this tolerance must be considered satisfactory. 

Apart from the difficulty of establishing what the ``right answer'' is, integrated-light observations of Galactic globular clusters face the difficulty that the clusters have large angular sizes on the sky. As we have seen, this means that uncertainties due to stochastic fluctuations in the numbers of stars sampled within the observed area can be dominant. 
This problem will usually be less acute in observations of extragalactic GCs, and our stochastic sampling experiments indicate that the stochastically induced (one-sigma) uncertainty on overall metallicities is less than 0.1 dex even for clusters as faint as $M_V\sim-5.5$ while the effect on most abundance ratios will be even smaller. Of course, larger deviations may still occur in cases that happen to be affected by particularly ``unlucky'' realisations of the HRD.
Accurate abundance analysis is thus expected to be possible for clusters well below the turn-over of the GC luminosity function in a typical galaxy, so that the majority of the GC population in a given galaxy will generally be amenable to such integrated-light analysis. 
 Studies that target significantly fainter (or younger) clusters should, however, pay close attention to the uncertainties introduced by stochastic sampling of the HRD. To some extent, such uncertainties may be alleviated if independent information about the CMD is available, for example via resolved imaging. Indeed, we find that differences between our integrated-light abundances and literature values tend to decrease when only stars located within the scanned parts of the GCs are included. An alternative approach, suggested by \citet{Colucci2011a}, is to generate large numbers of random CMD realisations and search for those that simultaneously reproduce the spectroscopic and photometric properties (e.g., integrated colours) of the clusters.  At any rate, we note that abundance \emph{ratios} are generally less sensitive to these effects.

In modelling the integrated light of the seven globular clusters in our sample, we found that an approach based on empirical information about the colour-magnitude diagrams of the clusters (from HST/ACS photometry) yields results that are of comparable accuracy to modelling based on theoretical isochrones. This was also found by \citet{McWilliam2008} for NGC~104. This is encouraging for applications of this technique to more distant systems, where resolved photometry may not be available. One potentially outstanding issue is how to select the proper isochrone for the modelling. The metallicity and $\alpha$-enhancement can be chosen to self-consistently match the values of these quantities derived from the spectroscopy (possibly requiring some iteration), but this leaves the age and (to some extent) the horizontal branch morphology as free parameters. However, moderate uncertainties on the age have only a relatively minor effect on the abundances. In L14 we found that the difference between using a 13 Gyr and an 8 Gyr isochrone changed $\mathrm{[Fe/H]}$ by less than 0.1 dex for the GC in the WLM galaxy, which has $\mathrm{[Fe/H]}\approx-2$.  Again, element abundance ratios were even less affected, by 0.05 dex or less. Similarly, \citet{McWilliam2008} found that the metallicity of NGC~104 was basically insensitive to the choice of isochrone for ages between 10 and 15 Gyr.  Such age differences may be constrained, for example, via Balmer line strengths measured on low-resolution spectra \citep[e.g.][]{Cenarro2007,Caldwell2011}.
Even younger ages would have a more noticeable effect on the abundances, but also on other properties such as colours and mass-to-light ratios, with the latter decreasing by a factor of about 2.6 from 13 Gyr to 4 Gyr for constant metallicity (according to simple stellar population models calculated via the PARSEC website\footnote{\texttt{http://stev.oapd.inaf.it/cgi-bin/cmd}}). The mass-to-light ratios could be constrained via velocity dispersion measurements that are typically obtained as a free by-product of the spectroscopic analysis.

\begin{table}
\caption{Mean differences of our abundance measurements with respect to other studies.
}
\label{tab:abuncmp}
\centering
\begin{tabular}{l r r r}
\hline\hline
  & C2009 & P2005 & R2014  \\
  & (1) & (2) & (3) \\
\hline
$\Delta$\mbox{[Na/Fe]} & $-0.14$ (0.08) & \ldots & $-0.08$ (0.11) \\
$\Delta$\mbox{[Mg/Fe]}$_\mathrm{std}$ & $-0.15$ (0.07) & $-0.01$ (0.19) & $-0.07$ (0.08) \\
$\Delta$\mbox{[Mg/Fe]}$_\mathrm{Kur}$ & $-0.08$ (0.08) & 0.06 (0.18) & $0.00$ (0.09) \\
$\Delta$\mbox{[Ca/Fe]} & \ldots & 0.06 (0.09) & 0.12 (0.10) \\
$\Delta$\mbox{[Cr/Fe]} & \ldots & \ldots & 0.01 (0.09) \\
$\Delta$\mbox{[Ti/Fe]} & \ldots & 0.10 (0.07) & 0.08 (0.08)  \\
$\Delta$\mbox{[Ba/Fe]} & \ldots & 0.15 (0.17) & \ldots \\
\hline
\end{tabular}
\tablefoot{The differences are in the sense (integrated light) - (literature). The numbers in parentheses are the r.m.s. of the differences.
}
\tablebib{
(1)~\citet{Carretta2009};
(2)~\citet{Pritzl2005};
(3)~\citet{Roediger2014}
}
\end{table}

\begin{table*}
\caption{Abundances of light elements for the four clusters in common between our study and those of \citet{Sakari2013} and \citet{Colucci2017}.
}
\label{tab:colcmp}
\centering
\begin{tabular}{rllllll}
\hline\hline
  & & NGC~104 & NGC~362 & NGC~6388 & NGC~6752 & NGC~7078 \\ \hline
\mbox{[Na/Fe]} &  C2009 (rms) & 0.53 (0.15) & 0.19 (0.19) & 0.59 (0.16) & 0.33 (0.27) & 0.20 (0.25) \\
& \citet{Colucci2017}  & $0.24\pm0.08$ & $-0.02\pm0.10$ & $-0.17\pm0.36$ & $0.04\pm0.07$ & \ldots \\
& \citet{Sakari2013} & $0.38\pm0.12$ & \ldots & \ldots & \ldots & $0.90\pm0.40$ \\
& Std analysis & 0.42 & $-0.05$ & 0.36 & 0.30 & 0.15 \\
\mbox{[Mg/Fe]} & C2009 (rms) & 0.52 (0.03) & 0.33 (0.04) & 0.21 (0.07) & 0.50 (0.05) & 0.45 (0.19) \\
& \citet{Colucci2017} & $0.29\pm0.08$ & $0.07\pm0.05$ & $-0.02\pm0.10$ & $0.06\pm0.07$ & \ldots \\
& \citet{Sakari2013} & $0.42\pm0.14$ & \ldots & \ldots & \ldots & $-0.15\pm0.21$ \\
& Std analysis & 0.44 & 0.15 & 0.11 & 0.39 & 0.18 \\
& Kurucz line list & 0.54 & 0.25 & 0.19 & 0.44 & 0.24 \\
\hline
\end{tabular}
\tablefoot{For convenience we repeat our abundance measurements from Table~\ref{tab:abun} (standard analysis) and \ref{tab:abun2} (Kurucz line list). For \citet{Carretta2009} we include the rms dispersions as given in their Table 10. \citep[Note that the data for NGC~362 are from][]{Carretta2013}.
}
\end{table*}

The comparison of our integrated-light abundance ratio measurements with literature data for individual stars is summarised in Table~\ref{tab:abuncmp}. Although we find good overall agreement with the literature data, small systematic offsets remain.  Our $\mathrm{[Na/Fe]}$ ratios are, on average about 0.14 dex lower than those reported by \citet{Carretta2009}, and they are about 0.08 dex lower than the average of the literature values in the compilation of \citet{Roediger2014} (but note that \citet{Roediger2014} also include the work of \citet{Carretta2009} as one of their sources.). When using up-to-date oscillator strengths, our $\mathrm{[Mg/Fe]}$ ratios are 0.08 dex lower on average than those found by \citet{Carretta2009}, but part of this offset may be attributed to the slightly different $\log gf$ values used by Carretta et al.\ (Sec.~\ref{sec:light}). Furthermore, when comparing with the literature compilations the offsets largely vanish. Our average $\mathrm{[Mg/Fe]}$ ratios (for the Kurucz line list) are actually slightly higher (by 0.06 dex) than those listed by \citet{Pritzl2005}, and the average difference with respect to those in \citet{Roediger2014} is less than 0.01 dex. 

For the reasons given at the beginning of this section, our main focus in this paper has been on comparing our integrated-light measurements with data for individual stars. However, it is worth discussing how our measurements compare with those obtained from other integrated-light analyses, in particular with regards to the light elements.
In Table~\ref{tab:colcmp}  we compare our measurements of Na and Mg with the work of \citet{Colucci2017} and \citet{Sakari2013}. For the four clusters in common with \citet{Colucci2017}, our $\mathrm{[Na/Fe]}$ and $\mathrm{[Mg/Fe]}$ ratios are on average 0.24 dex (r.m.s. 0.20 dex) and 0.26 dex (r.m.s. 0.08 dex) higher (when using our Mg abundances based on the Kurucz line list). The overall differences, as well as the spread, exceed the uncertainties due to stochastic fluctuations by a substantial margin.
\citet{Colucci2017} consider several effects that may cause systematic errors in their Mg abundances, such as the choice of damping constants, mismatch in the dwarf-to-giant ratios in the CMDs, or non-LTE effects, but conclude that none of them can fully explain their low [Mg/Fe] ratios. 
Our $\mathrm{[Na/Fe]}$ and $\mathrm{[Mg/Fe]}$ ratios for NGC~104 are in excellent agreement with those found by \citet{Sakari2013}, while the measurements of Sakari et al. for NGC~7078 differ more substantially from both ours and those of \citet{Carretta2009}. Sakari et al.\ suggest that their Mg and Na abundances may be affected by intra-cluster abundance variations, but it should also be noted that their measurements of Mg and Na in NGC~7078 are based on only a single line each, with large uncertainties (\ion{Mg}{I} 5528 \AA\ and \ion{Na}{I} 6154 \AA).
Overall, the results presented here, as well as the relatively normal [Mg/Fe] ratios found by \citet{Sakari2013}, suggest that the inherent uncertainties in measuring Mg abundances from integrated light are comparable to those for other elements. 

The normal $\mathrm{[Mg/Fe]}$ ratios found here are in contrast to the low $\mathrm{[Mg/Fe]}$ ratios that have been reported from integrated-light observations of some extragalactic GCs \cite[L12, L14,][]{Colucci2009,Colucci2014}. If these very low (in some cases significantly sub-solar) [Mg/Fe] ratios are caused by internal Mg abundance variations, then these clusters must have a much more extensive Mg/Al anti-correlation than those typical of Galactic GCs, with large numbers of very Mg-depleted stars. Of the clusters observed here, the largest internal spread in Mg abundance is found in NGC~7078, but we still measure a significantly super-solar Mg/Fe abundance ratio for this cluster ($\mathrm{[Mg/Fe]} = +0.24\pm0.05$ when using the Kurucz line list).  The origin of the light-element abundance spreads in GCs is poorly understood, and it is unclear why the extragalactic GCs should have a much larger Mg spread than the well-studied Galactic GCs. It is possible that environmental effects might play a role, and in this context it may be worth noting that the most extreme Mg spread among Galactic GCs is found in the remote cluster NGC~2419 \citep{Cohen2012}.

For  Ca, Ti, and Cr our average abundance ratios lie within 0.1 dex of those quoted in the literature compilations for individual stars. Our average abundances of Ca and Ti are also very similar to those measured in integrated light for five GCs by \citet{Sakari2014}, who find $\langle \mathrm{[Ca/Fe]\rangle} = +0.33$ (r.m.s. 0.06 dex) and $\langle \mathrm{[Ti/Fe]\rangle} = +0.34$ (r.m.s. 0.04 dex). \citet{Colucci2017} find slightly lower mean values of $\langle \mathrm{[Ca/Fe]\rangle} = +0.24$ (r.m.s. 0.09 dex) and $\langle \mathrm{[Ti/Fe]\rangle} = +0.21$ (r.m.s. 0.16 dex) for the 12 GCs in their sample. 
Since we do not measure \ion{Ti}{I} and \ion{Ti}{II} separately, we have used the average of the \ion{Ti}{I} and \ion{Ti}{II} abundances measured by Sakari et al. and Colucci et al.\ in this comparison. While the \ion{Ti}{I} and \ion{Ti}{II} abundance measurements can differ significantly for some individual clusters, the mean differences are only $\langle\mathrm{[\ion{Ti}{I}/Fe]}\rangle - \langle\mathrm{[\ion{Ti}{II}/Fe]}\rangle = -0.06\pm0.06$ dex  (for Sakari et al. study) and $\langle\mathrm{[\ion{Ti}{I}/Fe]}\rangle - \langle\mathrm{[\ion{Ti}{II}/Fe]}\rangle = +0.07\pm0.04$ dex  (Colucci et al.), i.e., only marginally significant and with opposite signs. Using the average values in the comparison thus appears justified.
 While a detailed comparison is somewhat hampered by the limited overlap between the different samples, differences of $\sim$0.10--0.15 dex between our average Ca and Ti abundances and those of Colucci et al. persist if we restrict the comparison to the four clusters that we have in common (NGC~104, NGC~362, NGC~6388, and NGC~6752). These differences are in the same sense as the offsets in Table~\ref{tab:abuncmp}.

For Ba, the average difference between our measurements and the literature data is slightly larger (0.15 dex) but we have noted that this average may hide a trend with metallicity, with our [Ba/Fe] ratios being about 0.2--0.4 dex higher than the literature values for the two most metal-poor clusters, NGC~7078 and NGC~7099.
In addition to the isotopic mixture, it may again be necessary to consider the role of non-LTE effects here. In their analysis of giants in NGC~104, \citet{Thygesen2014} found that NLTE corrections could reach up to $\sim+0.2$ dex in the coolest giants. According to the larger grid of non-LTE corrections for Ba that has recently been published by \citet{Korotin2015}, the corrections become negative at low metallicities, reaching up to $-0.3$ dex for some lines. At least qualitatively, this would be consistent with the trends observed here, although detailed modelling of non-LTE effects on the integrated light will be required in order to make more quantitative statements.

In summary, we conclude that it is indeed possible to measure chemical abundances reliably (to within $\sim0.1$ dex) for most elements from integrated-light spectra of GCs. This is in line with conclusions reached via similar comparisons by other authors \citep{McWilliam2008,Sakari2013,Sakari2014,Colucci2017}. Nevertheless, there is still room for refinement of the integrated-light analysis techniques. Abundance analysis of individual stars now commonly includes a treatment of non-LTE effects, which can be important for some elements. These have so far not been incorporated in our analysis, but might account for some of the remaining systematic differences with respect to studies of individual stars (e.g., for Na and Ba). When aiming for better than $\sim0.1$ dex accuracy, it may also be necessary to consider effects such as atomic diffusion, which can cause systematic variations in the surface abundances at the level of 0.1--0.2 dex as a function of evolutionary stage \citep{Korn2007,Nordlander2012,Husser2016}.

\section{Summary and conclusions}

We have presented integrated-light spectroscopic observations of seven Galactic globular clusters. From these observations, we have measured the abundances of several chemical elements and compared with results for individual stars in the clusters compiled from the literature. Our main findings are as follows:

\begin{itemize}
\item The iron abundances generally agree well (within $\sim$0.1 dex) with those in the literature, regardless of whether we base the modelling of the integrated light on empirical colour-magnitude diagrams of the clusters or on theoretical (Dartmouth) isochrones. 
\item A significant outlier is NGC~6752, for which we find a difference of about 0.3 dex between our integrated-light iron abundance and literature values. We attribute this to the difficulty of obtaining a good integrated-light spectrum of this very extended and relatively low surface brightness cluster, noting that the total luminosity sampled by the integrated-light spectrum ($M_V=-5.4$) is significantly lower for this cluster than for other clusters in our sample. Indeed, if we re-derive the abundances based on a CMD containing only those stars \emph{estimated} to be contained within the slit scan area, the discrepancy is reduced to 0.10--0.15 dex. 
\item Comparing our $\mathrm{[Na/Fe]}$ and $\mathrm{[Mg/Fe]}$ ratios with those found by \citet{Carretta2009} for individual stars, we find both abundance ratios to be lower (by about 0.15 dex) than the average values in the Carretta work. 
For $\mathrm{[Na/Fe]}$ we have not identified a clear cause of this difference, but we suggest that the role of non-LTE effects on integrated-light abundances may be worth investigating. For $\mathrm{[Mg/Fe]}$, most of the difference is attributable to the different line lists used in the different studies. When we use \ion{Mg}{I} oscillator strengths from the same source as \citet{Carretta2009}, the mean difference with respect to Carretta's work is only 0.04--0.07 dex (although this may be somewhat fortuitous, considering the limited overlap in the actual lines used). When comparing with Mg abundances from other literature sources, there is essentially no systematic difference with respect to our measurements, and our [Na/Fe] ratios are only $\sim0.08$ dex lower than the literature average.
\item The $\alpha$-element abundance ratios show the usual enhancement compared to the Solar-scaled abundance patterns ($\langle\mathrm{[Ca/Fe]}\rangle = +0.33$, $\langle\mathrm{[Ti/Fe]}\rangle=+0.36$). The Fe-peak element Cr is very close to the Solar-scaled ratio ($\langle\mathrm{[Cr/Fe]}\rangle = -0.06$), as observed for GCs and Milky Way stars in general, and we also find $\mathrm{[Sc/Fe]}$ and $\mathrm{[Mn/Fe]}$ ratios similar to those seen in Milky Way stars.
\item When comparing our $\mathrm{[Ba/Fe]}$ measurements with recent literature data, we find good agreement at the metal-rich end, whereas we find higher [Ba/Fe] ratios at low metallicities compared to measurements of individual stars. 
\end{itemize}

The overall agreement between published abundance measurements for individual stars in the clusters and our integrated-light measurements is, of course, encouraging, as is the relative insensitivity of the results to the exact approach used in modelling the Hertzsprung-Russell diagrams of the clusters. This adds confidence to the metallicities and chemical abundances derived for metal-poor GCs in our previous work on the Fornax and WLM dwarf galaxies. While a number of consistency checks were already carried out in those papers, the analysis presented here demonstrates that we can also extend such work into the more metal-rich regime.

For the majority of elements that we have compared here, our abundance ratios agree well with those derived from observations of individual stars.  Abundances of Mg agree well with those of other studies, once we adopt the most recent line lists, but the Mg abundances in our previous papers may be underestimated by $\sim0.1$ dex.
Future refinements to the modelling of integrated light that include a treatment of non-LTE effects and stellar evolutionary effects (such as atomic diffusion) may reduce the remaining systematic offsets for Na and Ba. 

With current 8--10 m telescopes it remains challenging to push this type of analysis significantly beyond the Local Group for old GCs, although it has been applied to GCs as far away as the M81 group (L14) and NGC~5128 \citep{Colucci2013}, at distances of $\sim4$ Mpc for more limited sets of elements. However, with the 30--40 m Extremely Large Telescopes of the next decade, it will be possible to obtain high S/N spectra for clusters fainter than $V=21$ in a few hours of integration time, which will be suitable for detailed abundance measurements such as those carried out in this paper. Combined with the multiplexing capability that is foreseen for ELT spectrographs \citep[e.g. MOSAIC on the European ELT;][]{Hammer2016}, this will allow efficient observations of more than a thousand GCs in NGC~5128 \citep{Woodley2010a,Harris2012,Taylor2016}, well over a hundred clusters in galaxies such as the ``Sombrero'' \citep{Spitler2006}, and the brightest GCs in galaxies as far away as the Virgo and Fornax clusters. At the same time, adaptive optics assisted imaging with ELTs will provide photometry of individual stars at similar distances  \citep{Deep2011}. This combination of detailed chemistry from GCs and resolved imaging of diffuse stellar populations will provide essential constraints on the assembly- and chemical enrichment histories of galaxies in a representative volume of the Universe.

\begin{acknowledgements}
We thank the referee, Charli M.\ Sakari, for a prompt and helpful report.
This research has made use of the NASA/IPAC Extragalactic Database (NED) which is operated by the Jet Propulsion Laboratory, California Institute of Technology, under contract with the National Aeronautics and Space Administration.
This work has made use of the VALD database, operated at Uppsala University, the Institute of Astronomy RAS in Moscow, and the University of Vienna. This research has made use of NASA's Astrophysics Data System Bibliographic Services. 
The Digitized Sky Surveys were produced at the Space Telescope Science Institute under U.S. Government grant NAG W-2166. The images of these surveys are based on photographic data obtained using the Oschin Schmidt Telescope on Palomar Mountain and the UK Schmidt Telescope. The plates were processed into the present compressed digital form with the permission of these institutions.
JB acknowledges support by NSF grant AST-1518294 and HST grant GO-13295.001-A.
JS acknowledges support by NSF grant AST-1514763 and a Packard Fellowship.
\end{acknowledgements}

\bibliographystyle{aa}
\bibliography{refs.bib}

\begin{appendix} 
\section{Individual abundance measurements}


\longtab[1]{
\begin{longtable}{lcc}
\caption{\label{tab:abN0104} Individual abundance measurements for NGC~104 (standard analysis)}\\
\hline\hline
Wavelength [\AA] & Value & Error \\ \hline
\endfirsthead
\caption{continued.}\\
\hline\hline
Wavelength [\AA] & Value & Error \\ \hline
\endhead
\hline
\endfoot
\mbox{[Fe/H]} \\
4200.0--4400.0 & $-0.794$ & 0.002 \\
4400.0--4600.0 & $-0.804$ & 0.001 \\
4600.0--4800.0 & $-0.835$ & 0.003 \\
4800.0--5000.0 & $-0.845$ & 0.001 \\
5000.0--5150.0 & $-0.885$ & 0.001 \\
5250.0--5400.0 & $-0.915$ & 0.002 \\
5400.0--5600.0 & $-0.943$ & 0.003 \\
5600.0--5800.0 & $-0.885$ & 0.002 \\
6000.0--6200.0 & $-0.862$ & 0.003 \\
\mbox{[Na/Fe]} \\
5677.0--5695.0 & $+0.412$ & 0.010 \\
6149.0--6166.0 & $+0.429$ & 0.007 \\
\mbox{[Mg/Fe]} \\
4347.0--4357.0 & $+0.289$ & 0.021 \\
4565.0--4576.0 & $+0.289$ & 0.009 \\
4700.0--4707.0 & $+0.481$ & 0.011 \\
5523.0--5531.5 & $+0.439$ & 0.005 \\
5705.0--5715.0 & $+0.538$ & 0.002 \\
\mbox{[Ca/Fe]} \\
4222.0--4232.0 & $+0.420$ & 0.007 \\
4280.0--4320.0 & $+0.488$ & 0.009 \\
4420.0--4460.0 & $+0.408$ & 0.004 \\
4575.0--4591.0 & $+0.408$ & 0.005 \\
5259.0--5268.0 & $+0.650$ & 0.011 \\
5580.0--5610.0 & $+0.378$ & 0.011 \\
6100.0--6175.0 & $+0.199$ & 0.007 \\
\mbox{[Sc/Fe]} \\
4290.0--4330.0 & $+0.501$ & 0.015 \\
4350.0--4440.0 & $+0.338$ & 0.010 \\
4665.0--4675.0 & $-0.070$ & 0.030 \\
5026.0--5036.0 & $+0.060$ & 0.020 \\
5521.0--5531.0 & $+0.130$ & 0.021 \\
5638.0--5690.0 & $+0.079$ & 0.011 \\
\mbox{[Ti/Fe]} \\
4292.0--4320.0 & $+0.550$ & 0.002 \\
4386.0--4420.0 & $+0.489$ & 0.002 \\
4440.0--4474.0 & $+0.330$ & 0.001 \\
4532.0--4574.0 & $+0.360$ & 0.002 \\
4587.0--4593.0 & $-0.010$ & 0.026 \\
4650.0--4715.0 & $+0.259$ & 0.001 \\
4750.0--4850.0 & $+0.259$ & 0.001 \\
4980.0--5045.0 & $+0.399$ & 0.002 \\
\mbox{[Cr/Fe]} \\
4250.0--4292.0 & $+0.030$ & 0.002 \\
4350.0--4400.0 & $-0.109$ & 0.012 \\
4520.0--4660.0 & $+0.030$ & 0.003 \\
5235.0--5330.0 & $-0.011$ & 0.003 \\
5342.0--5351.0 & $-0.340$ & 0.008 \\
5407.0--5413.0 & $-0.290$ & 0.027 \\
\mbox{[Mn/Fe]} \\
4750.0--4790.0 & $-0.160$ & 0.011 \\
6010.0--6030.0 & $-0.340$ & 0.015 \\
\mbox{[Ba/Fe]} \\
4551.0--4560.0 & $+0.102$ & 0.010 \\
4929.0--4939.0 & $+0.134$ & 0.016 \\
5849.0--5859.0 & $+0.348$ & 0.026 \\
6135.0--6145.0 & $+0.181$ & 0.021 \\
\end{longtable}
}

\longtab[2]{
\begin{longtable}{lcc}
\caption{\label{tab:abN0362} Individual abundance measurements for NGC~362 (standard analysis)}\\
\hline\hline
Wavelength [\AA] & Value & Error \\ \hline
\endfirsthead
\caption{continued.}\\
\hline\hline
Wavelength [\AA] & Value & Error \\ \hline
\endhead
\hline
\endfoot
\mbox{[Fe/H]} \\
4200.0--4400.0 & $-1.081$ & 0.005 \\
4400.0--4600.0 & $-1.061$ & 0.010 \\
4600.0--4800.0 & $-1.081$ & 0.011 \\
4800.0--5000.0 & $-1.102$ & 0.006 \\
5000.0--5150.0 & $-1.092$ & 0.006 \\
5250.0--5400.0 & $-1.002$ & 0.005 \\
5400.0--5600.0 & $-1.131$ & 0.005 \\
5600.0--5800.0 & $-1.121$ & 0.010 \\
6000.0--6200.0 & $-1.012$ & 0.006 \\
\mbox{[Na/Fe]} \\
5677.0--5695.0 & $-0.045$ & 0.015 \\
6149.0--6166.0 & $-0.045$ & 0.030 \\
\mbox{[Mg/Fe]} \\
4347.0--4357.0 & $-0.095$ & 0.030 \\
4565.0--4576.0 & $+0.196$ & 0.020 \\
4700.0--4707.0 & $+0.196$ & 0.020 \\
5523.0--5531.5 & $+0.135$ & 0.016 \\
5705.0--5715.0 & $+0.205$ & 0.021 \\
\mbox{[Ca/Fe]} \\
4222.0--4232.0 & $+0.196$ & 0.011 \\
4280.0--4320.0 & $+0.345$ & 0.011 \\
4420.0--4460.0 & $+0.325$ & 0.005 \\
4575.0--4591.0 & $+0.164$ & 0.015 \\
4873.0--4883.0 & $+0.814$ & 0.045 \\
5259.0--5268.0 & $+0.514$ & 0.016 \\
5580.0--5610.0 & $+0.276$ & 0.016 \\
6100.0--6175.0 & $+0.135$ & 0.006 \\
\mbox{[Sc/Fe]} \\
4290.0--4330.0 & $+0.276$ & 0.016 \\
4350.0--4440.0 & $+0.154$ & 0.016 \\
4665.0--4675.0 & $-0.005$ & 0.041 \\
5026.0--5036.0 & $+0.055$ & 0.030 \\
5521.0--5531.0 & $-0.016$ & 0.030 \\
5638.0--5690.0 & $+0.055$ & 0.011 \\
\mbox{[Ti/Fe]} \\
4292.0--4320.0 & $+0.445$ & 0.011 \\
4386.0--4420.0 & $+0.435$ & 0.011 \\
4440.0--4474.0 & $+0.295$ & 0.011 \\
4532.0--4574.0 & $+0.325$ & 0.005 \\
4587.0--4593.0 & $+0.526$ & 0.036 \\
4650.0--4715.0 & $+0.266$ & 0.011 \\
4750.0--4850.0 & $+0.295$ & 0.005 \\
4980.0--5045.0 & $+0.335$ & 0.005 \\
5152.5--5160.0 & $+0.225$ & 0.035 \\
\mbox{[Cr/Fe]} \\
4250.0--4292.0 & $-0.004$ & 0.010 \\
4350.0--4400.0 & $-0.134$ & 0.016 \\
4520.0--4660.0 & $+0.006$ & 0.005 \\
5235.0--5330.0 & $+0.026$ & 0.005 \\
5342.0--5351.0 & $-0.235$ & 0.025 \\
5407.0--5413.0 & $-0.344$ & 0.045 \\
\mbox{[Mn/Fe]} \\
4750.0--4790.0 & $-0.375$ & 0.015 \\
6010.0--6030.0 & $-0.364$ & 0.020 \\
\mbox{[Ba/Fe]} \\
4551.0--4560.0 & $+0.286$ & 0.015 \\
4929.0--4939.0 & $+0.365$ & 0.021 \\
5849.0--5859.0 & $+0.436$ & 0.030 \\
6135.0--6145.0 & $+0.306$ & 0.021 \\
\end{longtable}
}

\longtab[3]{
\begin{longtable}{lcc}
\caption{\label{tab:abN6254} Individual abundance measurements for NGC~6254 (standard analysis)}\\
\hline\hline
Wavelength [\AA] & Value & Error \\ \hline
\endfirsthead
\caption{continued.}\\
\hline\hline
Wavelength [\AA] & Value & Error \\ \hline
\endhead
\hline
\endfoot
\mbox{[Fe/H]} \\
4200.0--4400.0 & $-1.472$ & 0.005 \\
4400.0--4600.0 & $-1.492$ & 0.005 \\
4600.0--4800.0 & $-1.482$ & 0.005 \\
4800.0--5000.0 & $-1.503$ & 0.006 \\
5000.0--5150.0 & $-1.462$ & 0.005 \\
5250.0--5400.0 & $-1.423$ & 0.006 \\
5400.0--5600.0 & $-1.513$ & 0.006 \\
5600.0--5800.0 & $-1.542$ & 0.005 \\
6000.0--6200.0 & $-1.442$ & 0.005 \\
\mbox{[Na/Fe]} \\
5677.0--5695.0 & $-0.040$ & 0.025 \\
6149.0--6166.0 & $-0.032$ & 0.051 \\
\mbox{[Mg/Fe]} \\
4347.0--4357.0 & $+0.538$ & 0.051 \\
4565.0--4576.0 & $+0.270$ & 0.036 \\
4700.0--4707.0 & $+0.339$ & 0.025 \\
5523.0--5531.5 & $+0.379$ & 0.025 \\
5705.0--5715.0 & $+0.429$ & 0.025 \\
\mbox{[Ca/Fe]} \\
4222.0--4232.0 & $+0.379$ & 0.010 \\
4280.0--4320.0 & $+0.470$ & 0.020 \\
4420.0--4460.0 & $+0.410$ & 0.010 \\
4575.0--4591.0 & $+0.068$ & 0.021 \\
4873.0--4883.0 & $+0.960$ & 0.070 \\
5259.0--5268.0 & $+0.610$ & 0.016 \\
5580.0--5610.0 & $+0.458$ & 0.011 \\
6100.0--6175.0 & $+0.270$ & 0.005 \\
\mbox{[Sc/Fe]} \\
4290.0--4330.0 & $+0.049$ & 0.025 \\
4350.0--4440.0 & $+0.090$ & 0.021 \\
4665.0--4675.0 & $+0.229$ & 0.041 \\
5026.0--5036.0 & $+0.029$ & 0.036 \\
5521.0--5531.0 & $+0.139$ & 0.030 \\
5638.0--5690.0 & $+0.199$ & 0.016 \\
\mbox{[Ti/Fe]} \\
4292.0--4320.0 & $+0.470$ & 0.010 \\
4386.0--4420.0 & $+0.530$ & 0.016 \\
4440.0--4474.0 & $+0.349$ & 0.016 \\
4532.0--4574.0 & $+0.410$ & 0.005 \\
4587.0--4593.0 & $+0.869$ & 0.041 \\
4650.0--4715.0 & $+0.270$ & 0.016 \\
4750.0--4850.0 & $+0.358$ & 0.016 \\
4980.0--5045.0 & $+0.358$ & 0.011 \\
5152.5--5160.0 & $+0.538$ & 0.030 \\
\mbox{[Cr/Fe]} \\
4250.0--4292.0 & $-0.100$ & 0.020 \\
4350.0--4400.0 & $-0.231$ & 0.030 \\
4520.0--4660.0 & $-0.021$ & 0.011 \\
5235.0--5330.0 & $-0.051$ & 0.011 \\
5342.0--5351.0 & $+0.069$ & 0.030 \\
5407.0--5413.0 & $+1.159$ & 0.021 \\
\mbox{[Mn/Fe]} \\
4750.0--4790.0 & $-0.340$ & 0.020 \\
6010.0--6030.0 & $-0.610$ & 0.030 \\
\mbox{[Ba/Fe]} \\
4551.0--4560.0 & $+0.300$ & 0.021 \\
4929.0--4939.0 & $+0.399$ & 0.025 \\
5849.0--5859.0 & $+0.500$ & 0.035 \\
6135.0--6145.0 & $+0.500$ & 0.020 \\
\end{longtable}
}

\longtab[4]{
\begin{longtable}{lcc}
\caption{\label{tab:abN6388} Individual abundance measurements for NGC~6388 (standard analysis)}\\
\hline\hline
Wavelength [\AA] & Value & Error \\ \hline
\endfirsthead
\caption{continued.}\\
\hline\hline
Wavelength [\AA] & Value & Error \\ \hline
\endhead
\hline
\endfoot
\mbox{[Fe/H]} \\
4200.0--4400.0 & $-0.521$ & 0.002 \\
4400.0--4600.0 & $-0.472$ & 0.002 \\
4600.0--4800.0 & $-0.470$ & 0.004 \\
4800.0--5000.0 & $-0.521$ & 0.001 \\
5000.0--5150.0 & $-0.531$ & 0.001 \\
5250.0--5400.0 & $-0.502$ & 0.003 \\
5400.0--5600.0 & $-0.542$ & 0.001 \\
5600.0--5800.0 & $-0.573$ & 0.004 \\
6000.0--6200.0 & $-0.426$ & 0.005 \\
\mbox{[Na/Fe]} \\
5677.0--5695.0 & $+0.343$ & 0.010 \\
6149.0--6166.0 & $+0.369$ & 0.010 \\
\mbox{[Mg/Fe]} \\
4347.0--4357.0 & $-0.255$ & 0.030 \\
4565.0--4576.0 & $-0.006$ & 0.015 \\
4700.0--4707.0 & $+0.158$ & 0.005 \\
5523.0--5531.5 & $+0.149$ & 0.008 \\
5705.0--5715.0 & $+0.115$ & 0.009 \\
\mbox{[Ca/Fe]} \\
4222.0--4232.0 & $+0.033$ & 0.010 \\
4280.0--4320.0 & $+0.367$ & 0.005 \\
4420.0--4460.0 & $+0.189$ & 0.005 \\
4575.0--4591.0 & $+0.136$ & 0.016 \\
5259.0--5268.0 & $+0.683$ & 0.010 \\
5580.0--5610.0 & $+0.283$ & 0.007 \\
6100.0--6175.0 & $+0.127$ & 0.005 \\
\mbox{[Sc/Fe]} \\
4290.0--4330.0 & $+0.360$ & 0.016 \\
4350.0--4440.0 & $+0.355$ & 0.020 \\
4665.0--4675.0 & $-0.414$ & 0.035 \\
5026.0--5036.0 & $+0.026$ & 0.030 \\
5521.0--5531.0 & $+0.155$ & 0.026 \\
5638.0--5690.0 & $+0.026$ & 0.005 \\
\mbox{[Ti/Fe]} \\
4292.0--4320.0 & $+0.385$ & 0.012 \\
4386.0--4420.0 & $+0.336$ & 0.003 \\
4440.0--4474.0 & $+0.126$ & 0.008 \\
4532.0--4574.0 & $+0.197$ & 0.010 \\
4587.0--4593.0 & $-0.120$ & 0.036 \\
4650.0--4715.0 & $+0.317$ & 0.009 \\
4750.0--4850.0 & $+0.155$ & 0.001 \\
4980.0--5045.0 & $+0.406$ & 0.009 \\
\mbox{[Cr/Fe]} \\
4250.0--4292.0 & $+0.069$ & 0.004 \\
4350.0--4400.0 & $-0.016$ & 0.016 \\
4520.0--4660.0 & $-0.004$ & 0.004 \\
5235.0--5330.0 & $+0.074$ & 0.004 \\
5342.0--5351.0 & $-0.525$ & 0.009 \\
5407.0--5413.0 & $+0.255$ & 0.013 \\
\mbox{[Mn/Fe]} \\
4750.0--4790.0 & $-0.005$ & 0.010 \\
6010.0--6030.0 & $-0.304$ & 0.011 \\
\mbox{[Ba/Fe]} \\
4551.0--4560.0 & $-0.095$ & 0.011 \\
4929.0--4939.0 & $+0.334$ & 0.015 \\
5849.0--5859.0 & $+0.606$ & 0.021 \\
6135.0--6145.0 & $+0.257$ & 0.016 \\
\end{longtable}
}

\longtab[5]{
\begin{longtable}{lcc}
\caption{\label{tab:abN6752} Individual abundance measurements for NGC~6752 (standard analysis)}\\
\hline\hline
Wavelength [\AA] & Value & Error \\ \hline
\endfirsthead
\caption{continued.}\\
\hline\hline
Wavelength [\AA] & Value & Error \\ \hline
\endhead
\hline
\endfoot
\mbox{[Fe/H]} \\
4200.0--4400.0 & $-1.861$ & 0.005 \\
4400.0--4600.0 & $-1.841$ & 0.005 \\
4600.0--4800.0 & $-1.861$ & 0.005 \\
4800.0--5000.0 & $-1.881$ & 0.005 \\
5000.0--5150.0 & $-1.910$ & 0.005 \\
5250.0--5400.0 & $-1.941$ & 0.005 \\
5400.0--5600.0 & $-1.951$ & 0.011 \\
5600.0--5800.0 & $-1.861$ & 0.005 \\
6000.0--6200.0 & $-1.841$ & 0.011 \\
\mbox{[Na/Fe]} \\
5677.0--5695.0 & $+0.291$ & 0.011 \\
6149.0--6166.0 & $+0.352$ & 0.030 \\
\mbox{[Mg/Fe]} \\
4347.0--4357.0 & $+0.461$ & 0.026 \\
4565.0--4576.0 & $+0.152$ & 0.030 \\
4700.0--4707.0 & $+0.371$ & 0.015 \\
5523.0--5531.5 & $+0.262$ & 0.020 \\
5705.0--5715.0 & $+0.551$ & 0.016 \\
\mbox{[Ca/Fe]} \\
4222.0--4232.0 & $+0.242$ & 0.005 \\
4280.0--4320.0 & $+0.332$ & 0.005 \\
4420.0--4460.0 & $+0.392$ & 0.011 \\
4575.0--4591.0 & $+0.322$ & 0.016 \\
4873.0--4883.0 & $+0.772$ & 0.030 \\
5259.0--5268.0 & $+0.483$ & 0.015 \\
5580.0--5610.0 & $+0.493$ & 0.010 \\
6100.0--6175.0 & $+0.252$ & 0.010 \\
\mbox{[Sc/Fe]} \\
4290.0--4330.0 & $+0.203$ & 0.010 \\
4350.0--4440.0 & $+0.103$ & 0.010 \\
4665.0--4675.0 & $+0.002$ & 0.030 \\
5026.0--5036.0 & $+0.113$ & 0.026 \\
5521.0--5531.0 & $-0.038$ & 0.026 \\
5638.0--5690.0 & $+0.123$ & 0.010 \\
\mbox{[Ti/Fe]} \\
4292.0--4320.0 & $+0.452$ & 0.011 \\
4386.0--4420.0 & $+0.461$ & 0.010 \\
4440.0--4474.0 & $+0.402$ & 0.010 \\
4532.0--4574.0 & $+0.332$ & 0.011 \\
4587.0--4593.0 & $+0.683$ & 0.026 \\
4650.0--4715.0 & $+0.212$ & 0.011 \\
4750.0--4850.0 & $+0.371$ & 0.011 \\
4980.0--5045.0 & $+0.171$ & 0.005 \\
\mbox{[Cr/Fe]} \\
4250.0--4292.0 & $-0.087$ & 0.015 \\
4350.0--4400.0 & $-0.167$ & 0.016 \\
4520.0--4660.0 & $-0.018$ & 0.011 \\
5235.0--5330.0 & $-0.087$ & 0.011 \\
5342.0--5351.0 & $-0.218$ & 0.025 \\
5407.0--5413.0 & $-0.248$ & 0.030 \\
\mbox{[Mn/Fe]} \\
4750.0--4790.0 & $-0.337$ & 0.010 \\
6010.0--6030.0 & $-0.468$ & 0.025 \\
\mbox{[Ba/Fe]} \\
4551.0--4560.0 & $+0.313$ & 0.016 \\
4929.0--4939.0 & $+0.061$ & 0.020 \\
5849.0--5859.0 & $+0.112$ & 0.030 \\
6135.0--6145.0 & $+0.132$ & 0.025 \\
\end{longtable}
}

\longtab[6]{
\begin{longtable}{lcc}
\caption{\label{tab:abN7078} Individual abundance measurements for NGC~7078 (standard analysis)}\\
\hline\hline
Wavelength [\AA] & Value & Error \\ \hline
\endfirsthead
\caption{continued.}\\
\hline\hline
Wavelength [\AA] & Value & Error \\ \hline
\endhead
\hline
\endfoot
\mbox{[Fe/H]} \\
4200.0--4400.0 & $-2.405$ & 0.005 \\
4400.0--4600.0 & $-2.376$ & 0.011 \\
4600.0--4800.0 & $-2.435$ & 0.011 \\
4800.0--5000.0 & $-2.425$ & 0.005 \\
5000.0--5150.0 & $-2.315$ & 0.005 \\
5250.0--5400.0 & $-2.376$ & 0.006 \\
5400.0--5600.0 & $-2.425$ & 0.011 \\
5600.0--5800.0 & $-2.415$ & 0.015 \\
6000.0--6200.0 & $-2.325$ & 0.011 \\
\mbox{[Na/Fe]} \\
5677.0--5695.0 & $+0.154$ & 0.051 \\
\mbox{[Mg/Fe]} \\
4347.0--4357.0 & $+0.074$ & 0.045 \\
4565.0--4576.0 & $+0.414$ & 0.035 \\
4700.0--4707.0 & $+0.093$ & 0.026 \\
5523.0--5531.5 & $+0.135$ & 0.020 \\
5705.0--5715.0 & $+0.414$ & 0.060 \\
\mbox{[Ca/Fe]} \\
4222.0--4232.0 & $+0.305$ & 0.016 \\
4280.0--4320.0 & $+0.324$ & 0.021 \\
4420.0--4460.0 & $+0.295$ & 0.016 \\
4575.0--4591.0 & $+0.183$ & 0.036 \\
4873.0--4883.0 & $+0.535$ & 0.055 \\
5259.0--5268.0 & $+0.483$ & 0.020 \\
5580.0--5610.0 & $+0.364$ & 0.011 \\
6100.0--6175.0 & $+0.284$ & 0.010 \\
\mbox{[Sc/Fe]} \\
4290.0--4330.0 & $+0.364$ & 0.020 \\
4350.0--4440.0 & $+0.215$ & 0.020 \\
4665.0--4675.0 & $+0.234$ & 0.041 \\
5026.0--5036.0 & $+0.003$ & 0.040 \\
5521.0--5531.0 & $+0.114$ & 0.041 \\
5638.0--5690.0 & $+0.154$ & 0.025 \\
\mbox{[Ti/Fe]} \\
4292.0--4320.0 & $+0.525$ & 0.015 \\
4386.0--4420.0 & $+0.483$ & 0.016 \\
4440.0--4474.0 & $+0.414$ & 0.016 \\
4532.0--4574.0 & $+0.384$ & 0.011 \\
4587.0--4593.0 & $+0.715$ & 0.036 \\
4650.0--4715.0 & $+0.414$ & 0.021 \\
4750.0--4850.0 & $+0.445$ & 0.021 \\
4980.0--5045.0 & $+0.234$ & 0.016 \\
5152.5--5160.0 & $+0.464$ & 0.041 \\
\mbox{[Cr/Fe]} \\
4250.0--4292.0 & $-0.216$ & 0.020 \\
4350.0--4400.0 & $-0.325$ & 0.041 \\
4520.0--4660.0 & $-0.216$ & 0.015 \\
5235.0--5330.0 & $-0.085$ & 0.020 \\
5342.0--5351.0 & $-0.176$ & 0.041 \\
5407.0--5413.0 & $-0.115$ & 0.040 \\
\mbox{[Mn/Fe]} \\
4750.0--4790.0 & $-0.387$ & 0.030 \\
6010.0--6030.0 & $-0.125$ & 0.075 \\
\mbox{[Ba/Fe]} \\
4551.0--4560.0 & $+0.465$ & 0.025 \\
4929.0--4939.0 & $+0.375$ & 0.020 \\
5849.0--5859.0 & $+0.525$ & 0.041 \\
6135.0--6145.0 & $+0.465$ & 0.030 \\
\end{longtable}
}

\longtab[7]{
\begin{longtable}{lcc}
\caption{\label{tab:abN7099} Individual abundance measurements for NGC~7099 (standard analysis)}\\
\hline\hline
Wavelength [\AA] & Value & Error \\ \hline
\endfirsthead
\caption{continued.}\\
\hline\hline
Wavelength [\AA] & Value & Error \\ \hline
\endhead
\hline
\endfoot
\mbox{[Fe/H]} \\
4200.0--4400.0 & $-2.475$ & 0.011 \\
4400.0--4600.0 & $-2.416$ & 0.005 \\
4600.0--4800.0 & $-2.436$ & 0.005 \\
4800.0--5000.0 & $-2.466$ & 0.006 \\
5000.0--5150.0 & $-2.336$ & 0.005 \\
5250.0--5400.0 & $-2.376$ & 0.011 \\
5400.0--5600.0 & $-2.426$ & 0.005 \\
5600.0--5800.0 & $-2.336$ & 0.016 \\
6000.0--6200.0 & $-2.297$ & 0.011 \\
\mbox{[Na/Fe]} \\
5677.0--5695.0 & $+0.230$ & 0.046 \\
6149.0--6166.0 & $+0.250$ & 0.155 \\
\mbox{[Mg/Fe]} \\
4347.0--4357.0 & $-0.099$ & 0.055 \\
4565.0--4576.0 & $+0.479$ & 0.041 \\
4700.0--4707.0 & $+0.309$ & 0.026 \\
5523.0--5531.5 & $+0.230$ & 0.025 \\
5705.0--5715.0 & $+0.441$ & 0.051 \\
\mbox{[Ca/Fe]} \\
4222.0--4232.0 & $+0.121$ & 0.015 \\
4280.0--4320.0 & $+0.380$ & 0.016 \\
4420.0--4460.0 & $+0.301$ & 0.011 \\
4575.0--4591.0 & $+0.121$ & 0.036 \\
4873.0--4883.0 & $+0.380$ & 0.051 \\
5259.0--5268.0 & $+0.370$ & 0.025 \\
5580.0--5610.0 & $+0.410$ & 0.010 \\
6100.0--6175.0 & $+0.280$ & 0.010 \\
\mbox{[Sc/Fe]} \\
4290.0--4330.0 & $+0.320$ & 0.021 \\
4350.0--4440.0 & $+0.121$ & 0.021 \\
4665.0--4675.0 & $+0.030$ & 0.056 \\
5026.0--5036.0 & $+0.030$ & 0.041 \\
5521.0--5531.0 & $+0.179$ & 0.041 \\
5638.0--5690.0 & $+0.221$ & 0.025 \\
\mbox{[Ti/Fe]} \\
4292.0--4320.0 & $+0.470$ & 0.011 \\
4386.0--4420.0 & $+0.431$ & 0.010 \\
4440.0--4474.0 & $+0.380$ & 0.011 \\
4532.0--4574.0 & $+0.291$ & 0.005 \\
4587.0--4593.0 & $+0.561$ & 0.051 \\
4650.0--4715.0 & $+0.301$ & 0.021 \\
4750.0--4850.0 & $+0.470$ & 0.016 \\
4980.0--5045.0 & $+0.291$ & 0.010 \\
5152.5--5160.0 & $+0.380$ & 0.041 \\
\mbox{[Cr/Fe]} \\
4250.0--4292.0 & $-0.210$ & 0.020 \\
4350.0--4400.0 & $-0.579$ & 0.051 \\
4520.0--4660.0 & $-0.200$ & 0.016 \\
5235.0--5330.0 & $+0.010$ & 0.016 \\
5342.0--5351.0 & $-0.129$ & 0.025 \\
5407.0--5413.0 & $-0.149$ & 0.041 \\
\mbox{[Mn/Fe]} \\
4750.0--4790.0 & $-0.339$ & 0.020 \\
6010.0--6030.0 & $-0.489$ & 0.126 \\
\mbox{[Ba/Fe]} \\
4551.0--4560.0 & $+0.301$ & 0.030 \\
4929.0--4939.0 & $-0.089$ & 0.030 \\
5849.0--5859.0 & $-0.079$ & 0.045 \\
6135.0--6145.0 & $-0.030$ & 0.035 \\
\end{longtable}
}

\section{Abundance measurements for modified procedures}

\longtab[1]{
\begin{table*}
\caption{Abundance measurements using stars in slit scan areas. For each abundance ratio, the second
 line lists the weighted r.m.s. and the number of individual measurements.}
\label{tab:abun1}
\centering
\begin{tabular}{l r r r r r r r}
\hline\hline
 & NGC 104  & NGC 362  & NGC 6254  & NGC 6388  & NGC 6752  & NGC 7078  & NGC 7099 \\
\hline
\mbox{[Fe/H]}  & $-0.853$ \phantom{(9)}  & $-1.288$ \phantom{(9)}  & $-1.576$ \phantom{(9)}  & $-0.570$ \phantom{(9)}  & $-1.726$ \phantom{(9)}  & $-2.397$ \phantom{(9)}  & $-2.397$ \phantom{(9)} \\
 \; rms$_w$ (N)   & $0.037$ (9)  & $0.041$ (9)  & $0.027$ (9)  & $0.036$ (9)  & $0.041$ (9)  & $0.040$ (9)  & $0.057$ (9) \\
\mbox{[Na/Fe]}  & $0.421$ \phantom{(2)}  & $-0.002$ \phantom{(2)}  & $0.006$ \phantom{(2)}  & $0.376$ \phantom{(2)}  & $0.259$ \phantom{(2)}  & $0.162$ \phantom{(2)}  & $0.226$ \phantom{(2)} \\
 \; rms$_w$ (N)   & $0.007$ (2)  & $0.014$ (2)  & $0.012$ (2)  & $0.006$ (2)  & $0.046$ (2)  & \ldots\ (1)  & $0.003$ (2) \\
\mbox{[Mg/Fe]}  & $0.459$ \phantom{(5)}  & $0.211$ \phantom{(5)}  & $0.412$ \phantom{(5)}  & $0.143$ \phantom{(5)}  & $0.347$ \phantom{(5)}  & $0.181$ \phantom{(5)}  & $0.284$ \phantom{(5)} \\
 \; rms$_w$ (N)   & $0.076$ (5)  & $0.105$ (5)  & $0.069$ (5)  & $0.094$ (5)  & $0.107$ (5)  & $0.127$ (5)  & $0.147$ (5) \\
\mbox{[Ca/Fe]}  & $0.414$ \phantom{(7)}  & $0.276$ \phantom{(8)}  & $0.395$ \phantom{(8)}  & $0.243$ \phantom{(7)}  & $0.344$ \phantom{(8)}  & $0.336$ \phantom{(8)}  & $0.300$ \phantom{(8)} \\
 \; rms$_w$ (N)   & $0.108$ (7)  & $0.139$ (8)  & $0.140$ (8)  & $0.178$ (7)  & $0.112$ (8)  & $0.060$ (8)  & $0.101$ (8) \\
\mbox{[Sc/Fe]}  & $0.209$ \phantom{(6)}  & $0.078$ \phantom{(6)}  & $0.085$ \phantom{(6)}  & $0.068$ \phantom{(6)}  & $0.149$ \phantom{(6)}  & $0.194$ \phantom{(6)}  & $0.193$ \phantom{(6)} \\
 \; rms$_w$ (N)   & $0.174$ (6)  & $0.143$ (6)  & $0.068$ (6)  & $0.183$ (6)  & $0.059$ (6)  & $0.108$ (6)  & $0.106$ (6) \\
\mbox{[Ti/Fe]}  & $0.369$ \phantom{(8)}  & $0.313$ \phantom{(9)}  & $0.374$ \phantom{(9)}  & $0.228$ \phantom{(8)}  & $0.394$ \phantom{(8)}  & $0.405$ \phantom{(9)}  & $0.364$ \phantom{(9)} \\
 \; rms$_w$ (N)   & $0.115$ (8)  & $0.090$ (9)  & $0.100$ (9)  & $0.122$ (8)  & $0.095$ (8)  & $0.090$ (9)  & $0.077$ (9) \\
\mbox{[Cr/Fe]}  & $-0.068$ \phantom{(6)}  & $-0.051$ \phantom{(6)}  & $0.097$ \phantom{(6)}  & $0.008$ \phantom{(6)}  & $-0.070$ \phantom{(6)}  & $-0.181$ \phantom{(6)}  & $-0.141$ \phantom{(6)} \\
 \; rms$_w$ (N)   & $0.135$ (6)  & $0.106$ (6)  & $0.415$ (6)  & $0.203$ (6)  & $0.063$ (6)  & $0.068$ (6)  & $0.132$ (6) \\
\mbox{[Mn/Fe]}  & $-0.241$ \phantom{(2)}  & $-0.416$ \phantom{(2)}  & $-0.436$ \phantom{(2)}  & $-0.132$ \phantom{(2)}  & $-0.360$ \phantom{(2)}  & $-0.340$ \phantom{(2)}  & $-0.351$ \phantom{(2)} \\
 \; rms$_w$ (N)   & $0.090$ (2)  & $0.032$ (2)  & $0.128$ (2)  & $0.131$ (2)  & $0.029$ (2)  & $0.091$ (2)  & $0.033$ (2) \\
\mbox{[Ba/Fe]}  & $0.153$ \phantom{(4)}  & $0.229$ \phantom{(4)}  & $0.333$ \phantom{(4)}  & $0.194$ \phantom{(4)}  & $0.314$ \phantom{(4)}  & $0.402$ \phantom{(4)}  & $0.038$ \phantom{(4)} \\
 \; rms$_w$ (N)   & $0.072$ (4)  & $0.043$ (4)  & $0.063$ (4)  & $0.234$ (4)  & $0.091$ (4)  & $0.055$ (4)  & $0.179$ (4) \\
\hline
\end{tabular}
\end{table*}
}

\longtab[2]{
\begin{table*}
\caption{Abundance measurements using the Kurucz line list as of 18 Feb 2016. For each abundance ratio, the second
 line lists the weighted r.m.s. and the number of individual measurements.}
\label{tab:abun2}
\centering
\begin{tabular}{l r r r r r r r}
\hline\hline
 & NGC 104  & NGC 362  & NGC 6254  & NGC 6388  & NGC 6752  & NGC 7078  & NGC 7099 \\
\hline
\mbox{[Fe/H]}  & $-0.882$ \phantom{(9)}  & $-1.110$ \phantom{(9)}  & $-1.508$ \phantom{(9)}  & $-0.521$ \phantom{(9)}  & $-1.900$ \phantom{(9)}  & $-2.394$ \phantom{(9)}  & $-2.405$ \phantom{(9)} \\
 \; rms$_w$ (N)   & $0.053$ (9)  & $0.036$ (9)  & $0.047$ (9)  & $0.030$ (9)  & $0.045$ (9)  & $0.040$ (9)  & $0.047$ (9) \\
\mbox{[Na/Fe]}  & $0.424$ \phantom{(2)}  & $-0.030$ \phantom{(2)}  & $-0.016$ \phantom{(2)}  & $0.354$ \phantom{(2)}  & $0.322$ \phantom{(2)}  & $0.158$ \phantom{(2)}  & $0.255$ \phantom{(2)} \\
 \; rms$_w$ (N)   & $0.010$ (2)  & $0.009$ (2)  & $0.000$ (2)  & $0.005$ (2)  & $0.023$ (2)  & \ldots\ (1)  & $0.007$ (2) \\
\mbox{[Mg/Fe]}  & $0.539$ \phantom{(5)}  & $0.249$ \phantom{(5)}  & $0.427$ \phantom{(5)}  & $0.191$ \phantom{(5)}  & $0.442$ \phantom{(5)}  & $0.242$ \phantom{(5)}  & $0.334$ \phantom{(5)} \\
 \; rms$_w$ (N)   & $0.097$ (5)  & $0.167$ (5)  & $0.070$ (5)  & $0.130$ (5)  & $0.071$ (5)  & $0.099$ (5)  & $0.151$ (5) \\
\mbox{[Ca/Fe]}  & $0.411$ \phantom{(7)}  & $0.302$ \phantom{(8)}  & $0.421$ \phantom{(8)}  & $0.250$ \phantom{(7)}  & $0.363$ \phantom{(8)}  & $0.349$ \phantom{(8)}  & $0.305$ \phantom{(8)} \\
 \; rms$_w$ (N)   & $0.145$ (7)  & $0.183$ (8)  & $0.175$ (8)  & $0.249$ (7)  & $0.133$ (8)  & $0.061$ (8)  & $0.103$ (8) \\
\mbox{[Sc/Fe]}  & $0.308$ \phantom{(6)}  & $0.221$ \phantom{(6)}  & $0.248$ \phantom{(6)}  & $0.261$ \phantom{(6)}  & $0.198$ \phantom{(6)}  & $0.295$ \phantom{(6)}  & $0.262$ \phantom{(6)} \\
 \; rms$_w$ (N)   & $0.056$ (6)  & $0.052$ (6)  & $0.111$ (6)  & $0.069$ (6)  & $0.065$ (6)  & $0.087$ (6)  & $0.074$ (6) \\
\mbox{[Ti/Fe]}  & $0.393$ \phantom{(8)}  & $0.348$ \phantom{(9)}  & $0.397$ \phantom{(9)}  & $0.284$ \phantom{(8)}  & $0.341$ \phantom{(8)}  & $0.395$ \phantom{(9)}  & $0.375$ \phantom{(9)} \\
 \; rms$_w$ (N)   & $0.116$ (8)  & $0.086$ (9)  & $0.100$ (9)  & $0.120$ (8)  & $0.105$ (8)  & $0.110$ (9)  & $0.080$ (9) \\
\mbox{[Cr/Fe]}  & $-0.043$ \phantom{(6)}  & $-0.062$ \phantom{(6)}  & $0.093$ \phantom{(6)}  & $0.114$ \phantom{(6)}  & $-0.077$ \phantom{(6)}  & $-0.160$ \phantom{(6)}  & $-0.126$ \phantom{(6)} \\
 \; rms$_w$ (N)   & $0.115$ (6)  & $0.130$ (6)  & $0.399$ (6)  & $0.256$ (6)  & $0.023$ (6)  & $0.083$ (6)  & $0.123$ (6) \\
\mbox{[Mn/Fe]}  & $-0.191$ \phantom{(2)}  & $-0.283$ \phantom{(2)}  & $-0.363$ \phantom{(2)}  & $-0.116$ \phantom{(2)}  & $-0.304$ \phantom{(2)}  & $-0.295$ \phantom{(2)}  & $-0.327$ \phantom{(2)} \\
 \; rms$_w$ (N)   & $0.029$ (2)  & $0.095$ (2)  & $0.035$ (2)  & $0.017$ (2)  & $0.000$ (2)  & $0.142$ (2)  & $0.005$ (2) \\
\mbox{[Ba/Fe]}  & $0.220$ \phantom{(4)}  & $0.423$ \phantom{(4)}  & $0.508$ \phantom{(4)}  & $0.233$ \phantom{(4)}  & $0.288$ \phantom{(4)}  & $0.576$ \phantom{(4)}  & $0.148$ \phantom{(4)} \\
 \; rms$_w$ (N)   & $0.098$ (4)  & $0.034$ (4)  & $0.051$ (4)  & $0.074$ (4)  & $0.129$ (4)  & $0.041$ (4)  & $0.141$ (4) \\
\hline
\end{tabular}
\end{table*}
}

\longtab[3]{
\begin{table*}
\caption{Abundance measurements based on theoretical isochrones. For each abundance ratio, the second
 line lists the weighted r.m.s. and the number of individual measurements.}
\label{tab:abun3}
\centering
\begin{tabular}{l r r r r r r r}
\hline\hline
 & NGC 104  & NGC 362  & NGC 6254  & NGC 6388  & NGC 6752  & NGC 7078  & NGC 7099 \\
\hline
\mbox{[Fe/H]}  & $-0.864$ \phantom{(9)}  & $-1.170$ \phantom{(9)}  & $-1.623$ \phantom{(9)}  & $-0.710$ \phantom{(9)}  & $-1.797$ \phantom{(9)}  & $-2.417$ \phantom{(9)}  & $-2.350$ \phantom{(9)} \\
 \; rms$_w$ (N)   & $0.039$ (9)  & $0.044$ (9)  & $0.028$ (9)  & $0.044$ (9)  & $0.038$ (9)  & $0.041$ (9)  & $0.069$ (9) \\
\mbox{[Na/Fe]}  & $0.426$ \phantom{(2)}  & $-0.018$ \phantom{(2)}  & $0.012$ \phantom{(2)}  & $0.403$ \phantom{(2)}  & $0.287$ \phantom{(2)}  & $0.175$ \phantom{(2)}  & $0.219$ \phantom{(2)} \\
 \; rms$_w$ (N)   & $0.009$ (2)  & $0.013$ (2)  & $0.008$ (2)  & $0.025$ (2)  & $0.042$ (2)  & \ldots\ (1)  & $0.002$ (2) \\
\mbox{[Mg/Fe]}  & $0.436$ \phantom{(5)}  & $0.190$ \phantom{(5)}  & $0.370$ \phantom{(5)}  & $0.164$ \phantom{(5)}  & $0.369$ \phantom{(5)}  & $0.177$ \phantom{(5)}  & $0.264$ \phantom{(5)} \\
 \; rms$_w$ (N)   & $0.095$ (5)  & $0.088$ (5)  & $0.082$ (5)  & $0.107$ (5)  & $0.120$ (5)  & $0.115$ (5)  & $0.147$ (5) \\
\mbox{[Ca/Fe]}  & $0.375$ \phantom{(7)}  & $0.207$ \phantom{(8)}  & $0.344$ \phantom{(8)}  & $0.207$ \phantom{(7)}  & $0.361$ \phantom{(8)}  & $0.322$ \phantom{(8)}  & $0.296$ \phantom{(8)} \\
 \; rms$_w$ (N)   & $0.104$ (7)  & $0.136$ (8)  & $0.154$ (8)  & $0.196$ (7)  & $0.105$ (8)  & $0.080$ (8)  & $0.081$ (8) \\
\mbox{[Sc/Fe]}  & $0.230$ \phantom{(6)}  & $0.135$ \phantom{(6)}  & $0.176$ \phantom{(6)}  & $0.105$ \phantom{(6)}  & $0.040$ \phantom{(6)}  & $0.243$ \phantom{(6)}  & $0.138$ \phantom{(6)} \\
 \; rms$_w$ (N)   & $0.148$ (6)  & $0.054$ (6)  & $0.093$ (6)  & $0.143$ (6)  & $0.067$ (6)  & $0.096$ (6)  & $0.103$ (6) \\
\mbox{[Ti/Fe]}  & $0.369$ \phantom{(8)}  & $0.328$ \phantom{(9)}  & $0.392$ \phantom{(9)}  & $0.255$ \phantom{(8)}  & $0.316$ \phantom{(8)}  & $0.409$ \phantom{(9)}  & $0.341$ \phantom{(9)} \\
 \; rms$_w$ (N)   & $0.115$ (8)  & $0.057$ (9)  & $0.105$ (9)  & $0.088$ (8)  & $0.092$ (8)  & $0.113$ (9)  & $0.061$ (9) \\
\mbox{[Cr/Fe]}  & $-0.067$ \phantom{(6)}  & $-0.050$ \phantom{(6)}  & $0.077$ \phantom{(6)}  & $-0.041$ \phantom{(6)}  & $-0.089$ \phantom{(6)}  & $-0.183$ \phantom{(6)}  & $-0.130$ \phantom{(6)} \\
 \; rms$_w$ (N)   & $0.132$ (6)  & $0.101$ (6)  & $0.399$ (6)  & $0.196$ (6)  & $0.068$ (6)  & $0.071$ (6)  & $0.143$ (6) \\
\mbox{[Mn/Fe]}  & $-0.218$ \phantom{(2)}  & $-0.373$ \phantom{(2)}  & $-0.419$ \phantom{(2)}  & $-0.131$ \phantom{(2)}  & $-0.354$ \phantom{(2)}  & $-0.337$ \phantom{(2)}  & $-0.346$ \phantom{(2)} \\
 \; rms$_w$ (N)   & $0.068$ (2)  & $0.024$ (2)  & $0.123$ (2)  & $0.112$ (2)  & $0.039$ (2)  & $0.094$ (2)  & $0.026$ (2) \\
\mbox{[Ba/Fe]}  & $0.178$ \phantom{(4)}  & $0.376$ \phantom{(4)}  & $0.367$ \phantom{(4)}  & $0.193$ \phantom{(4)}  & $0.150$ \phantom{(4)}  & $0.438$ \phantom{(4)}  & $-0.014$ \phantom{(4)} \\
 \; rms$_w$ (N)   & $0.094$ (4)  & $0.043$ (4)  & $0.099$ (4)  & $0.228$ (4)  & $0.096$ (4)  & $0.070$ (4)  & $0.153$ (4) \\
\hline
\end{tabular}
\end{table*}
}

\longtab[4]{
\begin{table*}
\caption{Differences between standard analysis and abundances based on stars in slit scan areas.}
\label{tab:abdiff1}
\centering
\begin{tabular}{l r r r r r r r}
\hline\hline
 & NGC 104  & NGC 362  & NGC 6254  & NGC 6388  & NGC 6752  & NGC 7078  & NGC 7099 \\
\hline
$\Delta_{\mathrm{scn}-\mathrm{std}}$\mbox{[Fe/H]}  & $0.010$  & $-0.212$  & $-0.095$  & $-0.064$  & $0.157$  & $-0.009$  & $-0.001$ \\
$\Delta_{\mathrm{scn}-\mathrm{std}}$\mbox{[Na/Fe]}  & $-0.001$  & $0.043$  & $0.044$  & $0.020$  & $-0.043$  & $0.009$  & $-0.005$ \\
$\Delta_{\mathrm{scn}-\mathrm{std}}$\mbox{[Mg/Fe]}  & $0.017$  & $0.061$  & $0.034$  & $0.035$  & $-0.044$  & $0.004$  & $-0.003$ \\
$\Delta_{\mathrm{scn}-\mathrm{std}}$\mbox{[Ca/Fe]}  & $0.003$  & $0.003$  & $0.015$  & $-0.018$  & $-0.011$  & $0.006$  & $-0.006$ \\
$\Delta_{\mathrm{scn}-\mathrm{std}}$\mbox{[Sc/Fe]}  & $-0.010$  & $-0.039$  & $-0.045$  & $-0.050$  & $0.029$  & $-0.027$  & $0.001$ \\
$\Delta_{\mathrm{scn}-\mathrm{std}}$\mbox{[Ti/Fe]}  & $-0.001$  & $-0.025$  & $-0.038$  & $-0.031$  & $0.052$  & $-0.017$  & $-0.007$ \\
$\Delta_{\mathrm{scn}-\mathrm{std}}$\mbox{[Cr/Fe]}  & $-0.008$  & $-0.026$  & $-0.007$  & $0.027$  & $0.031$  & $0.004$  & $0.003$ \\
$\Delta_{\mathrm{scn}-\mathrm{std}}$\mbox{[Mn/Fe]}  & $-0.012$  & $-0.045$  & $-0.006$  & $0.022$  & $0.006$  & $0.007$  & $-0.007$ \\
$\Delta_{\mathrm{scn}-\mathrm{std}}$\mbox{[Ba/Fe]}  & $-0.002$  & $-0.101$  & $-0.079$  & $-0.006$  & $0.132$  & $-0.034$  & $-0.008$ \\
\hline
\end{tabular}
\end{table*}
}

\longtab[5]{
\begin{table*}
\caption{Differences between standard analysis and abundances based on Kurucz line list. 
}
\label{tab:abdiff2}
\centering
\begin{tabular}{l r r r r r r r}
\hline\hline
 & NGC 104  & NGC 362  & NGC 6254  & NGC 6388  & NGC 6752  & NGC 7078  & NGC 7099 \\
\hline
$\Delta_{\mathrm{kur}-\mathrm{std}}$\mbox{[Fe/H]}  & $-0.019$  & $-0.034$  & $-0.026$  & $-0.015$  & $-0.017$  & $-0.006$  & $-0.009$ \\
$\Delta_{\mathrm{kur}-\mathrm{std}}$\mbox{[Na/Fe]}  & $0.003$  & $0.015$  & $0.022$  & $-0.003$  & $0.020$  & $0.004$  & $0.023$ \\
$\Delta_{\mathrm{kur}-\mathrm{std}}$\mbox{[Mg/Fe]}  & $0.097$  & $0.100$  & $0.049$  & $0.083$  & $0.051$  & $0.065$  & $0.047$ \\
$\Delta_{\mathrm{kur}-\mathrm{std}}$\mbox{[Ca/Fe]}  & $-0.001$  & $0.029$  & $0.041$  & $-0.010$  & $0.009$  & $0.019$  & $-0.001$ \\
$\Delta_{\mathrm{kur}-\mathrm{std}}$\mbox{[Sc/Fe]}  & $0.089$  & $0.104$  & $0.118$  & $0.143$  & $0.079$  & $0.074$  & $0.070$ \\
$\Delta_{\mathrm{kur}-\mathrm{std}}$\mbox{[Ti/Fe]}  & $0.022$  & $0.009$  & $-0.016$  & $0.025$  & $-0.002$  & $-0.027$  & $0.004$ \\
$\Delta_{\mathrm{kur}-\mathrm{std}}$\mbox{[Cr/Fe]}  & $0.016$  & $-0.036$  & $-0.011$  & $0.133$  & $0.025$  & $0.025$  & $0.018$ \\
$\Delta_{\mathrm{kur}-\mathrm{std}}$\mbox{[Mn/Fe]}  & $0.038$  & $0.088$  & $0.067$  & $0.038$  & $0.063$  & $0.053$  & $0.017$ \\
$\Delta_{\mathrm{kur}-\mathrm{std}}$\mbox{[Ba/Fe]}   & $0.066$  & $0.093$  & $0.096$  & $0.033$  & $0.106$  & $0.140$  & $0.103$ \\
\hline
\end{tabular}
\end{table*}
}

\end{appendix}

\end{document}